\documentclass[10pt, journal, twocolumn]{article}

\usepackage{amsmath,amssymb,amsfonts}
\usepackage[pdftex]{graphicx}
\usepackage[font=small]{caption}
\usepackage{algorithm}
\usepackage{algorithmic}
\usepackage{array}
\usepackage[caption=false,font=footnotesize,labelfont=sf,textfont=sf]{subfig}
\usepackage{fixltx2e}
\usepackage{url}
\usepackage{varwidth}
\usepackage{textcomp}
\usepackage{csquotes}
\usepackage{afterpage}
\usepackage{placeins}
\usepackage{hyperref}
\usepackage[top=1in, bottom=1in, left=0.75in, right=0.75in]{geometry}
\usepackage{lipsum}
\usepackage{multirow}
\usepackage{tabularx,booktabs} 
\usepackage{array}
\usepackage{bm}
\usepackage{gensymb}
\usepackage{multirow}
\usepackage{soul}

\usepackage[backend=biber,style=numeric]{biblatex}
\usepackage[latin1]{inputenc}
\usepackage{tikz}
\usetikzlibrary{shapes,arrows}

\graphicspath{{./figures/}}
\usepackage{amsmath}
\usepackage{array}

\usepackage{dblfloatfix}

\usepackage{url}

\newcommand{\reviewermain}[1]{\textcolor{black}{#1}}

\newcommand{\reviewertwo}[1]{\textcolor{black}{#1}}
\newcommand{\reviewerthree}[1]{\textcolor{black}{#1}}
\newcommand{\secondreview}[1]{\textcolor{black}{#1}}

\begin{document}

\tikzstyle{decision} = [diamond, draw, fill=blue!20, 
    text width=4.5em, text badly centered, node distance=3cm, inner sep=0pt]
\tikzstyle{block} = [rectangle, draw, fill=blue!20, 
    text width=5em, text centered, rounded corners, minimum height=4em]
\tikzstyle{line} = [draw, -latex']
\tikzstyle{cloud} = [draw, ellipse,fill=red!20, node distance=3cm,
    minimum height=2em]

\title{Task-specific Performance Prediction and Acquisition Optimization for Anisotropic X-ray Dark-field Tomography}

\author{Theodor~Cheslerean-Boghiu,~
        Franz~Pfeiffer,~
                and~Tobias~Lasser,~
\thanks{T. Cheslerean-Boghiu and T. Lasser are with Computational Imaging and Inverse Problems, Department of Informatics, and Munich Institute of Biomedical Engineering, Technical University of Munich, Germany.}%
\thanks{F. Pfeiffer is with the Chair of Biomedical Physics, Department of Physics, and Munich Institute of Biomedical Engineering, Technical University of Munich, Germany, and with the Department of Diagnostic and Interventional Radiology, School of Medicine and Klinikum rechts der Isar, Technical University of Munich, Germany.}}

\date{}
\maketitle

\begin{abstract}
\textbf{
Anisotropic X-ray Dark-field Tomography (AXDT) is a recently developed imaging modality that enables the visualization of oriented microstructures using lab-based X-ray grating interferometer setups.
While there are very promising application scenarios, for example in materials testing of fibrous composites or in medical diagnosis of brain cell connectivity, AXDT faces challenges in practical applicability due to the complex and time-intensive acquisitions required to fully sample the anisotropic X-ray scattering functions.
However, depending on the specific imaging task at hand, a full sampling may not be required, allowing for reduced acquisitions.
In this work we are investigating a performance prediction approach for AXDT using task-specific detectability indices. 
Based on this approach we present a task-driven acquisition optimization method that enables reduced acquisition schemes while keeping the task-specific image quality high.
We demonstrate the feasibility and efficacy of the method in experiments with simulated and experimental data.}
\end{abstract}

\section{Introduction} \label{section:introduction}
\begin{figure}
\includegraphics[width=\linewidth]{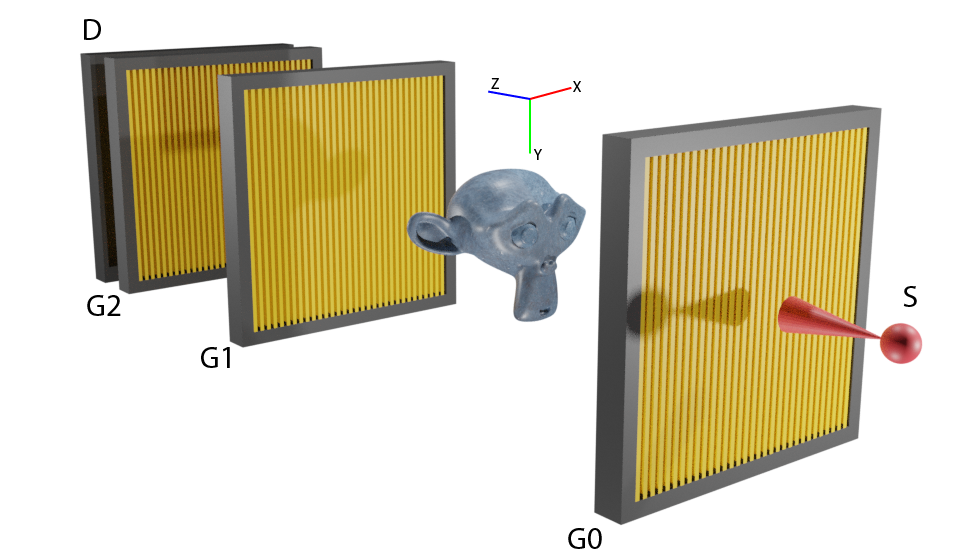}
\caption{Schematic of an X-ray grating interferometer. 
The lab X-ray source S illuminates the sample through the source grating G0, while the phase grating G1 creates an interference pattern that is sampled by the analyzer grating G2 in front of the X-ray detector D.
For Anisotropic X-ray Dark-field Tomography (AXDT), either the sample or the grating interferometer have to be rotated around all three axes in order to fully sample the scattering functions.  
\reviewerthree{This is achieved by placing the sample on an Eulerian cradle to perform the required rotations.}
Figure created in \textit{Blender} \cite{blender}.}
\label{fig:setup}
\end{figure}

\subsection*{Anisotropic X-ray Dark-field Tomography}
Talbot-Lau X-ray grating interferometers using lab X-ray sources (see Fig.~\ref{fig:setup}) allow for simultaneous acquisition of the conventional X-ray absorption contrast alongside the phase contrast \cite{Pfeiffer2006} and the dark-field contrast \cite{Pfeiffer2008}, which relate to the refraction and scattering of X-rays, respectively \reviewertwo{\cite{Momose2003}}.
The X-ray dark-field contrast is of particular interest, as it is induced by ultra-small angle scattering in the measured sample, and as such can reveal microstructures that cannot be directly resolved by the X-ray detector and that would be otherwise invisible.
Another unique property of the dark-field contrast is its directional anisotropy, meaning that the signal changes when the sample is rotated in the plane orthogonal to the incoming X-ray beam.
This anisotropy is particularly prominent in fibrous microstructures, as they cause scattering orthogonal to the fiber orientation, even though the microstructures themselves are too small to be resolved.
As grating interferometers are only sensitive to scattering that is orthogonal to the grating bars, to sample the full scattering function it is necessary to rotate the sample (or the grating interferometer) in all possible orientations, covering a full sphere.
Applying tomographic reconstruction techniques \secondreview{has enabled} the recovery of spherical scattering functions for each three-dimensional volume element in the sample, yielding the imaging modality called \enquote{Anisotropic X-ray Dark-field Tomography}, in short AXDT \cite{Wieczorek2016,Wieczorek2018}.

Imaging of the X-ray dark-field contrast has generated considerable interest in the medical context, for example in lung imaging, ranging from mouse models \cite{Gradl2019,Umkehrer2019}, pig models \cite{Hellbach2018,DeMarco2019}, to deceased human bodies \cite{Willer2018,Fingerle2019}.
The anisotropy of the dark-field contrast has first been used in planar radiographs in a technique called \enquote{X-ray Vector Radiography} \cite{Potdevin:2012is}, for example in the context of the analysis of bone microstructures \cite{Schaff:2014je,Baum:2015et,Jud2017} \reviewertwo{or reinforced carbon fiber materials \cite{Bayer2013}}.

In order to perform not just planar, but three-dimensional imaging of the anisotropic X-ray dark-field contrast, a suitable forward model is required.
One of the early approaches separated the isotropic and anisotropic parts of the dark-field signal \cite{Bayer2013, Bayer:2014ew}, restricting the anisotropic component to a vectorial entity instead of the full scattering function.
The first approach using Gaussian scattering functions, \enquote{X-ray Tensor Tomography}, was developed by our group, where the scattering functions were modeled using symmetric rank-2 tensors \cite{Malecki2014, Vogel2015}, with potential use-cases demonstrated for dental imaging \cite{Jud2016} and fibrous composite materials \cite{Sharma2016}. 
\reviewertwo{A similar approach has recently been developed by Felsner \textit{et al.} \cite{Felsner2019},  modeling the scattering profiles using 3D Gaussian functions.}
\secondreview{To overcome the limitations of rank-2 tensors, our group} derived a continuous model employing spherical functions to model the full scattering profiles \cite{Wieczorek2016}.
Using spherical harmonics for discretization and a column-block inverse problem, the new imaging modality \secondreview{was termed} \enquote{Anisotropic X-ray Dark-field Tomography}, or in short AXDT. 
Potential applications range from defect detection in fibrous composite materials \cite{Sharma2017} to diagnosis of neurodegenerative diseases through imaging brain connectivity \cite{Wieczorek2018}.

\subsection*{Acquisition Trajectories for AXDT}

A standard circular acquisition trajectory, as is typically used in conventional absorption X-ray computed tomography, would only measure one direction of scattering with fixed grating bars, and would only allow tomographic reconstruction of that particular scattering direction.
Hence, an ideal acquisition trajectory for AXDT should measure every direction of the scattering by rotating the sample (or the grating interferometer) in every possible direction with respect to the X-ray beam, and acquire a circular trajectory for each of those directions.
Such an ideal trajectory would be very complex and very time consuming to perform, while also administering a high radiation dose.

To increase the practical usefulness of AXDT, our group has studied global approaches to reduce the overall acquisition complexity \cite{Sharma2016, Sharma2017}.
By introducing a coverage metric in \cite{Sharma2016}, the overall quality of a trajectory can be judged, enabling the design of trajectories that use only two axes of rotation instead of three with only a slight degradation in imaging performance.
In \cite{Sharma2017}, we introduced a new method of designing trajectories using t-designs to sample scattering directions and generating circular orbits for each direction.
Using the coverage metric and null space analysis of the AXDT forward operator, we could show a five-fold reduction in measurement time while keeping comparable image quality by using diagonal grating alignment with a specifically designed trajectory.

\subsection*{\reviewerthree{Task-based Acquisition Optimization for AXDT}}

\reviewerthree{Such global optimization approaches, as introduced in \cite{Sharma2016,Sharma2017}}, are valid for any kind of sample, yielding nearly constant image quality for every region of the reconstructed object.
However, in many cases imaging is performed aiming at a specific task, such as \reviewerthree{lesion detection and lesion discrimination in a region of interest \cite{Richard2008}, not requiring high image quality in every region of the sample.}
Acquisition trajectories geared to a specific task could hence allow a further reduction of acquisition complexity and administered radiation dose.

Previous works in conventional absorption X-ray computed tomography have already shown that task-specific trajectories which incorporate prior knowledge of the sample are very advantageous \reviewerthree{over pre-defined ones which do not take in account the imaging task.}
In a materials testing context, it was shown that using prior knowledge from CAD models, task-specific trajectories with a drastically reduced amount of projections could be employed to reconstruct specific features, such as welded joints, with high quality \cite{Fischer2016}.
For an interventional medical context, task-specific orbits incorporating prior knowledge \reviewerthree{about the location and shape of metallic surgical devices obtained from diagnostic scans demonstrated a reduction in streak artifacts around the region of interest. \cite{Stayman2013, Stayman2015, Stayman2019}.}

\reviewerthree{For the imaging modality AXDT we propose in this work a similar concept of using prior knowledge about the sample and the imaging task at hand to compute an image quality metric, in this case a detectability index \cite{Gang2011}, to guide an acquisition optimization algorithm \cite{Fischer2016}.
The prior knowledge about the sample could come from a previous high-quality acquisition, while the imaging task will be the accurate visualization or detection of specific known features, i.e. a signal known exactly, background known exactly (SKE/BKE) task.}

\subsection*{\reviewerthree{Task-specific Performance Prediction for AXDT}}

\reviewerthree{We aim to predict the performance of an AXDT acquisition by the performance of a model observer in a binary hypothesis testing framework.
In this work, we introduce a detectability index for AXDT as a mathematical model for predicting task-based imaging performance based on prior knowledge of the sample.  
To compute the detectability index, we employ a non-prewhitening matched filter observer (NPWM), which has shown good performance in similar settings in conventional absorption X-ray computed tomography \cite{Gang2011, Fischer2016}.}

\reviewerthree{However, a NPWM observer does not accurately reflect human observer performance due to its underlying strategy of applying a template determined by the difference between the two hypotheses (signal present versus signal absent) without regard for the character of the background \cite{BarrettMyers:FIS,Barrett:98,Richard2008}.}
\reviewerthree{This problem could be overcome using a channelized hotelling observer (CHO) \cite{Barrett:90, Brankov2013}, which has been shown to be more consistent with the human visual system \cite{BarrettMyers:FIS,Barrett:95,Barrett:06}.
However, due to the spherical function-valued nature of AXDT and its novelty, there is currently no established human observer performance data available yet.
Hence, in this work, we have settled on a NPWM observer to compute the detectability index for AXDT.}

\subsection*{\reviewerthree{Our contribution}}

\reviewerthree{Building on this proposed detectability index using a NPWM observer model, we introduce a \reviewermain{greedy algorithm to generate optimized, task-specific acquisition trajectories for AXDT.}}
We demonstrate the efficacy of the algorithm in both a simulation study and an experimental study of a short fiber moulding part.
This work is building on our previous reports on task-specific trajectories in AXDT \cite{Boghiu2018,Boghiu2019}, with a comprehensive introduction on resolution properties of AXDT, as well as an improved detectability index and a much faster and more robust algorithm using sorted batches.

\section{Methods} \label{section:system_model}
In this section we first briefly \secondreview{recapitulate} the AXDT forward model and the resulting inverse problem \secondreview{from our previous work \cite{Wieczorek2016}}.
Then we introduce resolution properties for performance prediction in AXDT, which \reviewerthree{enables the definition of our proposed task-specific detectability index}.
Finally, we introduce our proposed algorithm for optimized task-specific trajectories in AXDT.

\subsection{The discrete AXDT forward model}
In AXDT we aim to reconstruct a field of spherical functions $\eta: \mathbb{S}_2 \times \mathbb{R}^3 \to \mathbb{R}$ from a set of dark-field measurements $\bm{d}=(d_i)$, $i=1,\ldots,I$, measured by an X-ray grating interferometer (see Figure~\ref{fig:setup}). Using real-valued spherical harmonics, we represent $\eta(\cdot, x)$ by the spherical harmonics coefficients $\eta_k^m(x)$ for $x\in\mathbb{R}^3$, where $k$ is the degree and $m$ the order of the respective spherical harmonics basis function.
To model a dark-field measurement $d_i$, we denote the corresponding X-ray path $L_i$ with direction $l_i\in\mathbb{S}_2$ and the corresponding grating sensitivity $s_i\in\mathbb{S}_2$.
The discretized forward model developed in our previous work \cite{Wieczorek2016} then reads
\begin{equation}
	d_i \approx \\ \exp\left( -\frac{1}{4\pi} \sum_{k=0}^4 \sum_{m=-k}^{k} h_k^m(s_i,l_i) \int_{L_i} \eta_k^m(x)dx \right),
\label{eq:forwardmodel}
\end{equation}
where $h_k^m:\mathbb{S}_2\times\mathbb{S}_2\to\mathbb{R}$ denotes the spherical harmonics coefficients of the weighting function $h:\mathbb{S}_2\times\mathbb{S}_2\times\mathbb{S}_2\to\mathbb{R}$ describing the interaction process of X-rays with the sample.

Discretizing our volume of interest into $J$ cubic voxels, we denote the discretized spherical harmonics coefficients as $\bm{\eta}_k^m\in\mathbb{R}^J$.
We formulate the system matrix $P\in\mathbb{R}^{I\times J}$ using the discretized line integrals for all the dark-field measurements $d_i$, $i=1,\ldots,I$, and summarize the $h_k^m/4\pi$ into a diagonal weighting matrix $W_k^m\in\mathbb{R}^{I\times I}$.
Then the fully discrete forward model reads
\begin{equation}
	\bm{d} \approx \exp\left(-\sum_{k=0}^4 \sum_{m=-k}^{k} W_k^m P \bm{\eta}_k^m\right),
\label{eq:forwardmodel2}
\end{equation}
for more details please see  \cite{Wieczorek2016}.
Finally, using $\bm{\eta}:=\big(\bm{\eta}_0^0,\ldots,\bm{\eta}_4^{-4},\ldots,\bm{\eta}_4^4\big)$ and $\bm{\mathcal{B}}:=\sum_{k=0}^4\sum_{-k}^k W_k^mP$, we summarize the discrete AXDT forward model as
\begin{equation}
    \bm{d} \approx \exp\big(-\bm{\mathcal{B}} \bm{\eta}\big).
\label{eq:forwardmodel3}
\end{equation}

\subsection{AXDT inverse problem}
\reviewertwo{The AXDT inverse problem is, like any tomographic inverse problem, ill-posed. 
To stabilize the problem, we introduce a penalized likelihood function using the forward model~(\ref{eq:forwardmodel3}),}
\begin{equation}
	L(\bm{\eta}, \bm{d}) := \frac{1}{2}\big\|\exp\big(-\bm{\mathcal{B}}\bm{\eta}\big) - \bm{d}\big\|_{2}^{2} + \frac{\beta}{2} R(\bm{\eta}),
\label{eq:likelihood}
\end{equation}
where $ \beta > 0 $, and $R(\bm{\eta})$ represents a quadratic roughness penalty \reviewertwo{\cite{Fessler1996srp,Fessler1996mav}}:
\begin{eqnarray}
R(\bm{\eta}) &:=& \sum_{k=0}^4 \sum_{m=-k}^{k} R(\bm{\eta}_{k}^{m}) \\
R(\bm{\eta}_{k}^{m}) &:=& \frac{1}{2}\sum_{i=1}^J\sum_{j=1}^Jw_{ij}\phi({[\bm{\eta}_{k}^{m}}]_{i}-{[\bm{\eta}_{k}^{m}}]_{j})
\end{eqnarray}
for $ k = 0,\dots,4$, and $ |m| \le k $, with $ \phi(x) = \frac{x^2}{2} $ a symmetric convex function. 
The weights $w_{ij}$ are \reviewertwo{the typical quadratic regularization weights \cite{Fessler1995mid,Zhang2018}} given by
\reviewertwo{
\begin{equation}
w_{ij} =
	\begin{cases}
		1 & \text{for first-order neighbors} \\
		\frac{1}{\sqrt{2}} & \text{for second-order neighbors} \\
		\frac{1}{\sqrt{3}} & \text{for third-order neighbors} \\
		0 & \text{otherwise,}
	\end{cases}
\end{equation}
where voxel $j$ is considered a \enquote{first-order neighbor} of voxel $j$ if their sides touch, a \enquote{second-order neighbor} if only their edges touch and a \enquote{third-order neighbor} if only their corners touch \cite{Fessler1996srp}.}

Given a realization $ \hat{\bm{d}} $ of dark-field measurements, we define the matching reconstruction $ \hat{\bm{\eta}} $ as the minimizer of the likelihood function,
\begin{equation} \label{eq:minimization}
	\hat{\bm{\eta}} := \underset{\bm{\eta}}{\arg\,\min}\, L(\bm{\eta},\hat{\bm{d}}).
\end{equation}

\subsection{Resolution properties} \label{section:resproperties}
Let $ \bm{\hat{d}} \in \mathbb{R}^I $ be a noisy dark-field measurement and let $ \bm{\hat{\eta}} = (\bm{\hat{\eta}}_k^m) \text{ , } \bm{\hat{\eta}}_k^m\in\mathbb{R}^J $, be a minimizer of the likelihood function $ L(\bm{\eta},\bm{\hat{d}}) $ as in eq.~(\ref{eq:minimization}). 
Assuming that $ \bm{\hat{\eta}} $ is unique for every $ \bm{\hat{d}} $, we seek to analyze \reviewertwo{the local resolution properties of the minimizer from eq.~(\ref{eq:minimization}).
We use the local impulse response $\lambda_j$, which describes the relative change in the reconstructed image given a small local perturbation in the imaged sample at location $j$ \cite{Fessler1996srp}. 
Moreover, the noise characteristics of the imaging system can be predicted by analyzing the covariance matrix of the estimator $\sigma_j$ at location $j$.}

\reviewertwo{For the penalized likelihood function as in eq.~(\ref{eq:likelihood}), we formulate the predictors for} the local impulse response $ \lambda_{j}: \mathbb{S}_2 \times \mathbb{R}^J \rightarrow \mathbb{S}_2 \times \mathbb{R}^J $ and the local covariance $ \sigma_j: \mathbb{S}_2 \times \mathbb{R}^J \rightarrow \mathbb{S}_2 \times \mathbb{R}^J $ as a function of $\hat{\bm{d}}$ and $\hat{\bm{\eta}}$,
\begin{equation} \label{eq:impulse1}
	\lambda_{j}(\bm{\hat{\eta}}) = [-\nabla^{20}L(\bm{\hat{\eta}}, \bm{\hat{d}}) ]^{-1} \nabla^{11}L(\bm{\hat{\eta}}, \bm{\hat{d}} )\frac{\partial}{\partial \bm{\eta_j}}\bm{\hat{d}},
\end{equation}
\begin{multline} \label{eq:covariance1}
	\sigma_j(\bm{\hat{\eta}}) \approx [-\nabla^{20}L(\bm{\hat{\eta}}, \bm{\hat{d}})]^{-1}[\nabla^{11}L(\bm{\hat{\eta}}, \bm{\hat{d}} )] \,\sigma(\bm{\hat{d}} ) \\ [\nabla^{11}L(\bm{\hat{\eta}}, \bm{\hat{d}} )]^{T}[-\nabla^{20}L(\bm{\hat{\eta}}, \bm{\hat{d}} )]^{-1} e_j^{\star},
\end{multline}
where $ e_{j}^{\star} \in \mathbb{S}_2 \times \mathbb{R}^J $ is a spherical perturbation (or impulse) modeling a scattering function at voxel $ j $. 

The derivatives $ \nabla^{20} =  \frac{\partial^{2}}{\partial \bm{\hat{\eta}}^2}, \nabla^{11} = \frac{\partial^{2}}{\partial \bm{\hat{\eta}} \partial \bm{\hat{d}}} $ are computed using eq.~(\ref{eq:likelihood}) and the abbreviation 
$ \varepsilon(\bm{\hat{\eta}}) = \exp(-\bm{\mathcal{B}} \bm{\hat{\eta}}) $,
\begin{eqnarray*}
\label{eq:nabla2}
\nabla^{20} L(\bm{\hat{\eta}}, \bm{\hat{d}}) &=& 2 \bm{\mathcal{B}}^{T} \varepsilon(\bm{\hat{\eta}}) \big(2\varepsilon(\bm{\hat{\eta}}) - \bm{\hat{d}}\big) \bm{\mathcal{B}}  - \beta \bm{\mathcal{R}}(\bm{\hat{\eta}}), \\
\label{eq:nabla3}
\nabla^{11} L(\bm{\hat{\eta}}, \bm{\hat{d}}) &=& 2 \bm{\mathcal{B}}^{T} \varepsilon(\bm{\hat{\eta}}), 
\end{eqnarray*}

with $ \bm{\mathcal{R}}(\bm{\hat{\eta}}) := \nabla^{20} R(\bm{\hat{\eta}}) $.
In total we receive
\begin{multline} \label{eq:impulse2}
	\lambda_{j}(\bm{\hat{\eta}}) = \big[-2\bm{\mathcal{B}}^{T}\varepsilon(\bm{\hat{\eta}})\big(2\varepsilon(\bm{\hat{\eta}}) - \hat{\bm{d}}\big)\bm{\mathcal{B}}e_{j}^{\star}  - \beta \bm{\mathcal{R}}(\bm{\hat{\eta}}) e_{j}^{\star}\big]^{-1} \\ 
	-2\bm{\mathcal{B}}^{T}\varepsilon(\bm{\hat{\eta}})\varepsilon(\bm{\hat{\eta}})\bm{\mathcal{B}}e_{j}^{\star},
\end{multline}
\begin{multline} \label{eq:covariance2}
	\sigma_j(\bm{\hat{\eta}}) \approx \big[-2\bm{\mathcal{B}}^{T} \varepsilon(\bm{\hat{\eta}})\big(2\varepsilon(\bm{\hat{\eta}}) - \hat{\bm{d}}\big) \bm{\mathcal{B}}e_{j}^{\star}  - \beta \bm{\mathcal{R}}(\bm{\hat{\eta}}) e_{j}^{\star}\big]^{-1} \\
	\big[2\bm{\mathcal{B}}^{T} \epsilon(\bm{\hat{\eta}})\big] \bm{\hat{d}} \big[2 \bm{\mathcal{B}}^{T} \epsilon(\bm{\hat{\eta}})\big]^{T} \\
	\big[-2\bm{\mathcal{B}}^{T} \varepsilon(\bm{\hat{\eta}}) \big(2\varepsilon(\bm{\hat{\eta}}) - \hat{\bm{d}}\big) \bm{\mathcal{B}}e_{j}^{\star}  - \beta \bm{\mathcal{R}}(\bm{\hat{\eta}}) e_{j}^{\star}\big]^{-1}e_j^{\star}.
\end{multline}

\subsection{Detectability index}
\label{sec:detindex}
Using the resolution properties from section~\ref{section:resproperties}, we can compute a detectability index $\delta_j^2$, which provides an estimate of how well a frequency template of a user-defined task ($ \mathcal{W}_{\text{ROI}} \in \mathbb{S}_2 \times \mathbb{C}^J $) can be discriminated from the noise in a penalized likelihood reconstruction. 
We use a non-prewhitening matched filter observer (NPWM) to compute the detectability index, \reviewerthree{which performs well in certain binary detection scenarios \cite{Fischer2016,Stayman2015,Gang2011},}
\begin{equation} \label{eq:detindex}
\delta_{j}^{2} = \frac {\big[\int_{\mathbb{R}^J}\int_{\mathbb{S}^2}(\text{MTF}_j(\bm{\hat{\eta}} ) \cdot \mathcal{W}_{\text{ROI}})^{2}\ dS dV\big]^{2}}{\int_{\mathbb{R}^J}\int_{\mathbb{S}^2} (\text{MTF}_j(\bm{\hat{\eta}} ) \cdot \mathcal{W}_{\text{ROI}})^{2} \cdot \text{NPS}_j(\bm{\hat{\eta}} )\,dS dV}
\end{equation}
where MTF is the modulation transfer function and NPS represents the noise power spectrum, which are estimated as outlined below. 
The integration is first done over the Fourier domain of the spherical harmonics coefficients and then over the spatial domain of the volume. 
We compute the MTF and NPS by taking the Fourier transform of the linear impulse response $\lambda_j$ in eq.~(\ref{eq:impulse2}) and the local covariance $\sigma_j$ in eq.~(\ref{eq:covariance2}), as in \cite{Boghiu2018}. 
Assuming local space invariance for both resolution properties, we can use a circulant approximation \cite{Stayman2004,Boghiu2018},
\reviewertwo{
\begin{multline}
\text{MTF}_{j}(\bm{\hat{\eta}}) = | \mathcal{F} \{ \lambda_{j}(\bm{\hat{\eta}}) \} | \approx \\ 
	\frac
	{ | \mathcal{F}\{-2\bm{\mathcal{B}}^T \varepsilon(\bm{\hat{\eta}})^2\bm{\mathcal{B}} e_{j}^{\star}\} | }
	{ | \mathcal{F}\{-2\bm{\mathcal{B}}^T (\varepsilon(\bm{\hat{\eta}}) \big( 2\varepsilon(\bm{\hat{\eta}})-\bm{\hat{d}}\big) \bm{\mathcal{B}}e_{j}^{\star}  - \beta \bm{\mathcal{R}}(\bm{\hat{\eta}}) e_{j}^{\star} \} | } ,
\label{eq:mtf}
\end{multline}
\begin{multline}
\text{NPS}_{j}(\bm{\hat{\eta}}) = | \mathcal{F} \{ \sigma_{j}(\bm{\hat{\eta}}) \} | \approx \\
	\frac
	{ | \mathcal{F}\{4\bm{\mathcal{B}}^T\mbox{diag}\big(\varepsilon(\bm{\hat{\eta}})\bm{\hat{d}}\varepsilon(\bm{\hat{\eta}})\big)\bm{\mathcal{B}}e_{j}^{\star}  \}  | }
	{\Big| \mathcal{F}\{-2\bm{\mathcal{B}}^T \varepsilon(\bm{\hat{\eta}}) \big( 2\varepsilon(\bm{\hat{\eta}})-\bm{\hat{d}}\big) \bm{\mathcal{B}} e_{j}^{\star} - \beta \bm{\mathcal{R}}(\bm{\hat{\eta}}) e_{j}^{\star}\}\Big|^2 },
\label{eq:nps}
\end{multline}
}
where the division is element-by-element, \reviewerthree{$ \mbox{diag}(\cdot) $ is an operator that creates a matrix with its argument on the main diagonal, and $ | \cdot | $ computes the element-wise absolute value of a complex vector.}

\subsection{Prior knowledge}
The computation of the detectability index $\delta_j^2$ in eq.~(\ref{eq:detindex}) relies on the resolution properties in eqs.~(\ref{eq:impulse2}) and (\ref{eq:covariance2}), which require prior knowledge, as they use $ \bm{\hat{\eta}} $, the reconstruction of the imaged sample matching the measured data $ \bm{\hat{d}} $.
In conventional X-ray computed tomography (CT), different methods to obtain that prior knowledge have been employed.
Fischer et al. \cite{Fischer2016}, for example, used a CAD model of the sample in industrial CT applications, while Stayman et al. \cite{Stayman2013} used a high-quality pre-operative scan in interventional CT to provide the set of measurements which encoded the object-dependency of the local impulse response and the local covariance. 

In the case of AXDT, the reconstructed quantity $\bm{\hat{\eta}}$ is a volume of spherical scattering functions discretized using spherical harmonics, which precludes the use of simple CAD models.
In this work, we instead used existing dark-field measurements of the imaged sample (simulated or acquired experimentally) to obtain the respective reconstruction ($ \bm{\hat{\eta}} $), which in turn encoded the required prior knowledge for the computation of the detectability index.

\subsection{Task-driven path optimization}
We employ the detectability index from eq.~(\ref{eq:detindex}) as a fitness metric for a task-driven acquisition optimization algorithm to characterize the imaging performance in a specific region of interest, given a certain data acquisition.
Existing methods for finding optimal task-driven acquisition trajectories rely on greedy-search algorithms \cite{Fischer2016,Stayman2013}, where a set of optimal acquisition poses is iteratively computed from a set of existing poses. 
In each iteration, a detectability index is computed for each potential individual pose to be added to the acquisition trajectory, and the one with the highest respective detectability index is chosen and added to the optimized acquisition trajectory. 
The continuous re-computation of the detectability index in each iteration for each potential pose is extremely computationally intensive, in particular for AXDT with its very complex forward model~(\ref{eq:forwardmodel2}).
In our previous work \cite{Boghiu2019}, such computations took two weeks to complete on a high-performance computer, yielding an optimized trajectory of only 100 poses for a strongly downsampled reconstruction problem of $160^3$ voxels.
For realistic scenarios matching conventional CT applications (for example with $>500^3$ voxels and $>2000$ poses), such iterative greedy-search algorithms are currently computationally infeasible.

In this work, we propose an improved algorithm with sorted batches, where at each iteration we select a batch of \reviewerthree{$ bN $ acquisition poses with the highest detectability index (with $b\in(0,1]$ denoting the batch size as a fraction of the expected geometry size $N$), which are then appended to the set of already chosen poses.  
Throughout this work, we will use percentages for $b$ for simplicity, but the actual batch size $bN$ has to be an integer and, therefore, will be computed as $\left \lceil{bN} \right \rceil$.}
We call the proposed method \textit{Accelerated Greedy Search with Sorted Batches} (AGS), the pseudocode is shown in Algorithm~\ref{alg:pathopt}.

\begin{algorithm}
 \caption{\reviewerthree{Accelerated Greedy Search with Sorted Batches: AGS(N, b)}}
  \label{alg:pathopt}
\begin{algorithmic}[1]
\STATE Let $ \mathcal{P}_{\text{all}} $ be the set of all possible poses
\STATE Let $ \mathcal{P}_{\text{sel}} = \emptyset $ be the set of selected poses
\STATE Let $ \mathcal{W}_{\text{ROI}} $ be a user-defined task
\STATE Let $ N $ be the final size of the geometry
\STATE Let $ bN$ be the batch size for $b\in(0,1]$
\WHILE {$|\mathcal{P}_{\text{sel}}| < N  $ } 
	\STATE $ D = \{ \delta^{2}(p \cup \mathcal{P}_{\text{sel}} ,\mathcal{W}_{\text{ROI}}) | \ p \in \mathcal{P}_{\text{all}} \setminus \mathcal{P}_{\text{sel}} \} $
	\STATE Sort $ D $ $ \rightarrow $ $ D_{\text{sort}} $ (in descending order)
	\STATE Take first $ bN$ elements: $  D_{\text{sort}}= D_{\text{sort}} [1:(bN)] $
	\STATE $ \mathcal{P}_{\text{sel}} = \{ p | p \in \mathcal{P}_{\text{all}} \wedge  \delta^{2}(p \cup \mathcal{P}_{\text{sel}} ,\mathcal{W}_{\text{ROI}}) \in D_{\text{sort}} \} \cup  \mathcal{P}_{\text{sel}}$
\ENDWHILE
\RETURN $ \mathcal{P}_{\text{sel}} $
\end{algorithmic}
\end{algorithm}

\section{Experiments and Results} \label{sec:exp}

We first describe an experiment to validate the performance of the proposed task-based detectability index (\ref{eq:detindex}) for the use case of AXDT in section~\ref{sec:detindexmaps} using simulated data.
In particular, we aim to show that the chosen observer model (NPWM), which has only been used in scalar-valued conventional X-ray CT so far, correctly reflects the ability of the system to discriminate signal from noise in our spherical function-valued imaging modalities AXDT.

In the second step we investigate the performance of our proposed batched AGS algorithm, first using simulated data (section~\ref{sec:algosim}) and then using experimental data of a thermoplastic fiber mould (section~\ref{sec:algoexp}).
The settings common to all experiments and the quality metrics employed are outlined in section~\ref{sec:settings}.

\subsection{Experimental Settings} \label{sec:settings}

The forward model in eq.~(\ref{eq:forwardmodel2}) and the reconstruction (\ref{eq:minimization}), as well as the  resolution properties in eqs.~(\ref{eq:mtf}) and (\ref{eq:nps}), the detectability index in eq.~(\ref{eq:detindex}), and the AGS algorithm (Algorithm~\ref{alg:pathopt}) were implemented using C++ in the open-source image reconstruction framework \textit{elsa} \cite{elsa}. 
All computations were performed on a computer equipped with dual Intel Xeon E5-2687W v2 processors and 128 GB of RAM, coupled with two Nvidia GeForce RTX 2080Ti's graphical processing units accelerating the forward- and backward-projection operations. 

All reconstructions throughout this work, whether from simulated or experimental data, were obtained by running \reviewermain{$20$} iterations of a conjugate gradient method on the likelihood function (\ref{eq:likelihood}), guaranteeing a residual error smaller than \reviewermain{$10^{5}$}. 
In all experiments, the regularization parameter $ \beta $ was empirically set to $ 10^{3} $, which ensured smooth enough detectability index maps for both simulations and real-data experiments.

To compare AXDT reconstructions of the same sample acquired using an acquisition trajectory $X$ and a reference trajectory $R$, we extracted the \secondreview{main} microstructure orientation from the respective reconstructed spherical harmonics coefficients as in \cite{Wieczorek2018}, and then used the experimental metric from \cite{Sharma2016} to compare the quality,
\begin{equation} \label{eq:expmetric}
\text{EM}(X) = \frac{1}{J_{ROI}}\sum_{j=1}^{J_{ROI}} \big| \langle U_j(X), U_j(R) \rangle \big|
\end{equation}
where $ J_{ROI} $ is the number of voxels in a region of interest, $ U_j $ the extracted \secondreview{main} microstructure orientation at voxel location $ j = 1, \dots , J_{ROI} $ inside the region of interest, and $ \langle \, , \rangle $ denotes the standard scalar product. 

For the simulations, \secondreview{besides} the fiber extraction step, we directly compared the reconstructed spherical harmonics coefficients of $\bm{\eta}=(\bm{\eta}_k^m)$ with a reference $\bm{\eta}_R=((\bm{\eta}_k^m)_R)$ using the root mean squared error at the location $j$ of one impulse~$e_j^{\star}$:
\begin{equation} \label{eq:rmse}
\text{RMSE}(\bm{\eta},\bm{\eta}_{\text{R}}) = \sqrt{\underset{k,m}{\sum}((\bm{\eta}_k^m)^j - (\bm{\eta}_k^m)_{\text{R}}^j)^2},
\end{equation}
where $ k $ and $ m $ are the order and degree of the spherical harmonics coefficients, respectively.

\begin{figure}
	\begin{varwidth}{0.45\linewidth}		
	  	\centering
		\centerline{\includegraphics[width=4cm,height=4cm]{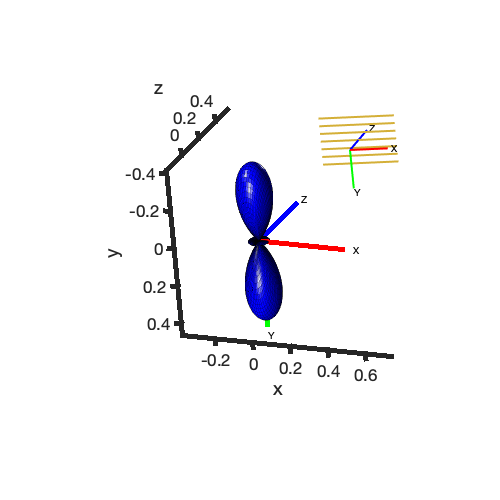}}
  		\centerline{(A)}\medskip
	\end{varwidth}	
	\begin{varwidth}{0.45\linewidth}
  		\centering
  		\centerline{\includegraphics[width=4cm,height=4cm]{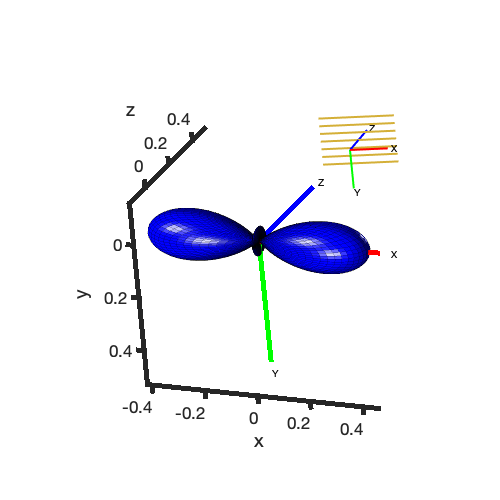}}
  		\centerline{(B)}\medskip
	\end{varwidth}
	\\
	\begin{varwidth}{0.45\linewidth}		
	  	\centering
		\centerline{\includegraphics[width=4cm,height=4cm]{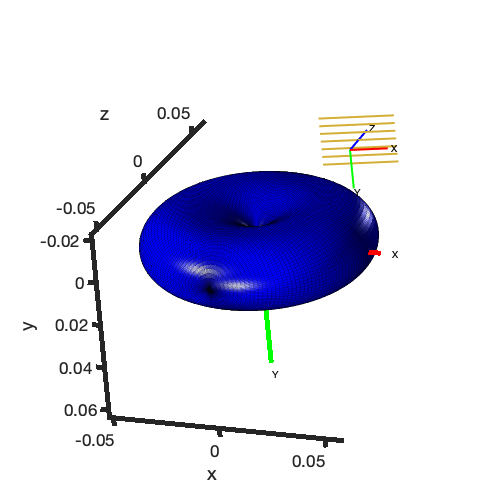}}
  		\centerline{(C)}\medskip
	\end{varwidth}	
	\begin{varwidth}{0.45\linewidth}
  		\centering
  		\centerline{\includegraphics[width=4cm,height=4cm]{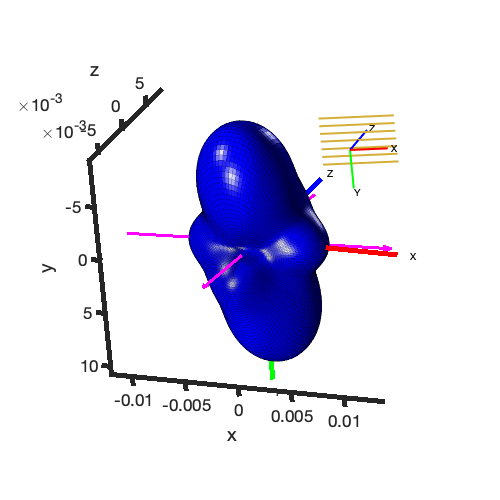}}
  		\centerline{\secondreview{(D)}}\medskip
	\end{varwidth}
	\\
	\caption{Examples of spherical functions. The grating orientation relative to the used coordinate system is given by the golden bars in the top right of each diagram; it is fixed while the sample is rotated around its axes. }
	\label{fig:impulse}
\end{figure}

\subsection{Detectability index validation} \label{sec:detindexmaps}

\begin{figure}[h]
	\begin{varwidth}{0.48\linewidth}		
	  	\centering
		\centerline{\includegraphics[width=4cm,height=2cm]{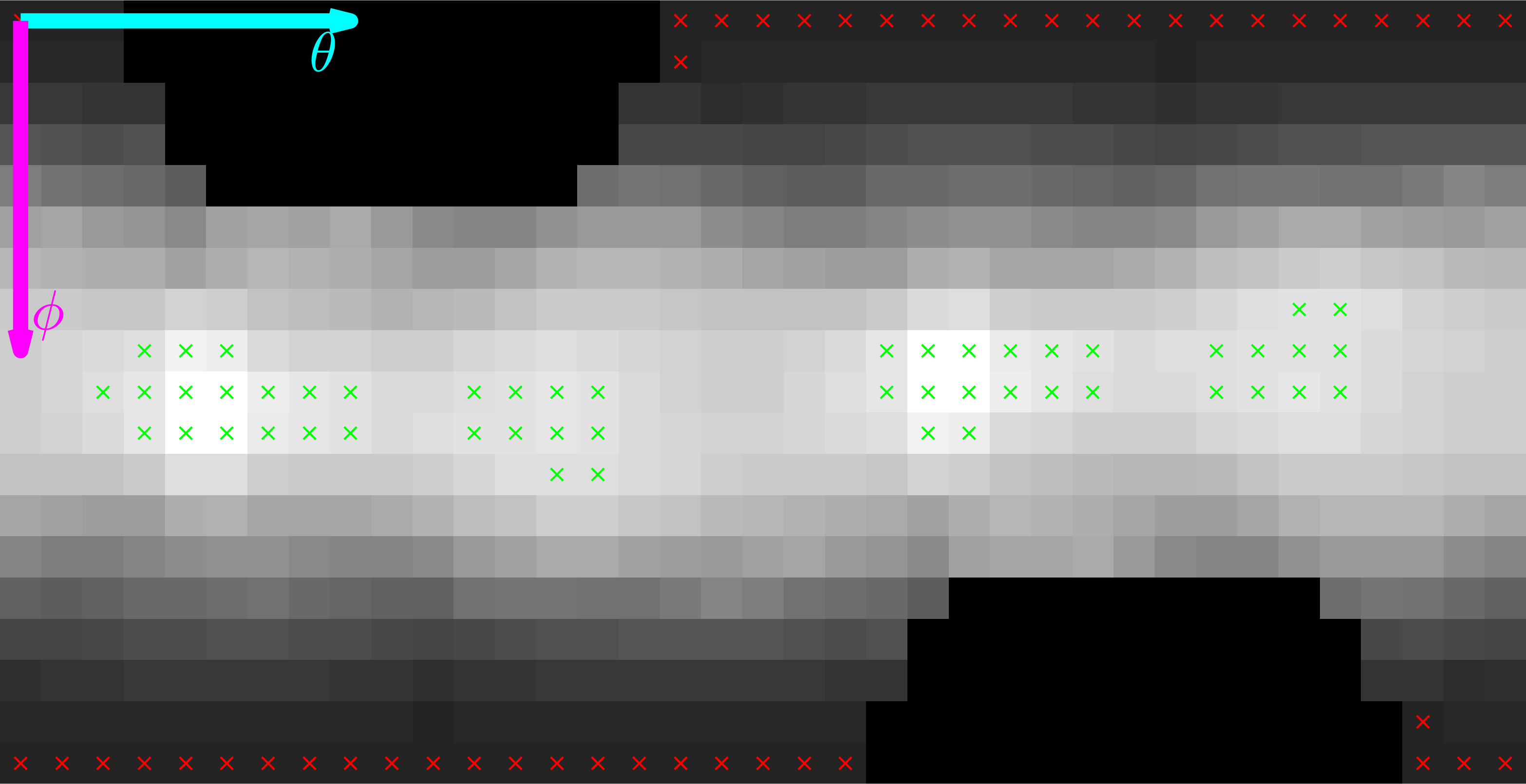}}
  		\centerline{(A)}\medskip
	\end{varwidth}
	\begin{varwidth}{0.48\linewidth}		
	  	\centering
		\centerline{\includegraphics[width=4cm,height=2cm]{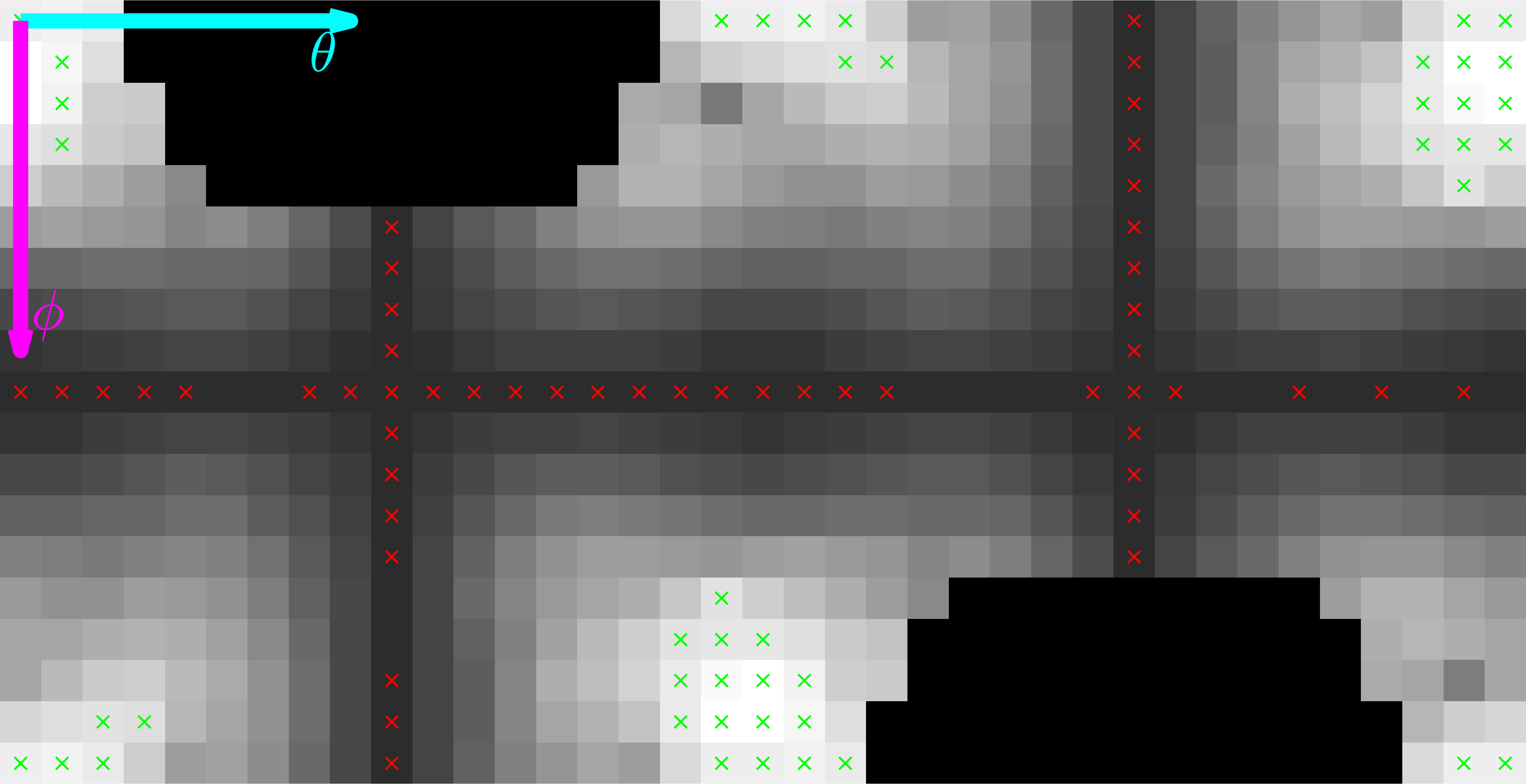}}
  		\centerline{(B)}\medskip
	\end{varwidth}
	\\	
	\begin{varwidth}{0.48\linewidth}		
	  	\centering
		\centerline{\includegraphics[width=4cm,height=2cm]{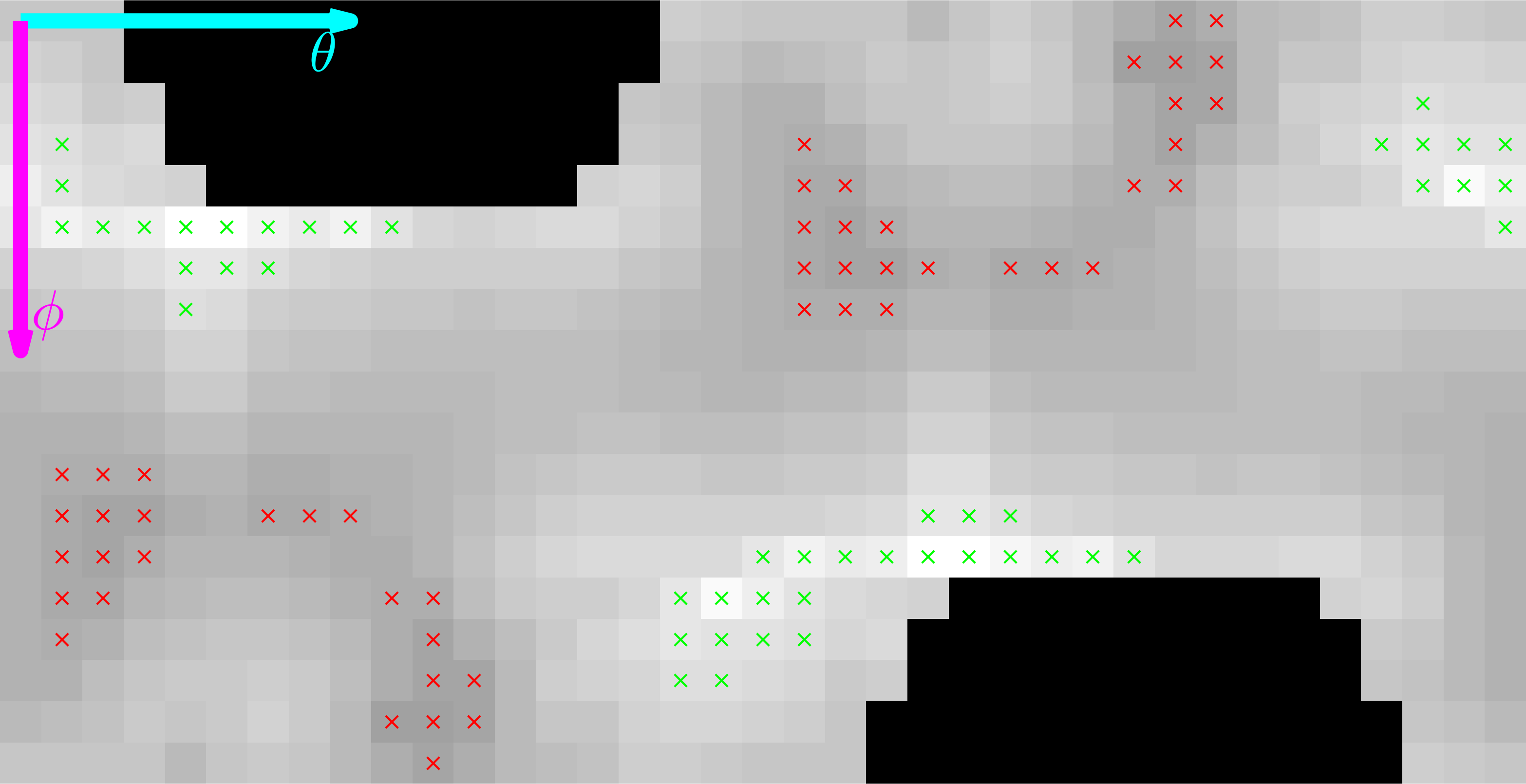}}
		\centerline{(C)}\medskip
	\end{varwidth}
	\begin{varwidth}{0.48\linewidth}		
	  	\centering
		\centerline{\includegraphics[width=4cm,height=2cm]{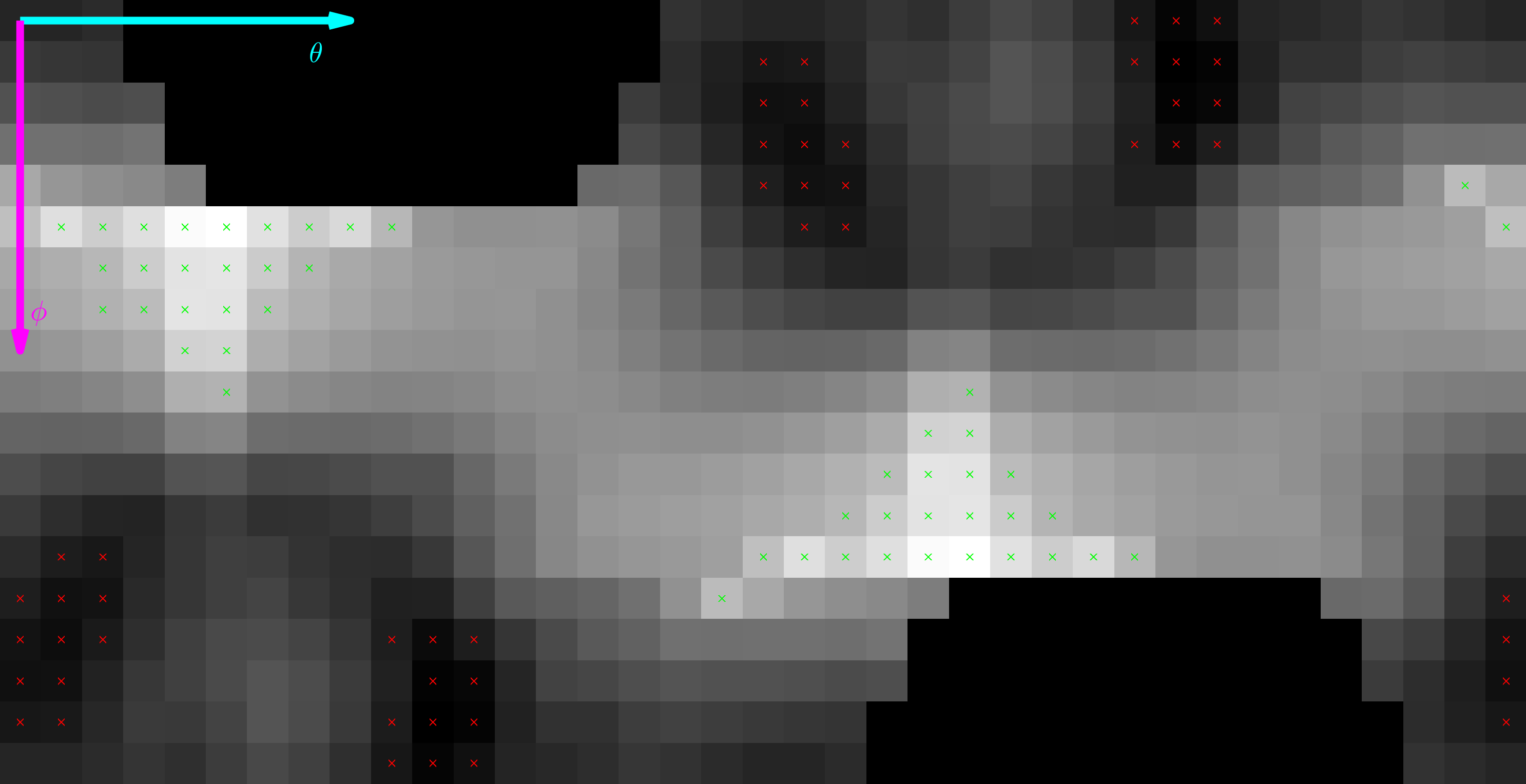}}
		\centerline{\secondreview{(D)}}\medskip
	\end{varwidth}
	\caption{Detectability index maps of four simulated phantoms consisting of one spherical function each, with the same function also acting as the task template, see Fig.~\ref{fig:impulse}. The detectability index was computed using exactly one acquisition pose corresponding to an azimuth-inclination pair $(\theta,\phi)$. Highlighted in each map are $ 100 $ poses with the lowest (red crosses) and highest (green crosses) detectability index value.}
	\label{fig:detindex_angle_map}
\end{figure}

To confirm that the proposed detectability index is valid for AXDT, we simulated measurements of a simple phantom consisting of one spherical function placed exactly in the middle of the volume, while everything else was set to~$0$.
The volume used $ 40^3 $ isotropic voxels of size $1$, with a matching detector of $40^2$ pixels of size $1$ in a parallel-beam setting.

We considered four different spherical functions, as depicted in Fig.~\ref{fig:impulse}, for a total of four phantoms, and in each case used the same spherical function located at the same spot again as the task template to compute the detectability index.
\secondreview{In Fig.~\ref{fig:impulse}, impulses (A) and (B) are simple spherical Dirac impulses, while (C) and (D) relate to more complex scattering profiles from experimental data.
(D) in particular corresponds to a case representing two fiber orientations at once, as indicated by the pink arrows.}
We computed a detectability index map for each of the four phantoms, each time using exactly one acquisition pose corresponding to an azimuth-inclination angle pair $ (\theta,\phi)$, where we allowed all possible combinations of $\theta = [0\degree,10\degree,...,360\degree]$ and $ \phi = [0\degree,10\degree,...,180\degree]$,  with a fixed grating orientation parallel to the $x$-axis.
The results are shown in Fig.~\ref{fig:detindex_angle_map}.

The 100 acquisition poses with the lowest and highest detectability index have been marked in the maps using red and green crosses, respectively.
The two black areas in each of the detectability index maps reflect areas which our experimental setup cannot measure due to limited rotational freedom (for more details see \cite{Sharma2017}), hence we also ignored them in this simulation study.

\begin{table}[h]
\resizebox{0.5\textwidth}{!}{
\begin{tabular}{| c | c | c |}
\hline
\begin{minipage}{0.48\linewidth}
\includegraphics[width=\linewidth]{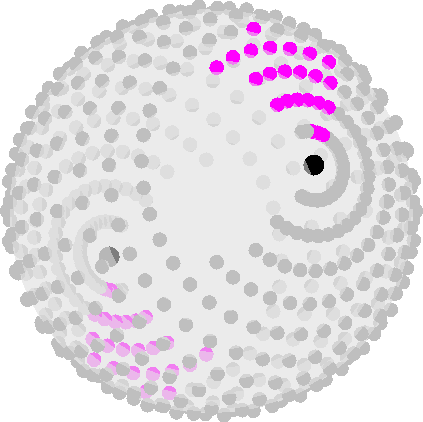}

\end{minipage}&

\begin{minipage}{0.48\linewidth}
\includegraphics[width=\linewidth]{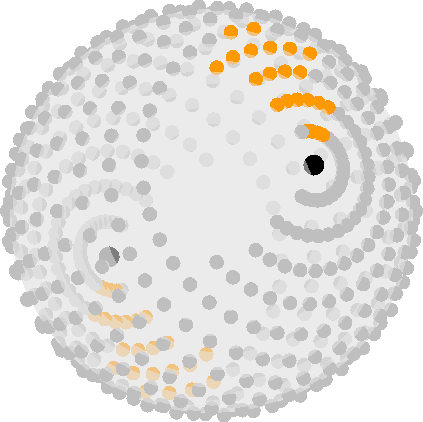}

\end{minipage}&

\begin{minipage}{0.48\linewidth}
\includegraphics[width=\linewidth]{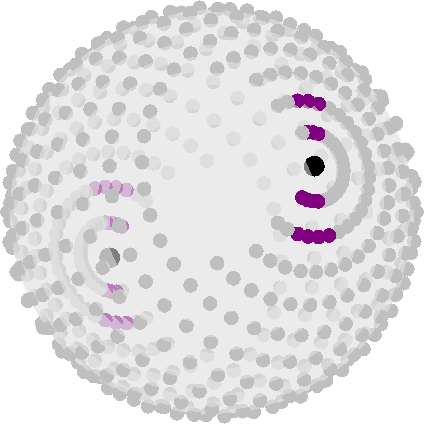} 

\end{minipage}\\
\hline
\begin{minipage}{0.48\linewidth}
\vfill
\centering
\large{$ \text{AGS}(50,10\%) $}
\end{minipage}&
\begin{minipage}{0.48\linewidth}
\vfill
\centering
\large{$ \text{AGS}(50,20\%) $}
\end{minipage} & 
\begin{minipage}{0.48\linewidth}
\vfill
\centering
\large{$ \text{AGS}(50,50\%) $}
\end{minipage} \\
\hline
\begin{minipage}{0.48\linewidth}
\includegraphics[width=\linewidth]{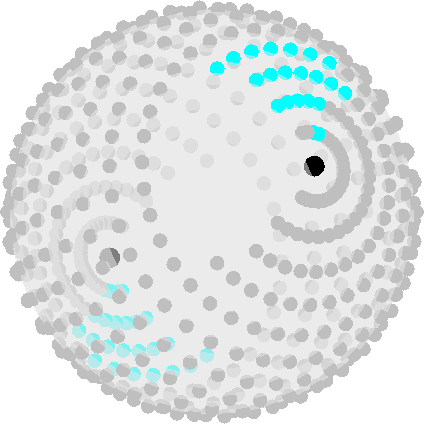}

\end{minipage}&

\begin{minipage}{0.48\linewidth}
\includegraphics[width=\linewidth]{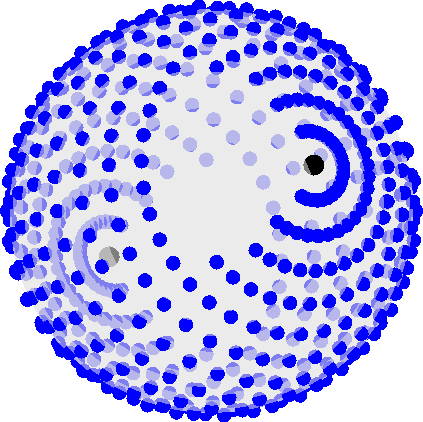} 

\end{minipage}&

\begin{minipage}{0.48\linewidth}
\includegraphics[width=\linewidth]{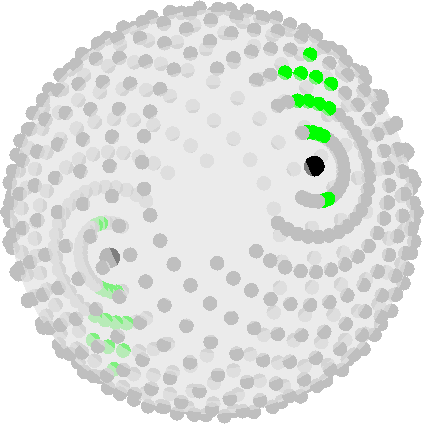} 

\end{minipage}\\
\hline
\begin{minipage}{0.48\linewidth}
\vfill
\centering
\large{$ \text{AGS}(50,2\%) $}
\end{minipage} &
\begin{minipage}{0.48\linewidth}
\vfill
\centering
\large{Reference}
\end{minipage} &
\begin{minipage}{0.48\linewidth}
\vfill
\centering
\large{$ \text{AGS}(50,100\%) $}
\end{minipage}\\
\hline
\begin{minipage}{0.48\linewidth}
\includegraphics[width=\linewidth]{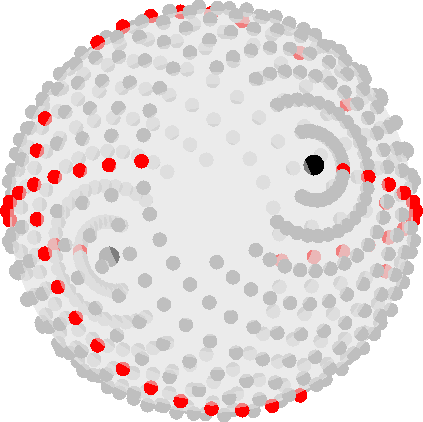} 

\end{minipage}& 
\begin{minipage}{0.48\linewidth}
\includegraphics[width=\linewidth]{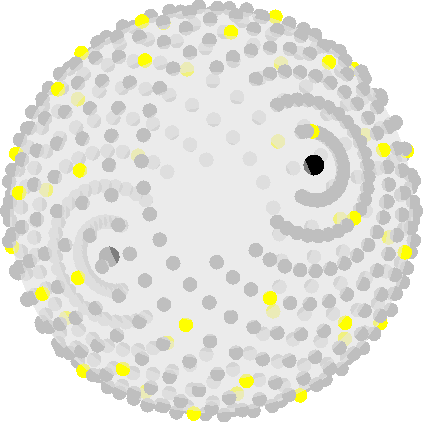} 

\end{minipage}&
\begin{minipage}{0.48\linewidth}
\includegraphics[width=\linewidth]{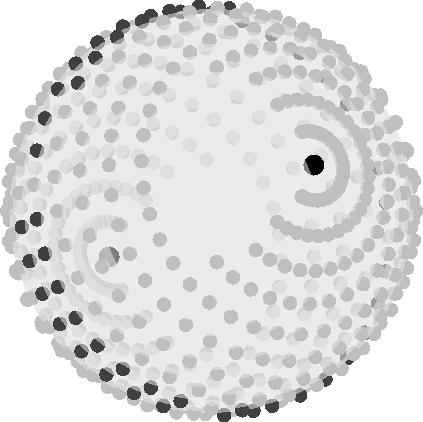} 

\end{minipage}\\
\hline
\begin{minipage}{0.48\linewidth}
\vfill
\centering
\large{Non-optimal}
\end{minipage} &
\begin{minipage}{0.48\linewidth}
\vfill
\centering
\large{t-design}
\end{minipage} &
\begin{minipage}{0.48\linewidth}
\vfill
\centering
\large{Circular}
\end{minipage}\\
\hline
\end{tabular}
}

\caption{Different acquisition trajectories \reviewerthree{ computed for the spherical function (B) from Fig.~\ref{fig:impulse}} containing a subset of poses from the \enquote{Reference} trajectory. 
The north ($ \phi = 0 $) and south ($ \phi = 180 $) poles are marked by black dots on each sphere; they lie on the $y$-axis shown in Table~\ref{table:simulation_reconstruction}. 
The perspective is adjusted for visualization purposes.}
\label{table:simulation_geometry}
\end{table}

\begin{table}[h]
\resizebox{0.5\textwidth}{!}{
\begin{tabular}{| c | c | c |}
\hline
\begin{minipage}{0.48\linewidth}
\includegraphics[width=\linewidth]{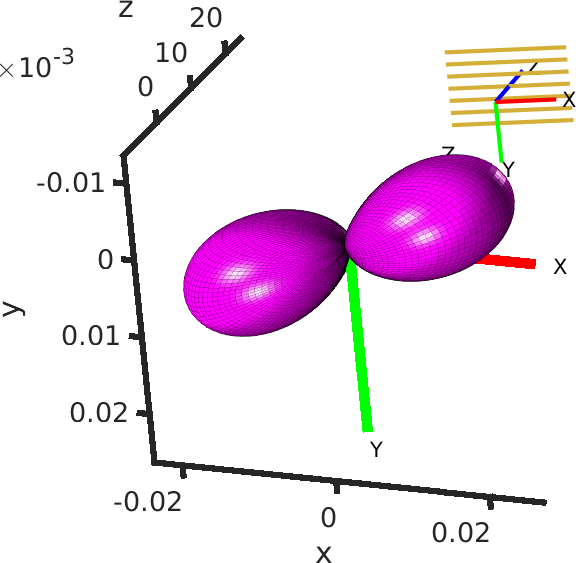}

\end{minipage}&

\begin{minipage}{0.48\linewidth}
\includegraphics[width=\linewidth]{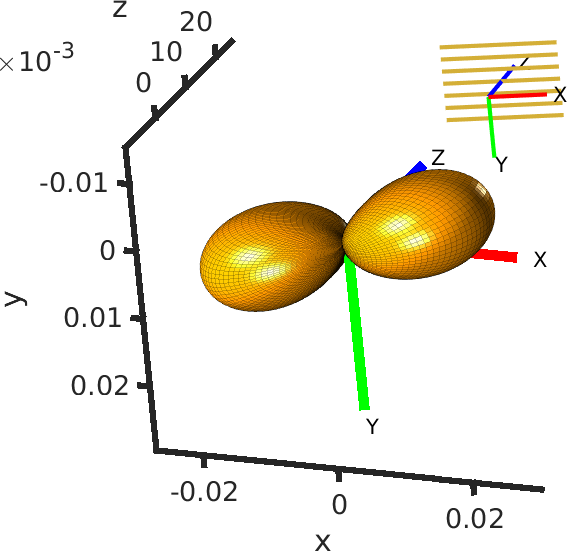}

\end{minipage}&

\begin{minipage}{0.48\linewidth}
\includegraphics[width=\linewidth]{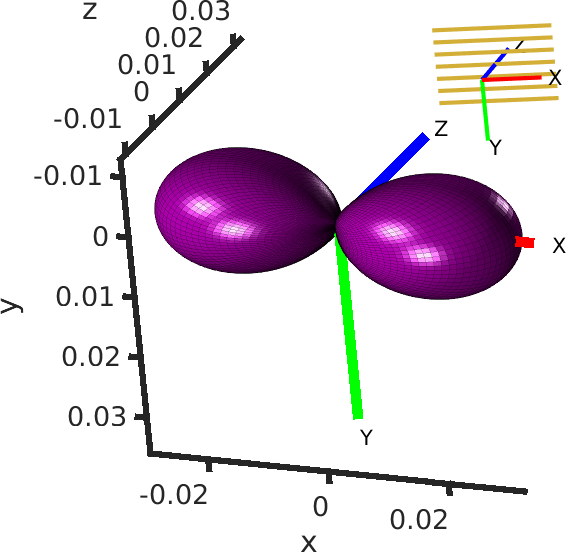} 

\end{minipage}\\
\hline
\begin{minipage}{0.48\linewidth}
\vfill
\centering
\large{$ \text{AGS}(50,10\%) $}
\end{minipage}&
\begin{minipage}{0.48\linewidth}
\vfill
\centering
\large{$ \text{AGS}(50,20\%) $}
\end{minipage} & 
\begin{minipage}{0.48\linewidth}
\vfill
\centering
\large{$ \text{AGS}(50,50\%) $}
\end{minipage} \\
\hline
\begin{minipage}{0.48\linewidth}
\includegraphics[width=\linewidth]{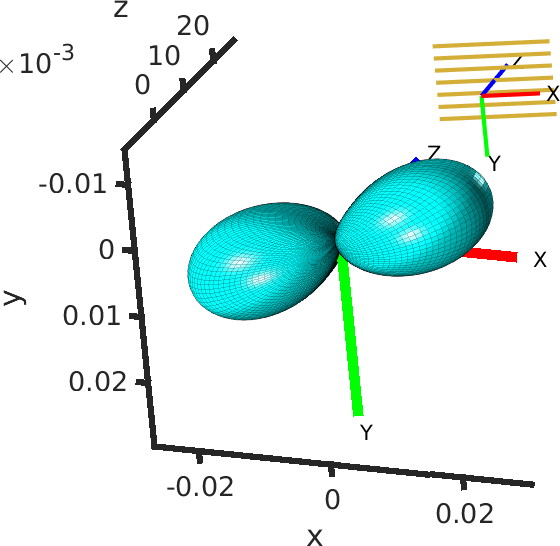}

\end{minipage}&

\begin{minipage}{0.48\linewidth}
\includegraphics[width=\linewidth]{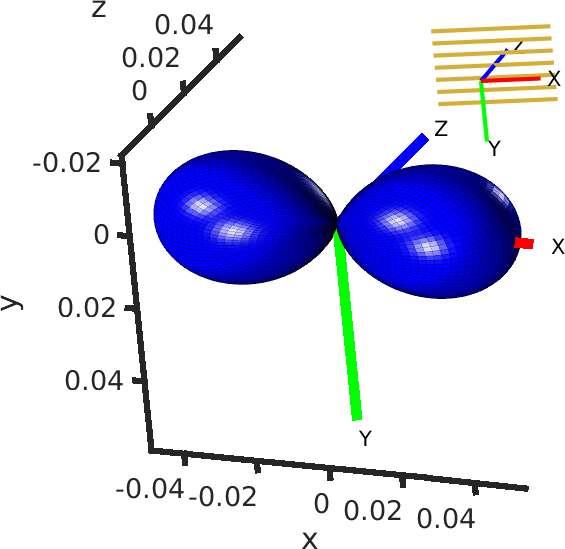} 

\end{minipage}&

\begin{minipage}{0.48\linewidth}
\includegraphics[width=\linewidth]{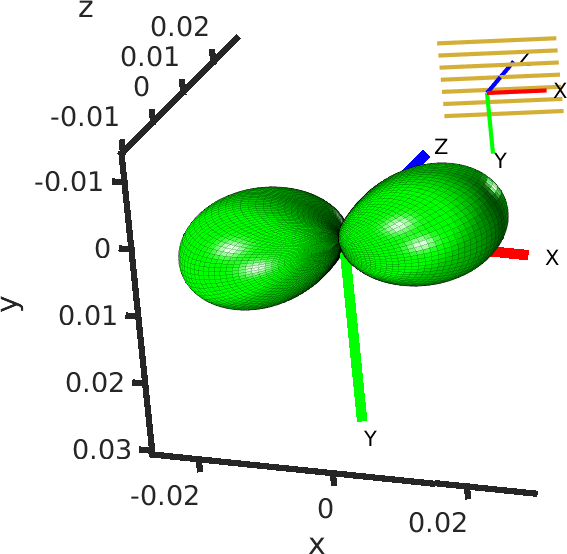} 

\end{minipage}\\
\hline
\begin{minipage}{0.48\linewidth}
\vfill
\centering
\large{$ \text{AGS}(50,2\%) $}
\end{minipage} &
\begin{minipage}{0.48\linewidth}
\vfill
\centering
\large{Reference}
\end{minipage} &
\begin{minipage}{0.48\linewidth}
\vfill
\centering
\large{$ \text{AGS}(50,100\%)  $}
\end{minipage}\\
\hline
\begin{minipage}{0.48\linewidth}
\includegraphics[width=\linewidth]{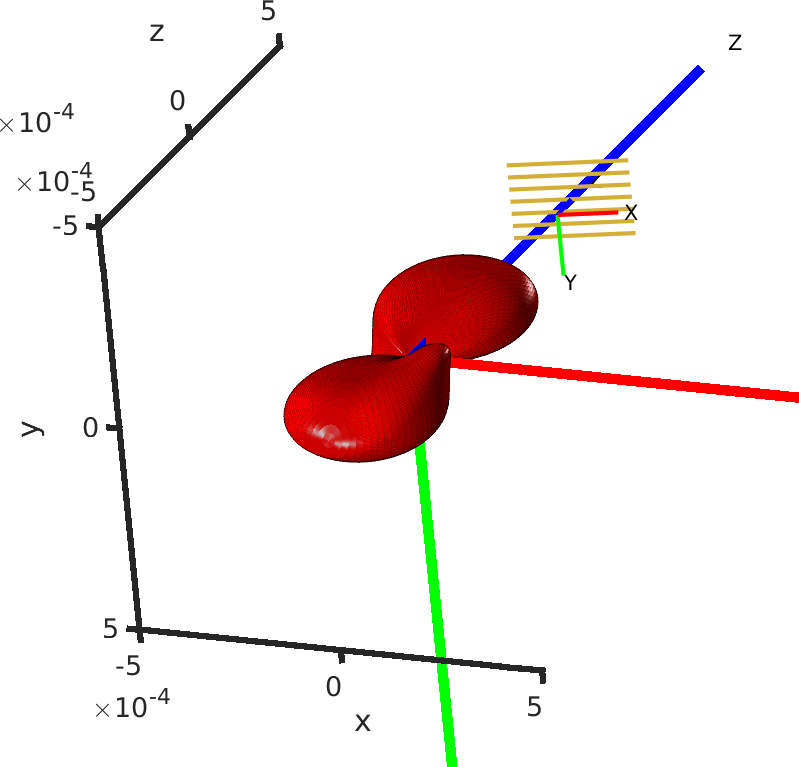} 

\end{minipage}& 
\begin{minipage}{0.48\linewidth}
\includegraphics[width=\linewidth]{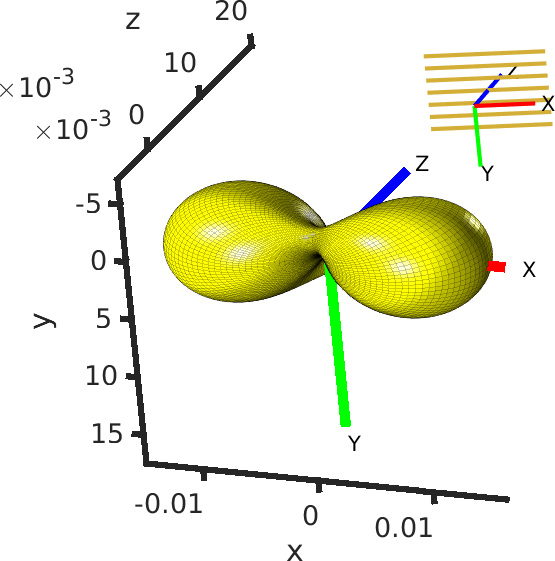} 

\end{minipage}&
\begin{minipage}{0.48\linewidth}
\includegraphics[width=\linewidth]{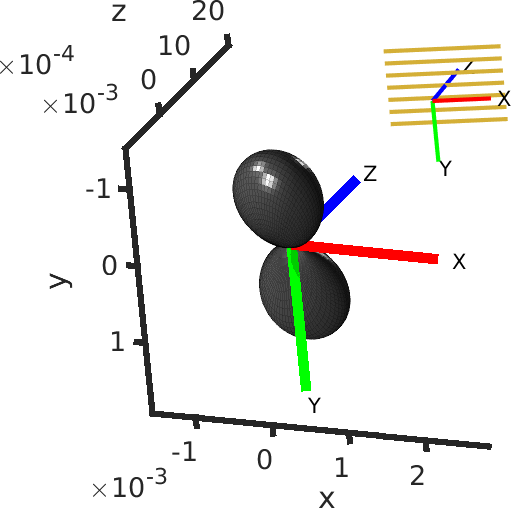} 

\end{minipage}\\
\hline
\begin{minipage}{0.48\linewidth}
\vfill
\centering
\large{Non-optimal}
\end{minipage} &
\begin{minipage}{0.48\linewidth}
\vfill
\centering
\large{t-design}
\end{minipage} &
\begin{minipage}{0.48\linewidth}
\vfill
\centering
\large{Circular}
\end{minipage}\\
\hline
\end{tabular}
}

\caption{Reconstructions of the phantom containing the spherical function (B) from Fig.~\ref{fig:impulse} using simulated measurements of the corresponding trajectories from \ref{table:simulation_geometry}. 
The color matches that of the trajectory. }
\label{table:simulation_reconstruction}
\end{table}

\subsection{Algorithm validation with simulated data} \label{sec:algosim}

We investigated the performance of our proposed algorithm using the phantoms from section~\ref{sec:detindexmaps} corresponding to the spherical functions \reviewerthree{(A) and} (B) in Fig.~\ref{fig:impulse}.
We defined a \enquote{Reference} trajectory containing all possible azimuth-inclination angle pairs $ (\theta,\phi)$, with $\theta = [0\degree,10\degree,...,360\degree]$ and $ \phi = [0\degree,10\degree,...,180\degree]$, minus those pairs that our experimental setup cannot measure as detailed in section~\ref{sec:detindexmaps}, resulting in a total of $589$ acquisition poses.
The resulting trajectory \reviewerthree{for the spherical function (B)} is shown in the center of Table~\ref{table:simulation_geometry}.

Using Algorithm~\ref{alg:pathopt} with $\mathcal{P}_{\text{all}}$ set to the \enquote{Reference} trajectory and batch sizes \reviewerthree{$b=100\%,50\%,20\%,10\%,2\%$ (here $b=2\%$ means that at each iteration we choose only one pose, which is equivalent to the greedy approach from~\cite{Fischer2016})} we computed five \enquote{optimal} trajectories containing $N=50$ poses, denoted as $\text{AGS}(50, b)$, \reviewerthree{for both spherical functions}.
The computed trajectories \reviewerthree{for (B)} are shown in Table~\ref{table:simulation_geometry}.
For comparison, we created three additional trajectories also containing $N=50$ poses: \enquote{Non-optimal} contains the $50$ angles from  \enquote{Reference} which had the lowest detectability indices, \enquote{t-design} is a geometry that is uniformly sampling the sphere \cite{Sloane1996}, and \enquote{Circular} contains $50$ angles restricted to inclination angles $ \phi = -10,0,10 $.

\begin{figure*}[t]
\begin{varwidth}{0.5\linewidth}  
\centerline{\includegraphics[width=\linewidth, height=0.45\linewidth]{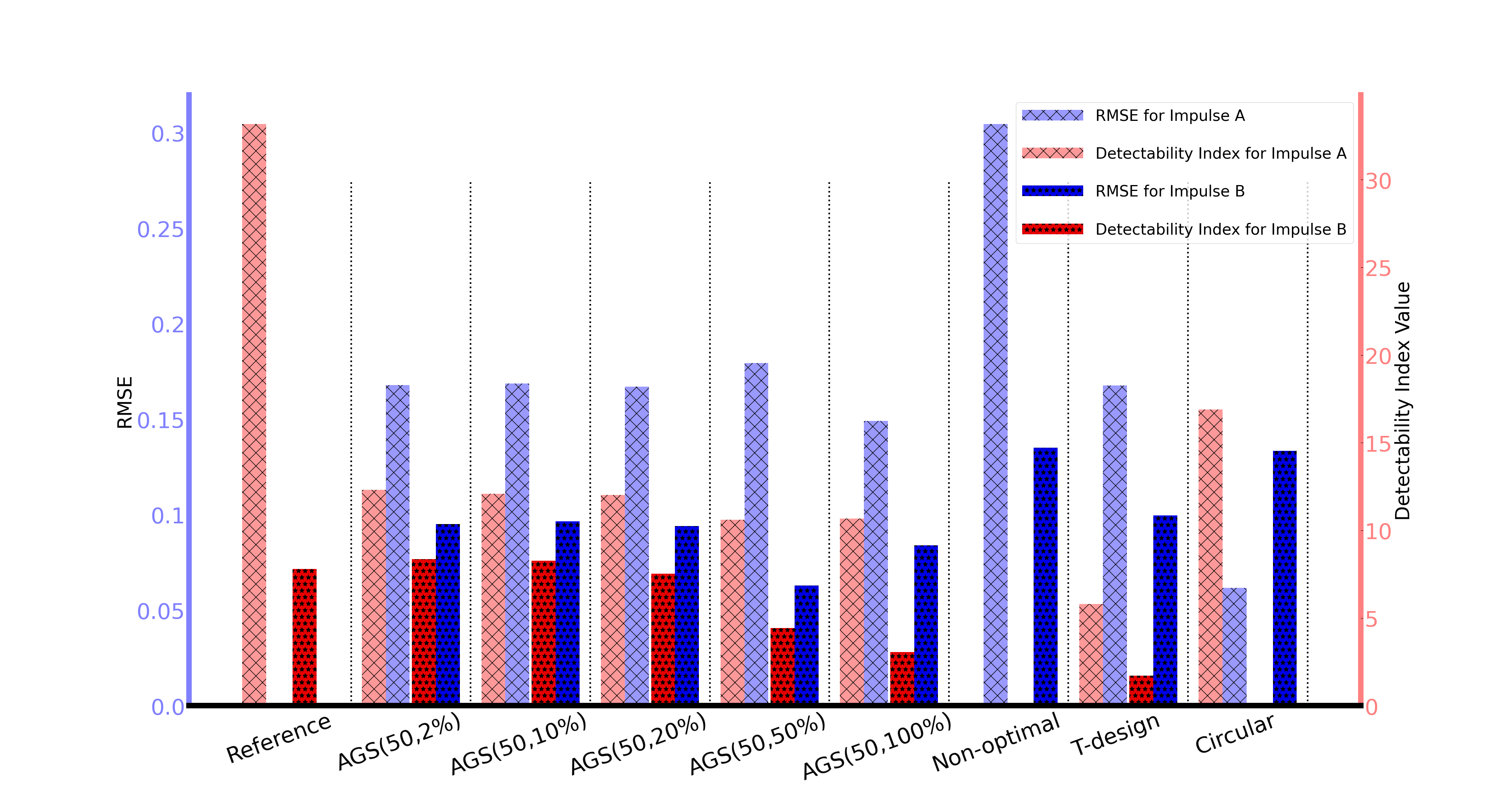}}
\centerline{(I)}\medskip
\end{varwidth}
\hfill
\begin{varwidth}{0.5\linewidth}  
\centerline{\includegraphics[width=\linewidth, height=0.45\linewidth]{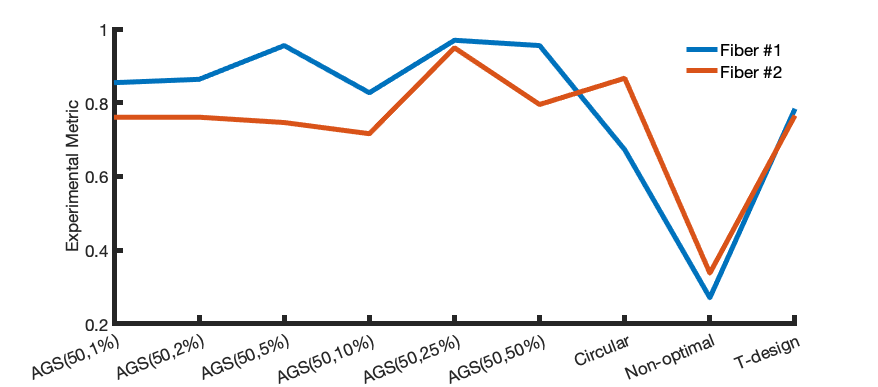}}
\centerline{\secondreview{(II)}}\medskip
\end{varwidth}
\caption{(I) Quantitative measures of the reconstructions \reviewerthree{of the spherical phantoms (A) (marked with hatches) and (B) (marked with stars)} as in Table~\ref{table:simulation_reconstruction} compared to the \enquote{Reference} reconstructions. Left axis (in blue) is the RMSE, while the right axis (in red) is the sum of the detectability index values \reviewerthree{for the entire trajectory}.
 \secondreview{(II) Experimental metric plotted for the first and second main fiber orientation extracted from the reconstructions of the spherical phantom (D) using different optimized and static acquisition trajectories as in Table~\ref{table:simulation_reconstruction} compared to the \enquote{Reference} reconstruction}.}
\label{fig:simulation_results}
\end{figure*}

For each of these trajectories, noise-free measurements of the \reviewerthree{phantoms} were simulated, which in turn were used compute reconstructions, as shown in Table~\ref{table:simulation_reconstruction} \reviewerthree{for the spherical function (B)}.
In order to compare the reconstruction quality, the RMSE was computed according to eq.~(\ref{eq:rmse}) between the \enquote{Reference} reconstruction and the other reconstructions. 
\reviewerthree{We ran the same set of experiments for both impulses (A) and (B) and plotted the results in Fig.~\ref{fig:simulation_results}(I) along with the value of the detectability index for the entire trajectory.}

\secondreview{For a more realistic scenario, we repeated the experiment for scattering profile (D) from Fig.~\ref{fig:impulse}, which is taken from experimental data and was produced by two fibers perpendicular to one another (the two fibers are represented by pink arrows).  
To compare the results obtained with the different optimized trajectories, we computed the experimental metric from eq.~(\ref{eq:expmetric}) for each individual extracted fiber orientation and plotted the values in Fig.~\ref{fig:simulation_results}(II).}

\newcommand\x{0.3}

\begin{figure}[h]
\begin{center}
\begin{varwidth}{\linewidth}  		
  		\includegraphics[width=\x\linewidth,height=\x\linewidth]{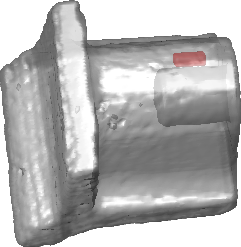}
  		\includegraphics[width=\x\linewidth,height=\x\linewidth]{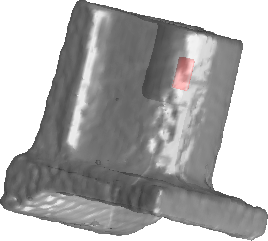}
  		\includegraphics[width=\x\linewidth,height=\x\linewidth]{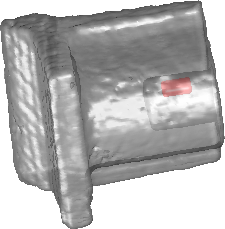}
  		\\
  		\vfill
  		\includegraphics[width=\x\linewidth,height=\x\linewidth]{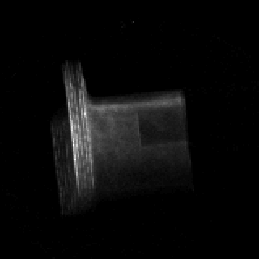}
  		\includegraphics[width=\x\linewidth,height=\x\linewidth]{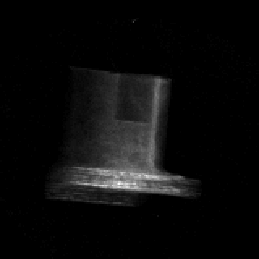}
  		\includegraphics[width=\x\linewidth,height=\x\linewidth]{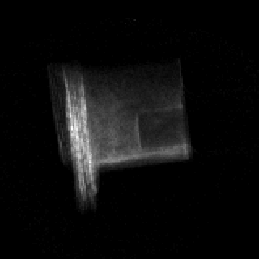}
\end{varwidth}
\end{center}
	\caption{Images of the thermoplastic fiber mould sample used in section~\ref{sec:algoexp}. \textit{Top row:} renderings from different view points of the X-ray absorption contrast reconstruction of the sample, with the region of interest highlighted in red. \textit{Bottom row:} experimental X-ray dark-field measurements of the sample corresponding to the view points shown above.}
	\label{fig:cfkrender}
\end{figure}

\begin{figure*}[t]
\begin{varwidth}{0.2\linewidth}
  	\centering
	\centerline{\includegraphics[width=\linewidth,height=\linewidth]{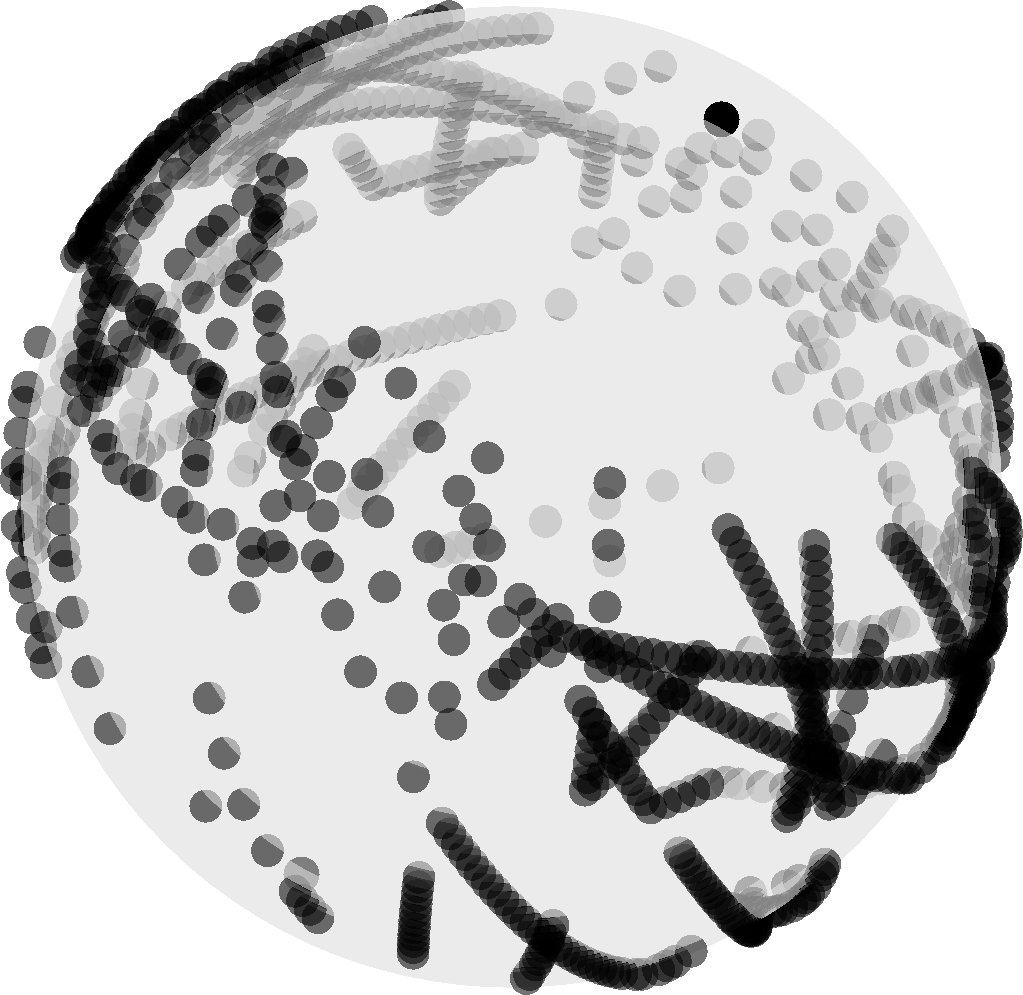}}
\end{varwidth}
\hfill
\begin{varwidth}{0.3\linewidth}  
	\centering
	\centerline{\includegraphics[width=1.25\linewidth,height=\linewidth]{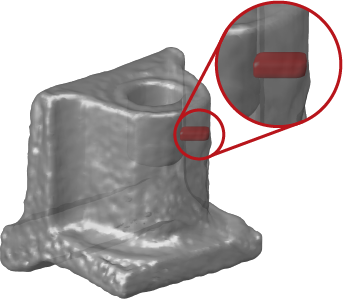}}
\end{varwidth}
\hfill
\begin{varwidth}{0.4\linewidth}
	\centering
  	\centerline{\includegraphics[width=\linewidth,height=0.75\linewidth]{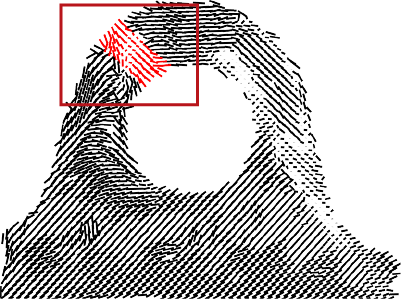}}
\end{varwidth}
\caption{Reconstructions of the experimental data of the thermoplastic fiber mould sample. \textit{Left:} visualization of the acquisition trajectory with $966$ poses.
	\textit{Middle:} rendering of the X-ray absorption contrast reconstruction, highlighting in red the region of interest containing a weld line, which is our imaging task here.
	\textit{Right:} slice showing the \secondreview{main} fiber orientations extracted from the AXDT reconstruction of the sample, using the acquisition trajectory on the left. The red area highlights the imaging task, the weld line.}
\label{fig:cfk_render_slice}
\end{figure*}

\subsection{Algorithm validation with experimental data} \label{sec:algoexp}

In this experiment, we investigate the performance of our proposed algorithm using experimental data of a thermoplastic fiber mould, for an overview of the sample see Fig.~\ref{fig:cfkrender}.
\reviewerthree{We measured the object using a setup as shown in Fig. \ref{fig:setup}, consisting of an X-ray WorX micro-focus X-ray tube at a voltage of 60kVp and 25W power, and a Varian PaxScan 2520DX detector with a pixel size of $127\mu m$. 
The gratings have periods of $10 \mu m$ for G0, $ 5 \mu m$ for G1 and $10 \mu m $ for G2 respectively, and were arranged in the first fractional Talbot configuration at a design energy of 45 keV. 
For every measurement an acquisition with $1s$ exposure time was performed for each of the 7 individual phase steps.}
The \reviewermain{measured} sample consists of fibers that have a thickness of approximately $7\mu m$, and has a notable weld line feature in the region of interest, highlighted in red in Fig.~\ref{fig:cfk_render_slice}.
For AXDT reconstruction we used a volume of $160^3$ isotropic voxels with size $508 \mu m$.
Even though this cannot directly resolve the fibers, it is possible to extract the \secondreview{main} fiber orientations from the AXDT reconstruction, displaying the weld line prominently as shown in Fig.~\ref{fig:cfk_render_slice}. 
Additional details of the experimental setup and acquisition parameters can be found in Sharma \textit{et al.} \cite{Sharma2017}.

We use the measurements from a high-quality trajectory with $966$ acquisition poses as prior knowledge to compute the detectability index, while the weld line region, highlighted in red in Fig.~\ref{fig:cfk_render_slice}, served as the task template.
We ran Algorithm~\ref{alg:pathopt} with $\mathcal{P}_{\text{all}}$ set to the high-quality trajectory (the \enquote{Reference}) and batch sizes \reviewerthree{$b=10\%,20\%,25\%,33\%,50\%,100\%$} to produce \enquote{optimal} acquisition trajectories with $N=100,150,\ldots,300$ poses.

\begin{figure*}[!ht]
\begin{varwidth}{0.19\linewidth}
  		\centering
  		\centerline{\includegraphics[width=\linewidth]{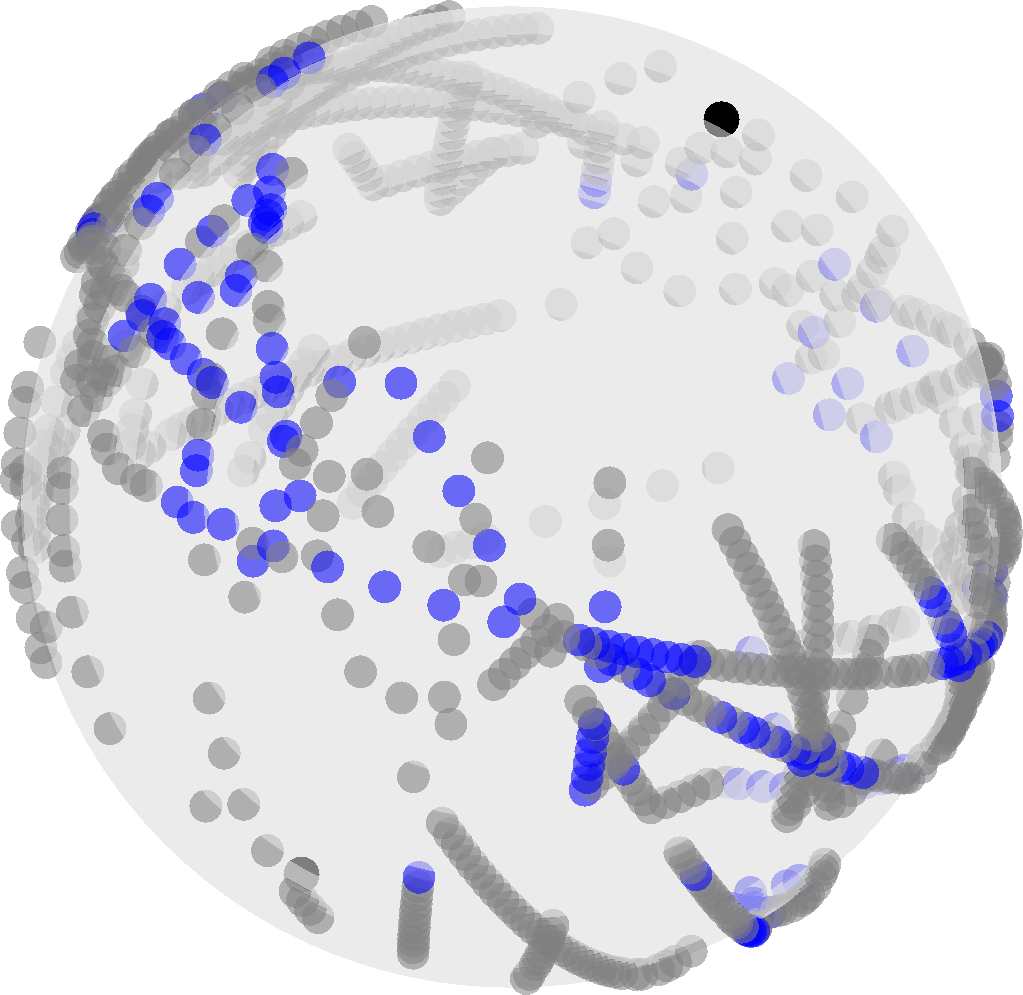}}
  		\centerline{AGS(200,100\%)}\medskip
\end{varwidth}
\hfill
\begin{varwidth}{0.19\linewidth}
  		\centering
  		\centerline{\includegraphics[width=\linewidth]{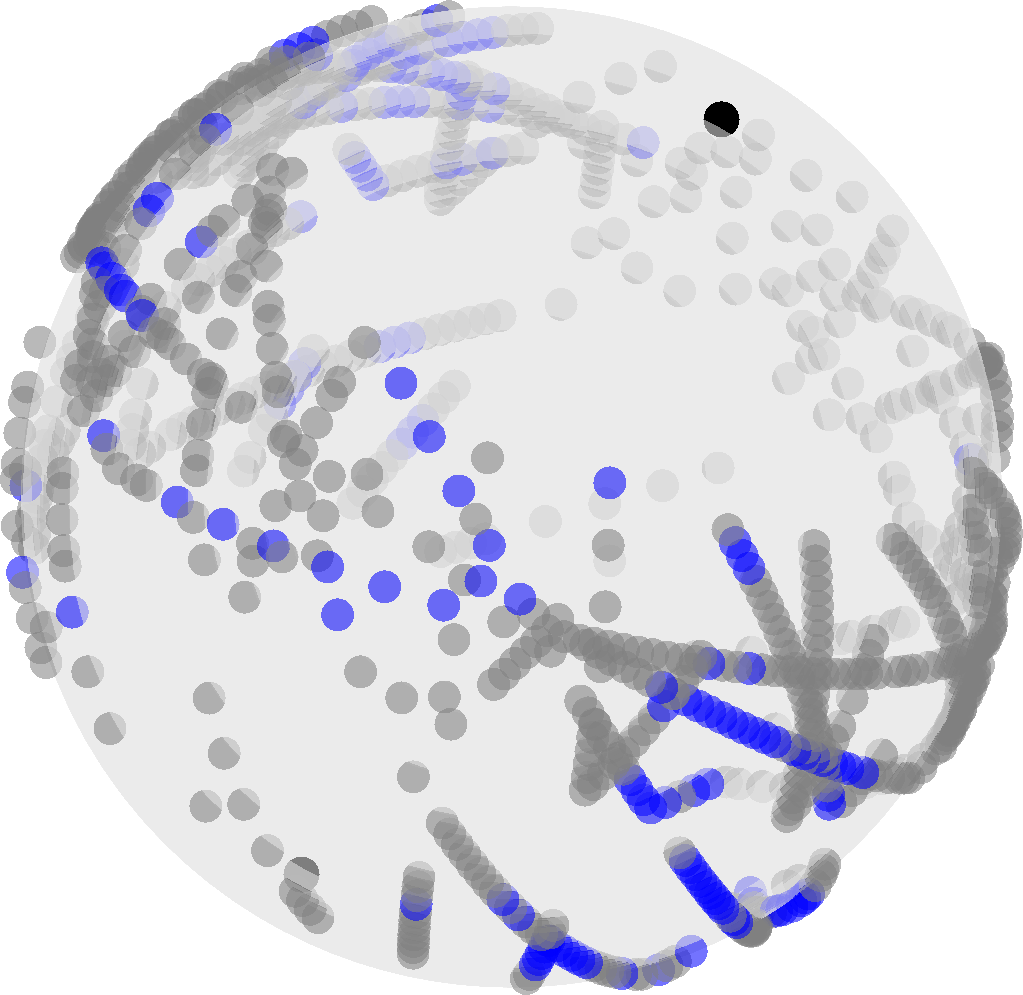}}
  		\centerline{AGS(200,50\%)}\medskip
\end{varwidth}
\hfill
\begin{varwidth}{0.19\linewidth}
  		\centering
  		\centerline{\includegraphics[width=\linewidth]{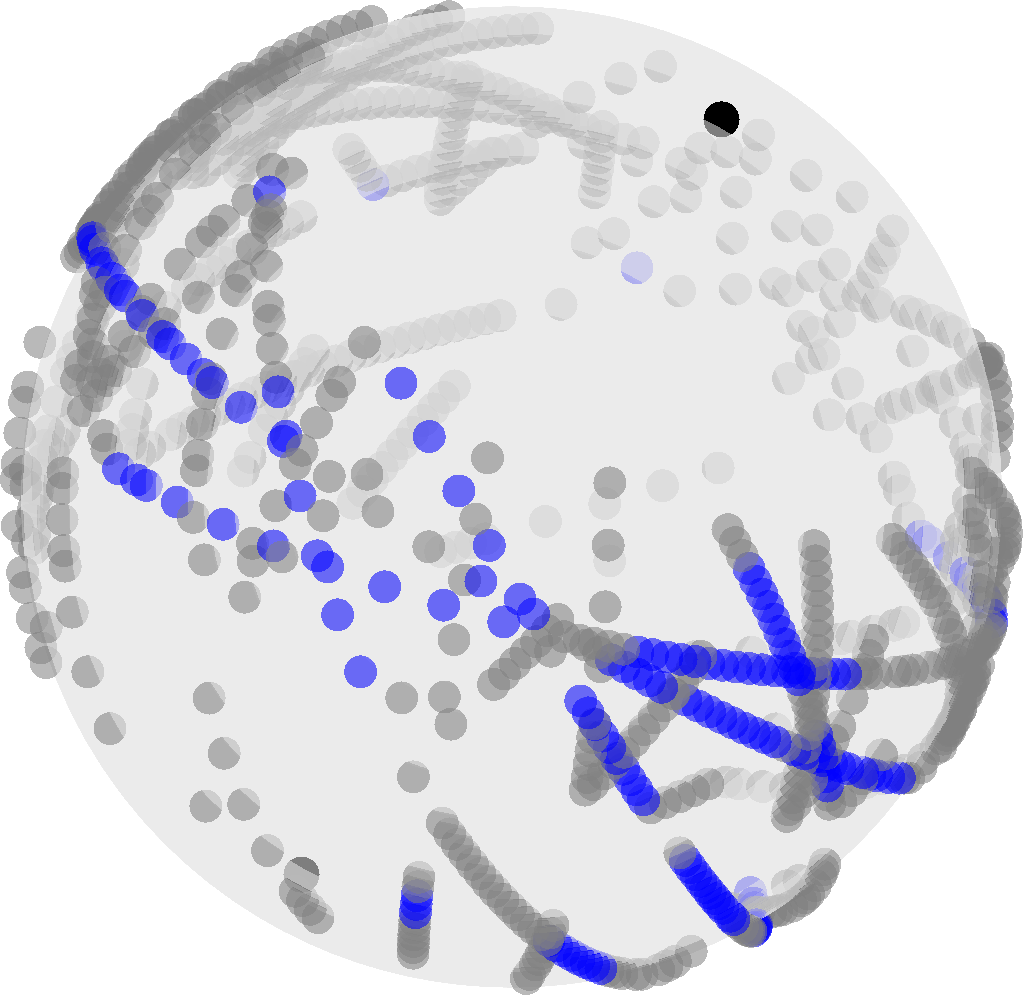}}
  		\centerline{AGS(200,33\%)}\medskip
\end{varwidth}
\hfill
\begin{varwidth}{0.19\linewidth}
  		\centering
  		\centerline{\includegraphics[width=\linewidth]{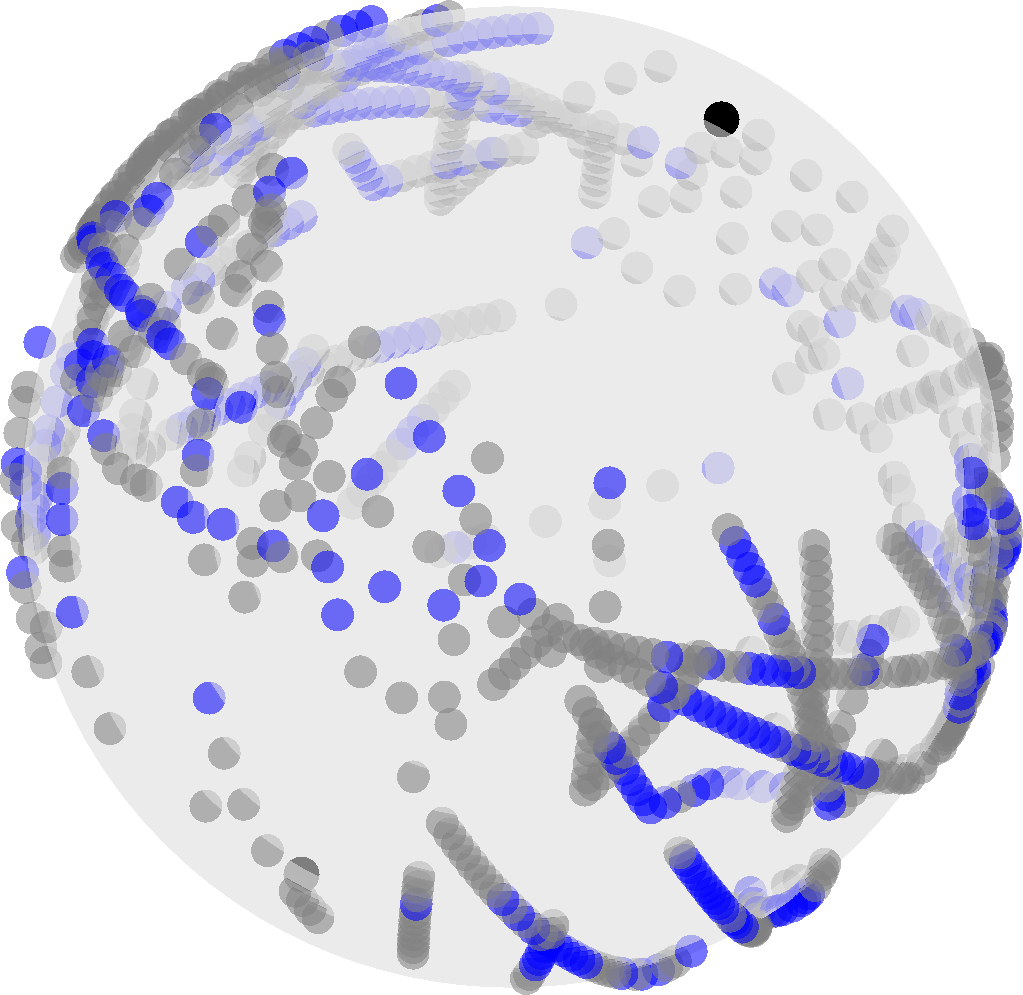}}
  		\centerline{AGS(200,25\%)}\medskip
\end{varwidth}
\hfill
\begin{varwidth}{0.19\linewidth}
  		\centering
  		\centerline{\includegraphics[width=\linewidth]{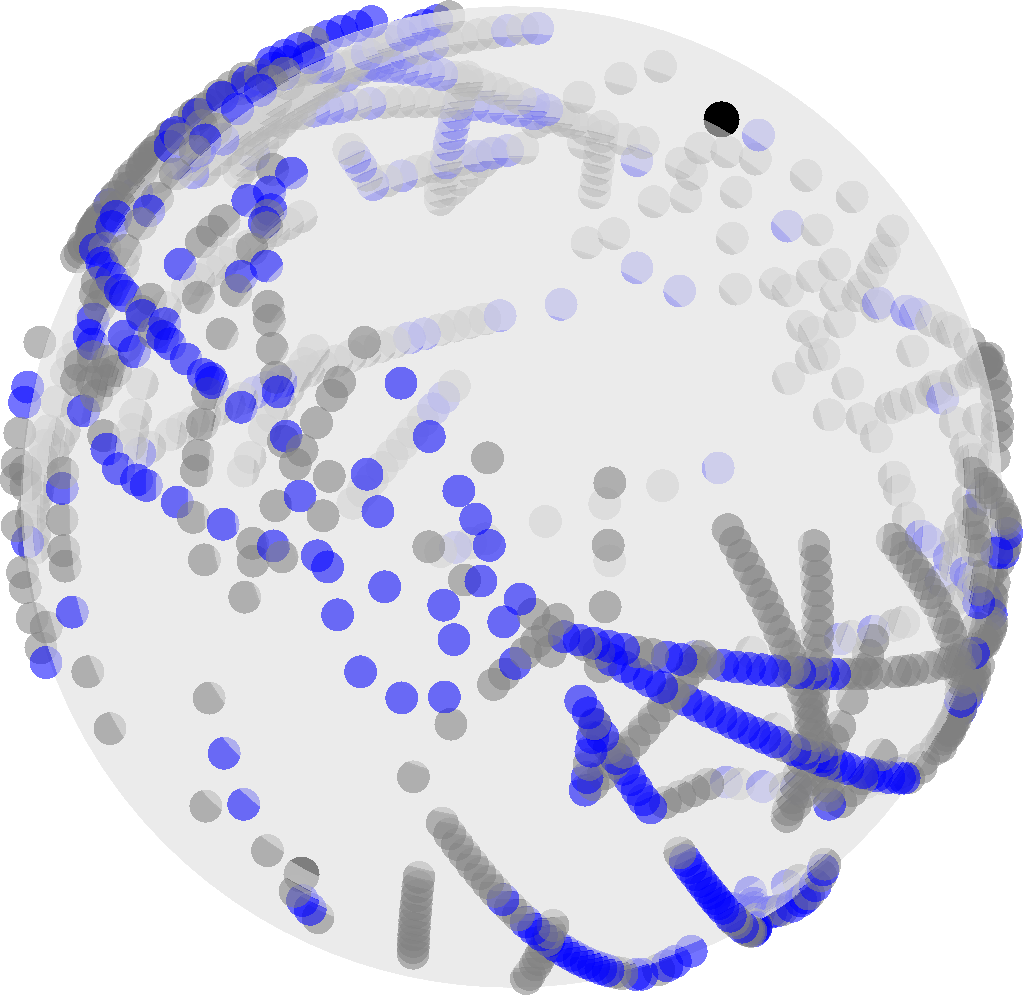}}
  		\centerline{AGS(200,10\%)}\medskip
\end{varwidth}
\caption{\reviewerthree{Results of the experimental data using the thermoplastic fiber mould sample.  
Shown are plots of selected acquisition trajectories computed by $\text{AGS}(N,b)$. 
Dots in blue are the poses chosen by the algorithm, while dots in gray represent the original pool of available poses, corresponding to the high-quality trajectory.}}
\label{fig:cfk_result_geometries}
\end{figure*}

\begin{figure*}[!h]
\begin{varwidth}{0.19\linewidth}
\hfill
\end{varwidth}
\begin{varwidth}{0.19\linewidth}
\hfill
\end{varwidth}
\begin{varwidth}{0.19\linewidth}
  		\centering
  		\centerline{\includegraphics[width=\linewidth]{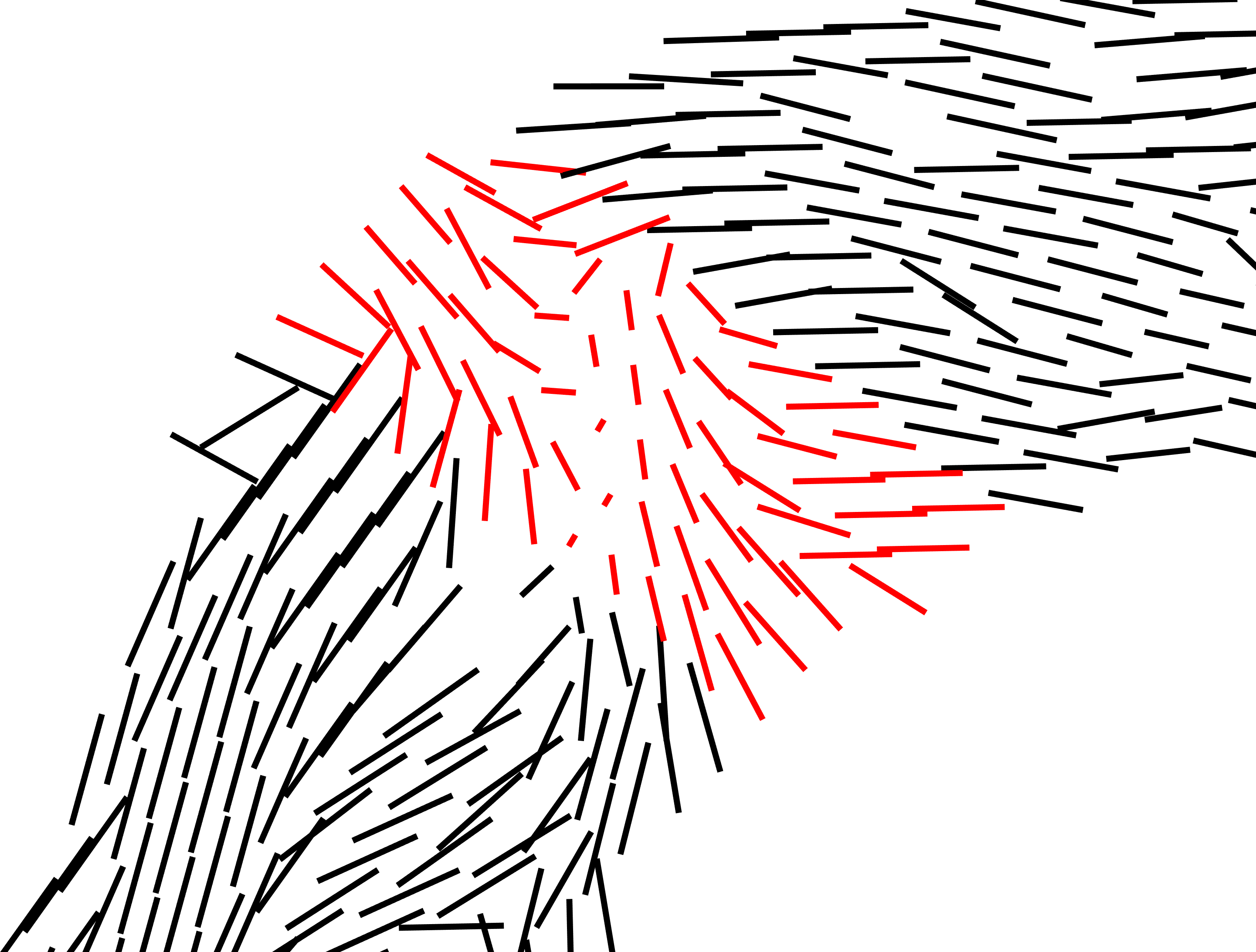}}
  		\centerline{Reference}\medskip
\end{varwidth}
\begin{varwidth}{0.19\linewidth}
\hfill
\end{varwidth}
\begin{varwidth}{0.19\linewidth}
\hfill
\end{varwidth}
\\
\begin{varwidth}{0.19\linewidth}
  		\centering
  		\centerline{\includegraphics[width=\linewidth]{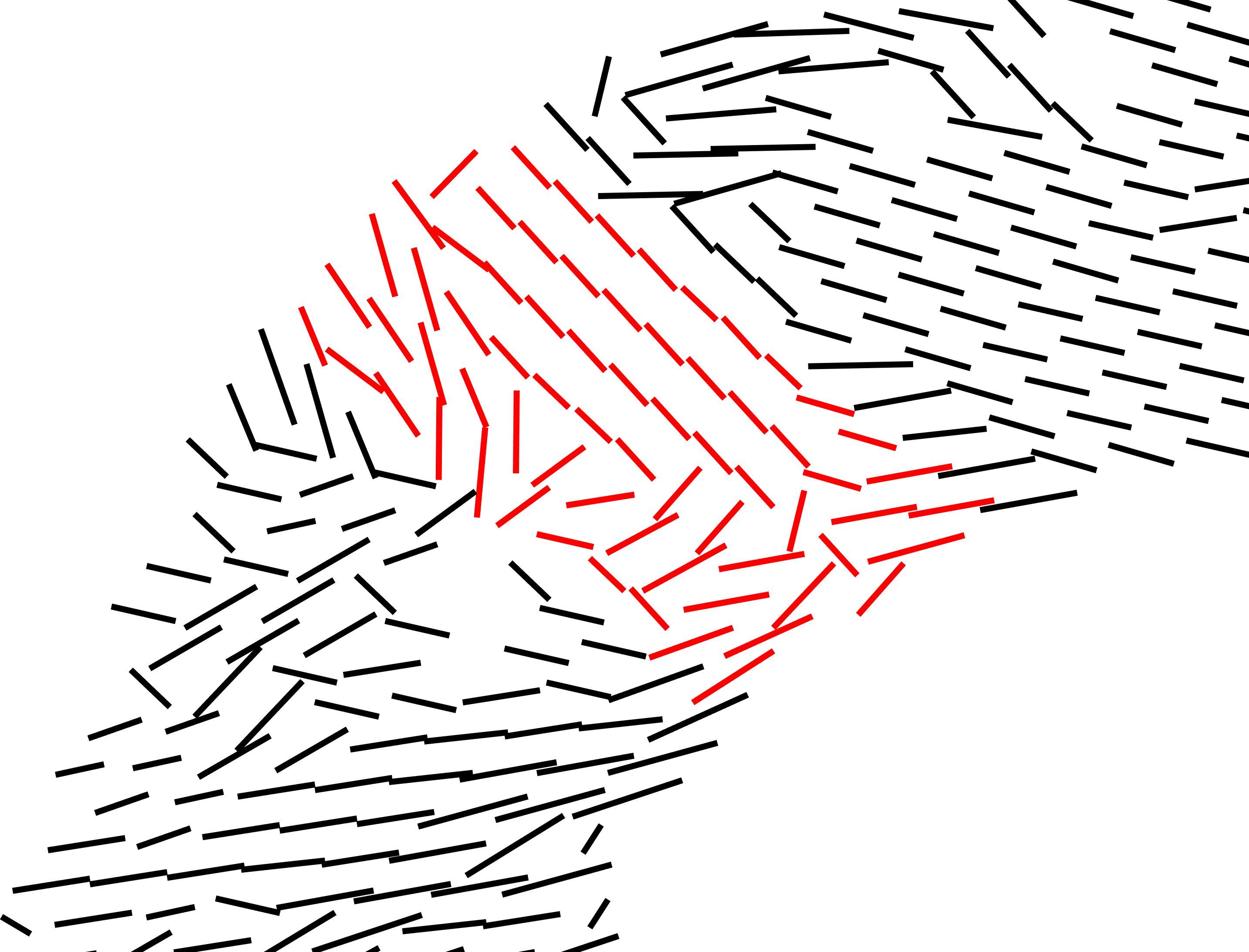}}
  		\centerline{AGS(100,100\%)}\medskip
\end{varwidth}
\begin{varwidth}{0.19\linewidth}
  		\centering
  		\centerline{\includegraphics[width=\linewidth]{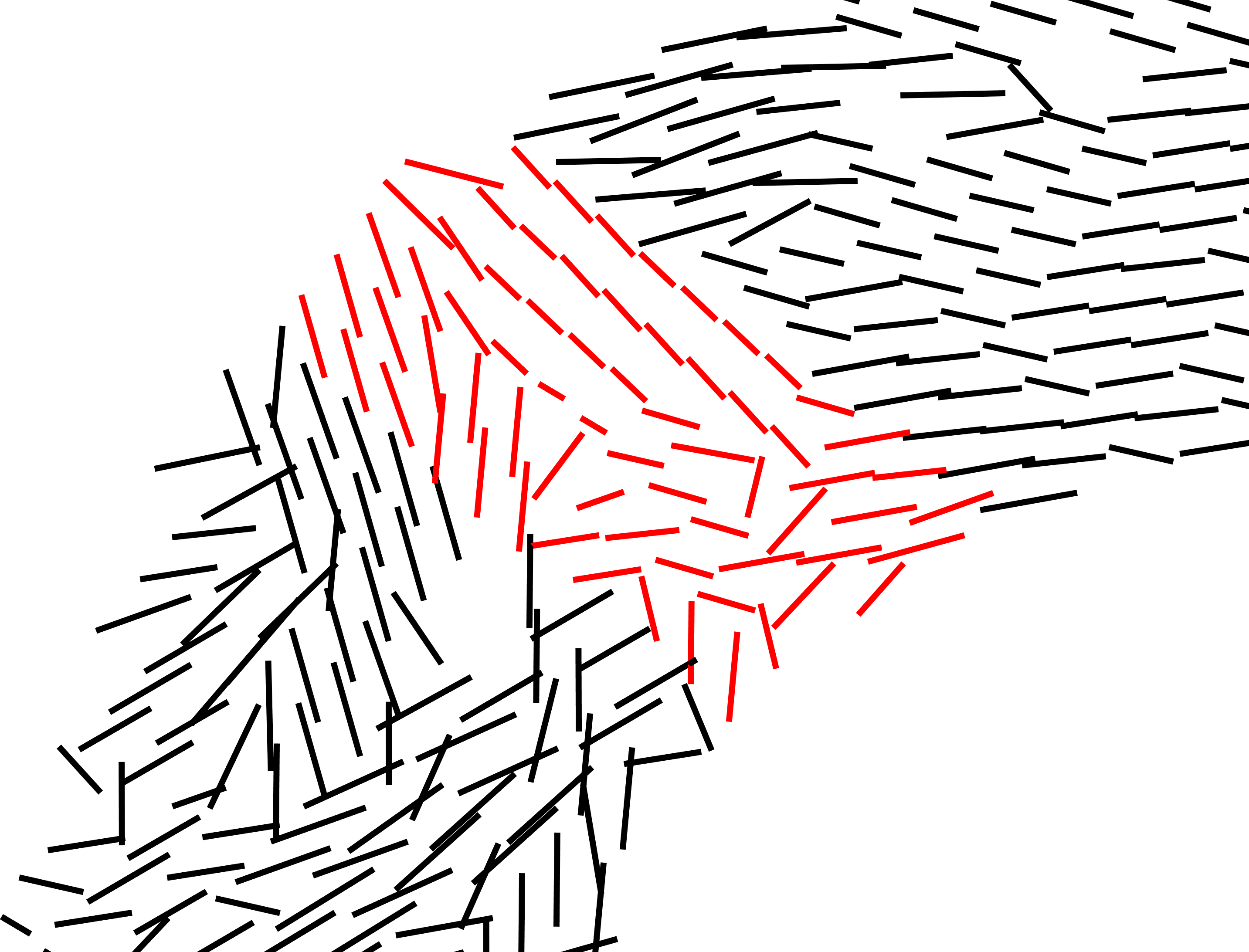}}
  		\centerline{AGS(150,100\%)}\medskip
\end{varwidth}
\begin{varwidth}{0.19\linewidth}
  		\centering
  		\centerline{\includegraphics[width=\linewidth]{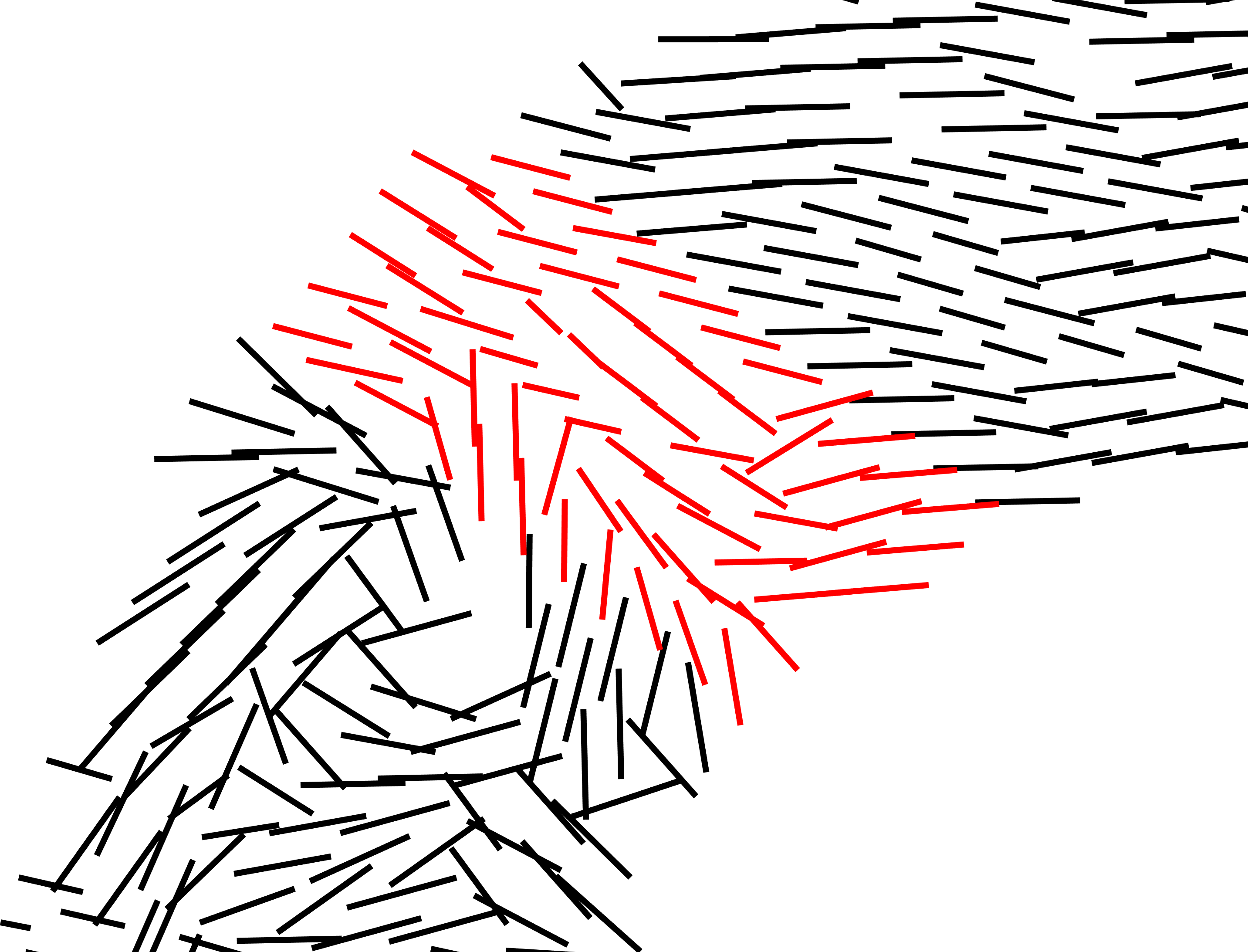}}
  		\centerline{AGS(200,100\%)}\medskip
\end{varwidth}
\begin{varwidth}{0.19\linewidth}
  		\centering
  		\centerline{\includegraphics[width=\linewidth]{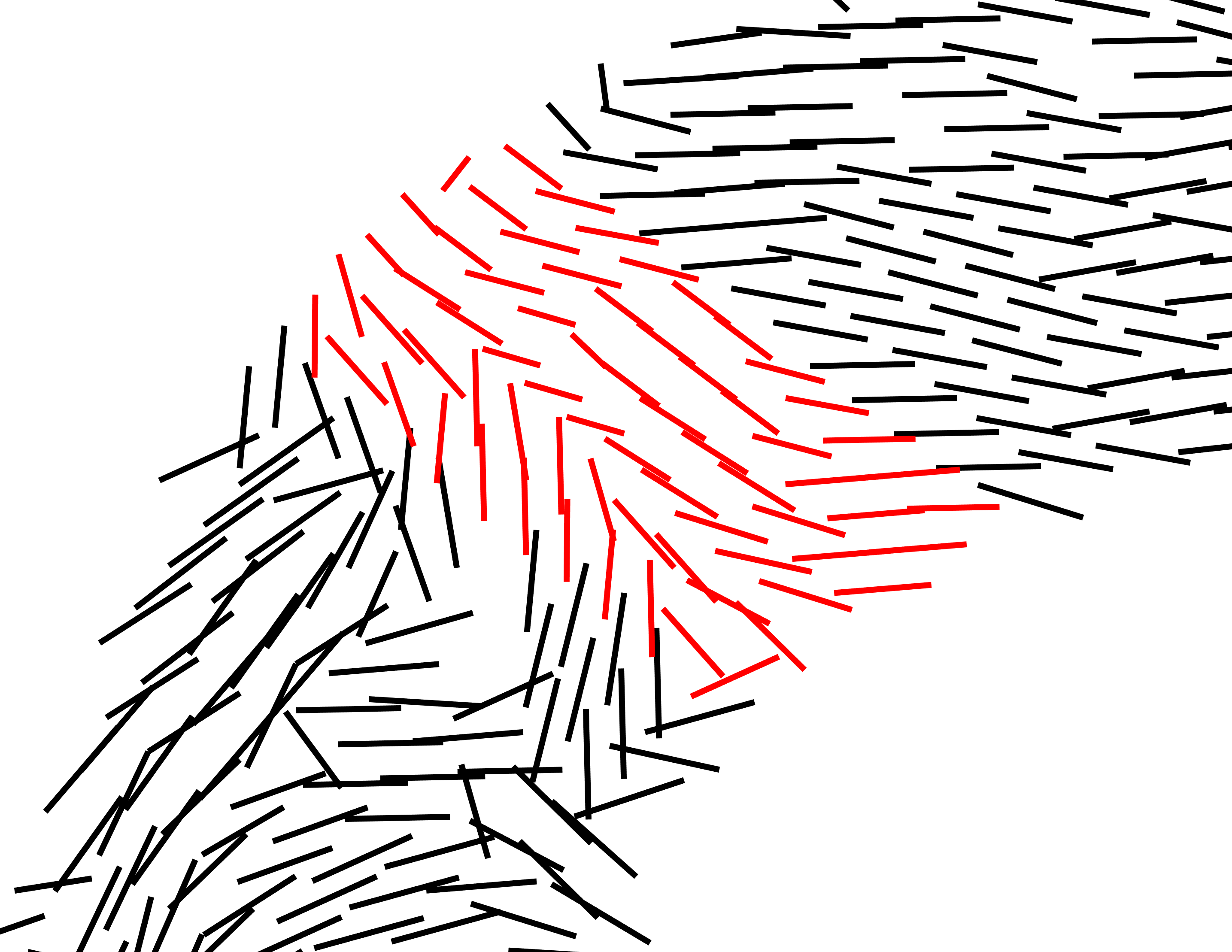}}
  		\centerline{AGS(250,100\%)}\medskip
\end{varwidth}
\begin{varwidth}{0.19\linewidth}
  		\centering
  		\centerline{\includegraphics[width=\linewidth]{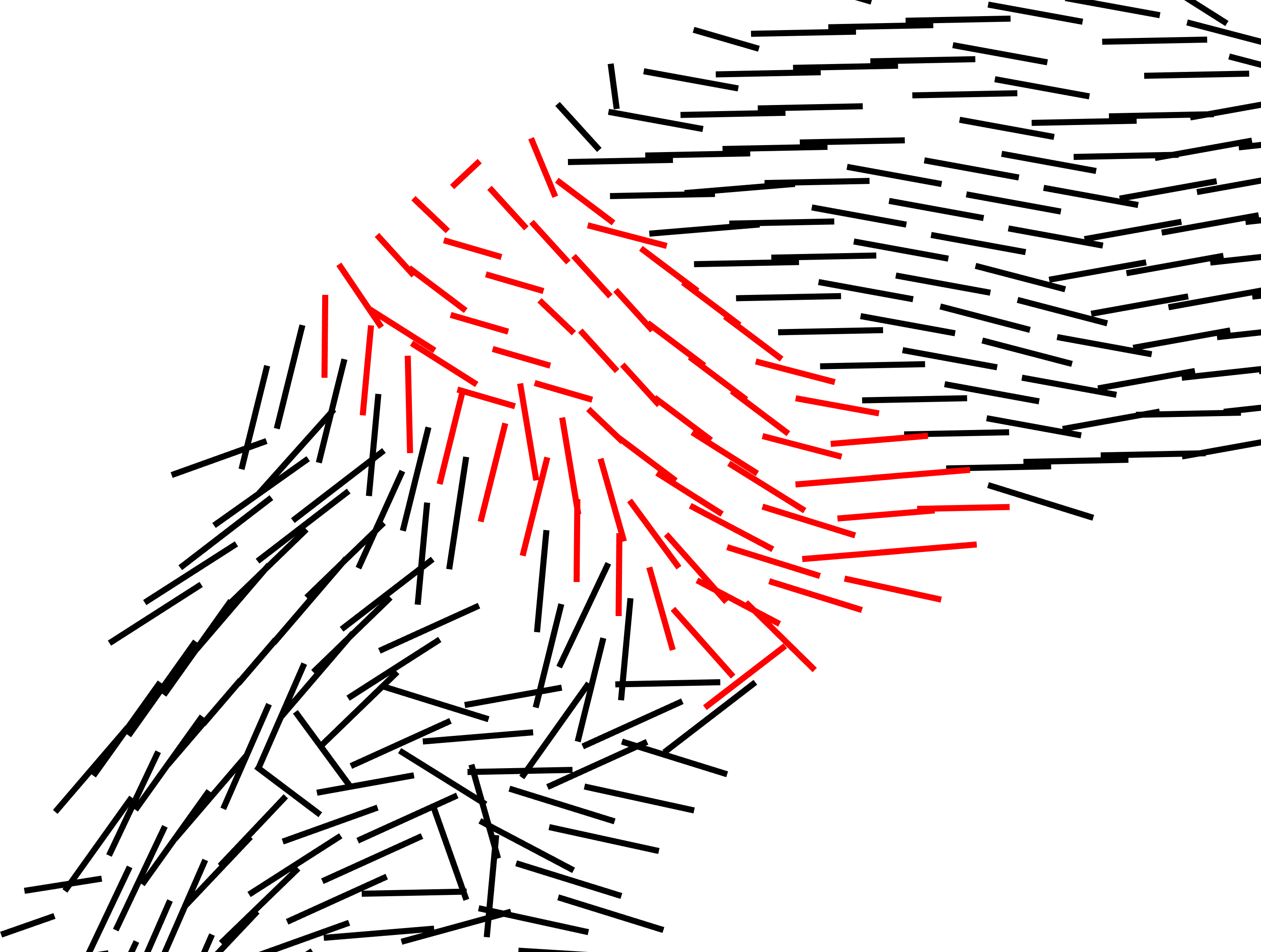}}
  		\centerline{AGS(300,100\%)}\medskip
\end{varwidth}
\\
\begin{varwidth}{0.19\linewidth}
  		\centering
  		\centerline{\includegraphics[width=\linewidth]{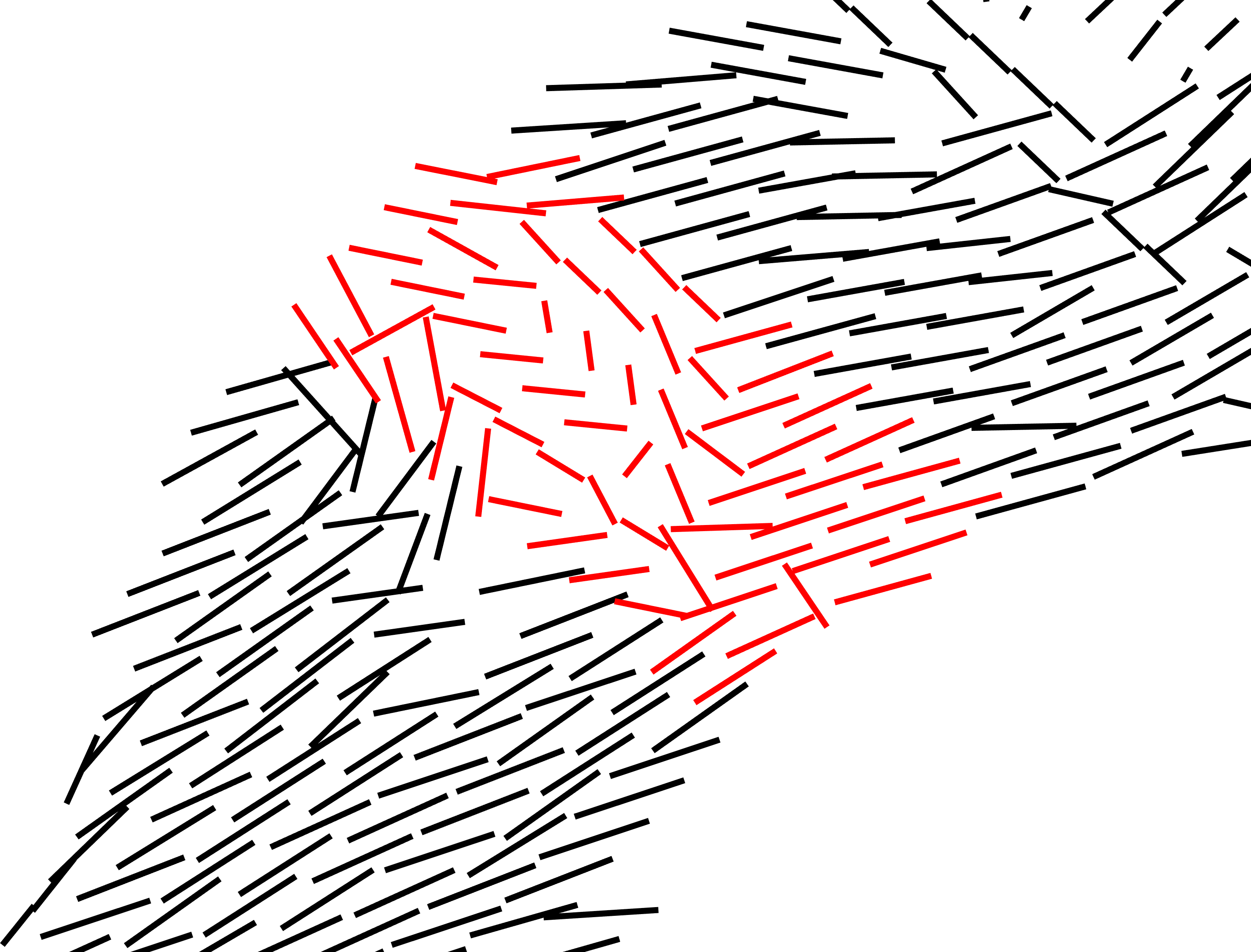}}
  		\centerline{AGS(100,50\%)}\medskip
\end{varwidth}
\begin{varwidth}{0.19\linewidth}
  		\centering
  		\centerline{\includegraphics[width=\linewidth]{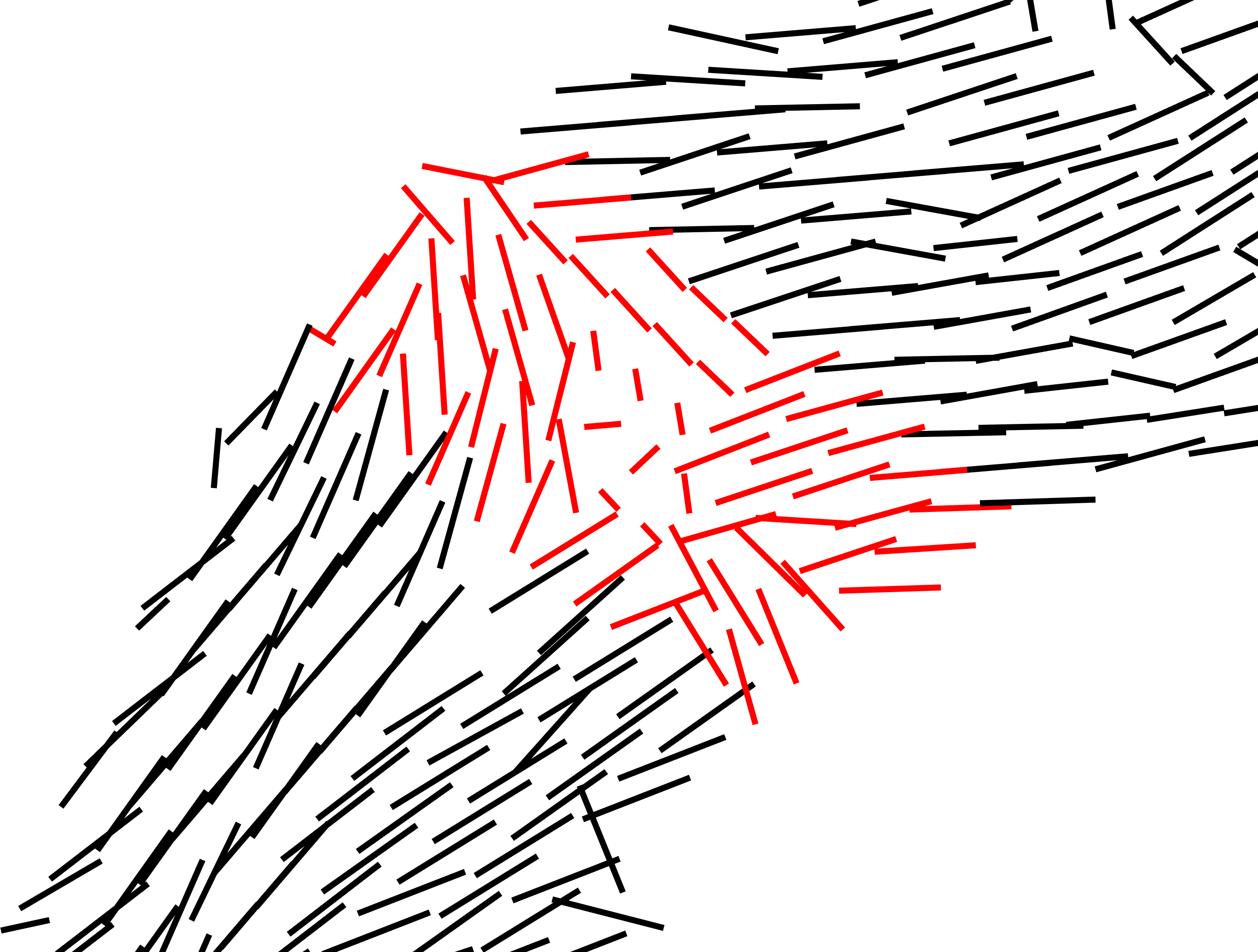}}
  		\centerline{AGS(150,50\%)}\medskip
\end{varwidth}
\begin{varwidth}{0.19\linewidth}
  		\centering
  		\centerline{\includegraphics[width=\linewidth]{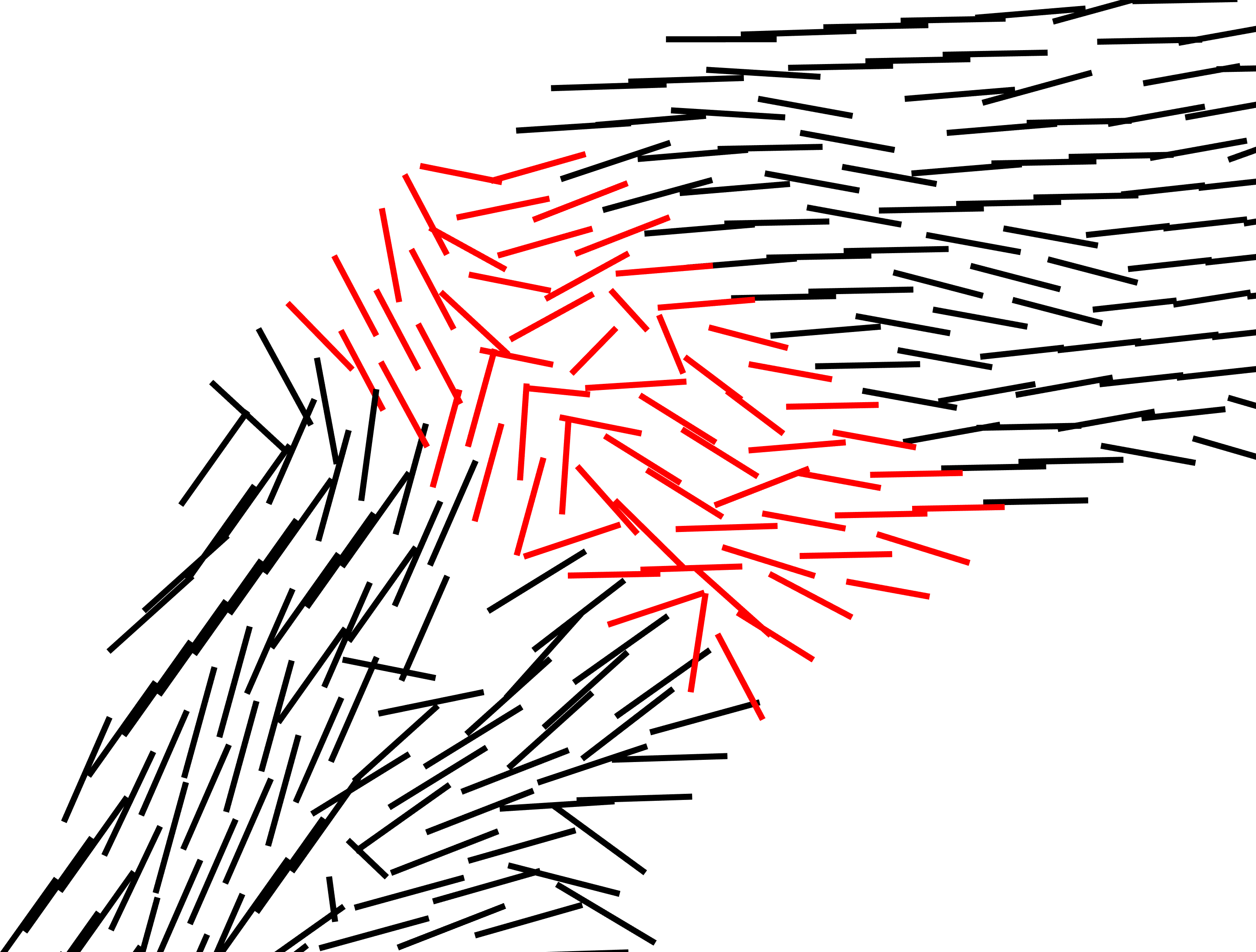}}
  		\centerline{AGS(200,50\%)}\medskip
\end{varwidth}
\begin{varwidth}{0.19\linewidth}
  		\centering
  		\centerline{\includegraphics[width=\linewidth]{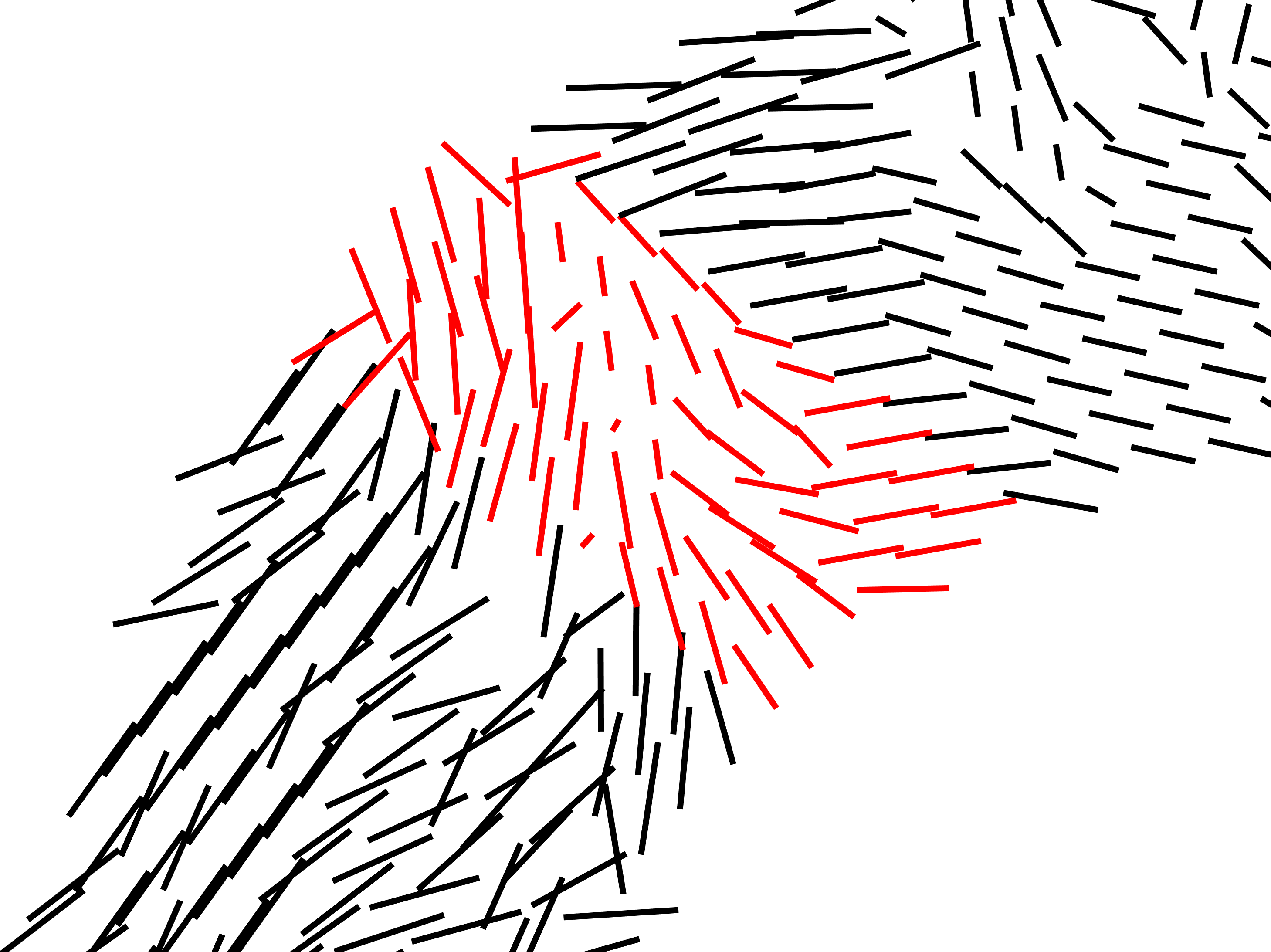}}
  		\centerline{AGS(250,50\%)}\medskip
\end{varwidth}
\begin{varwidth}{0.19\linewidth}
  		\centering
  		\centerline{\includegraphics[width=\linewidth]{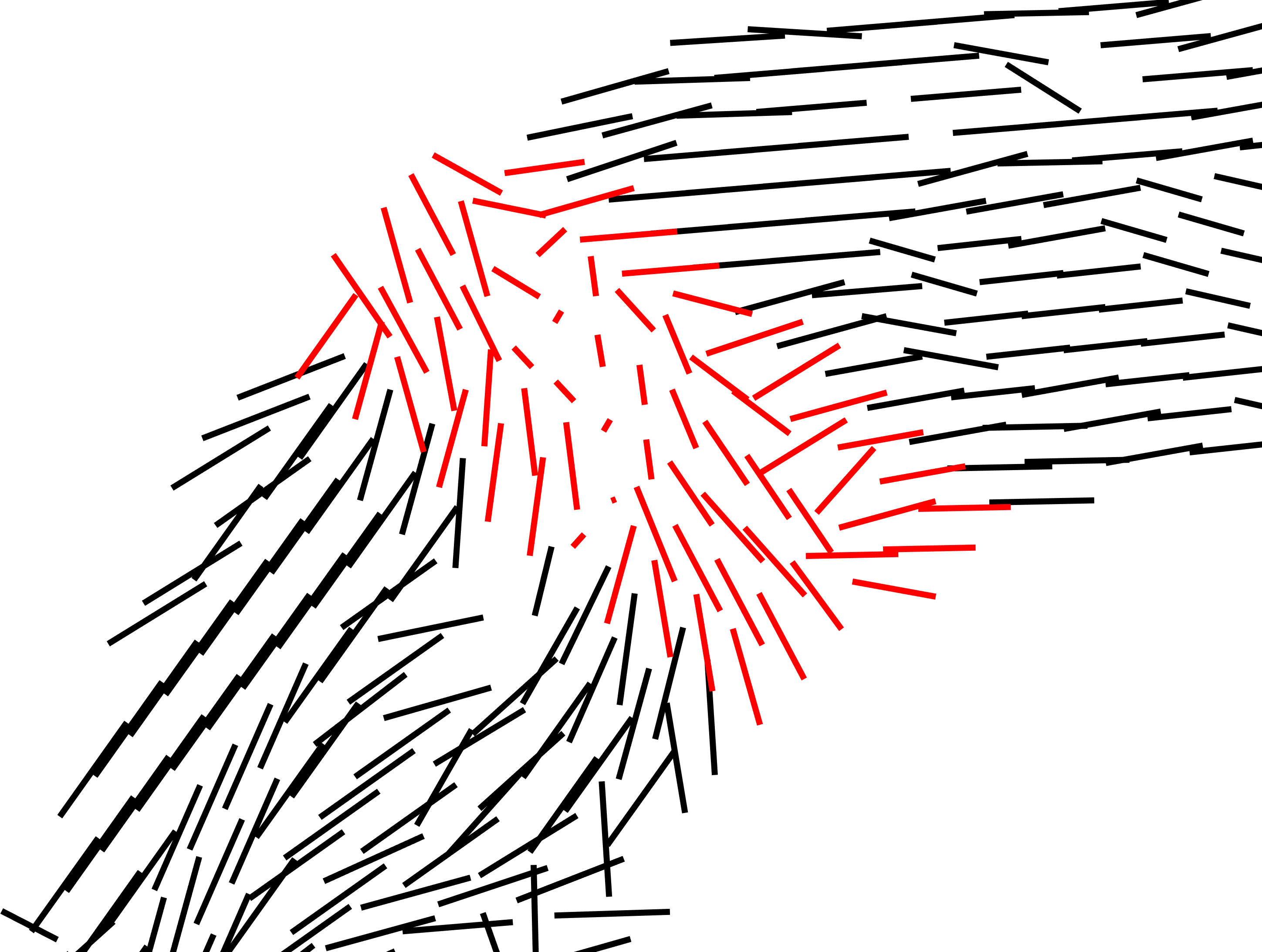}}
  		\centerline{AGS(300,50\%)}\medskip
\end{varwidth}
\\
\begin{varwidth}{0.19\linewidth}
  		\centering
  		\centerline{\includegraphics[width=\linewidth]{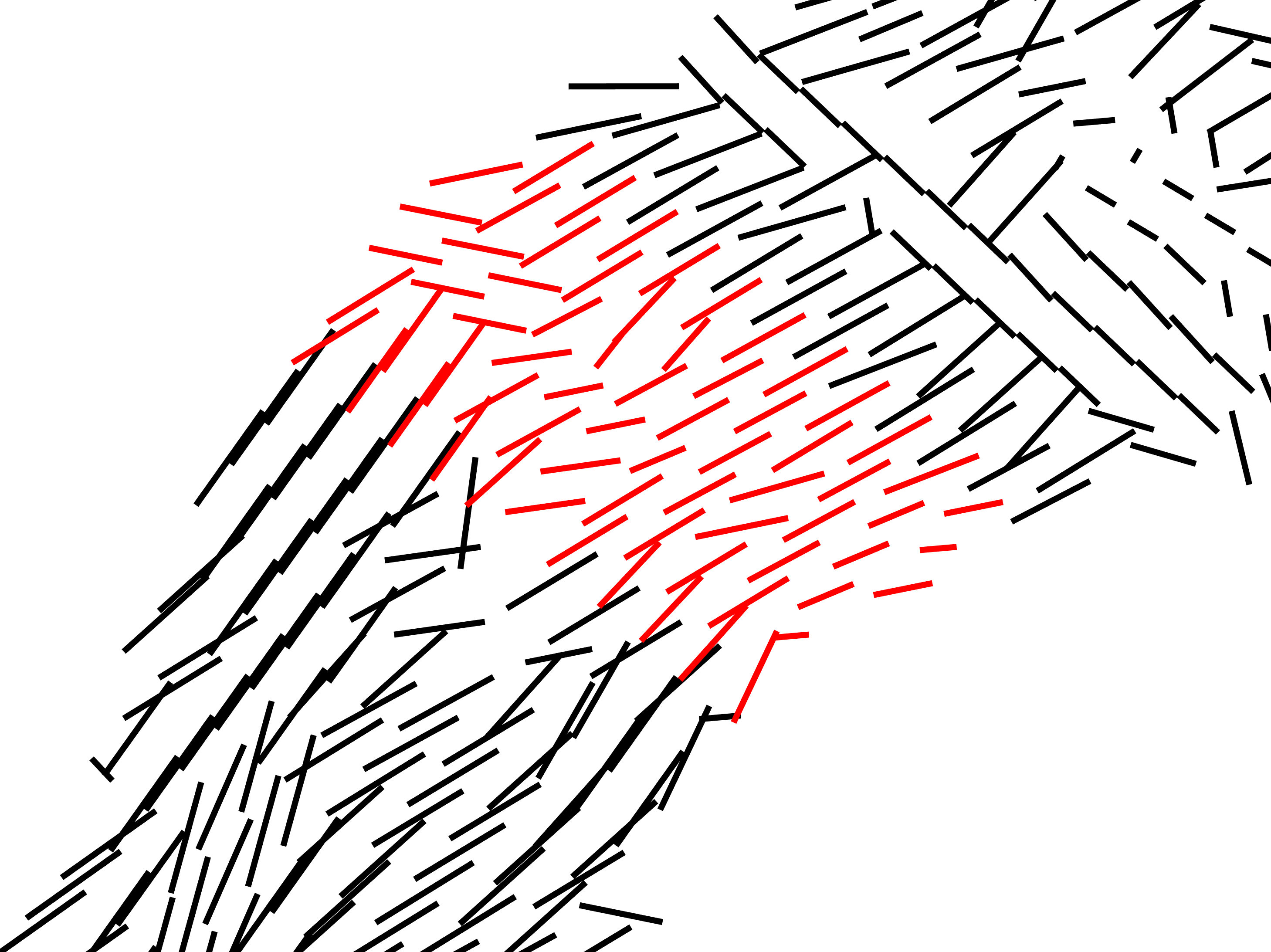}}
  		\centerline{AGS(100,33\%)}\medskip
\end{varwidth}
\begin{varwidth}{0.19\linewidth}
  		\centering
  		\centerline{\includegraphics[width=\linewidth]{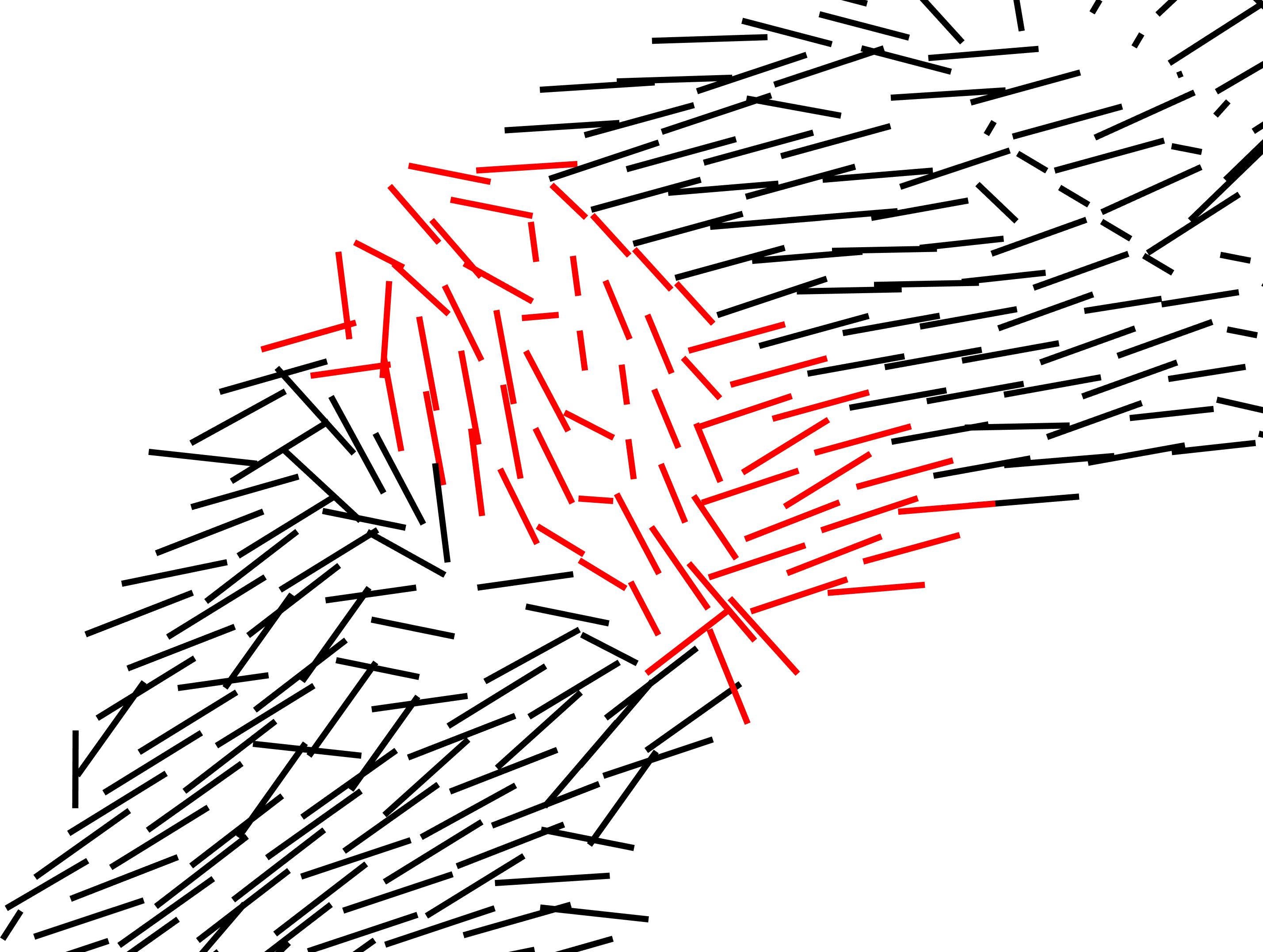}}
  		\centerline{AGS(150,33\%)}\medskip
\end{varwidth}
\begin{varwidth}{0.19\linewidth}
  		\centering
  		\centerline{\includegraphics[width=\linewidth]{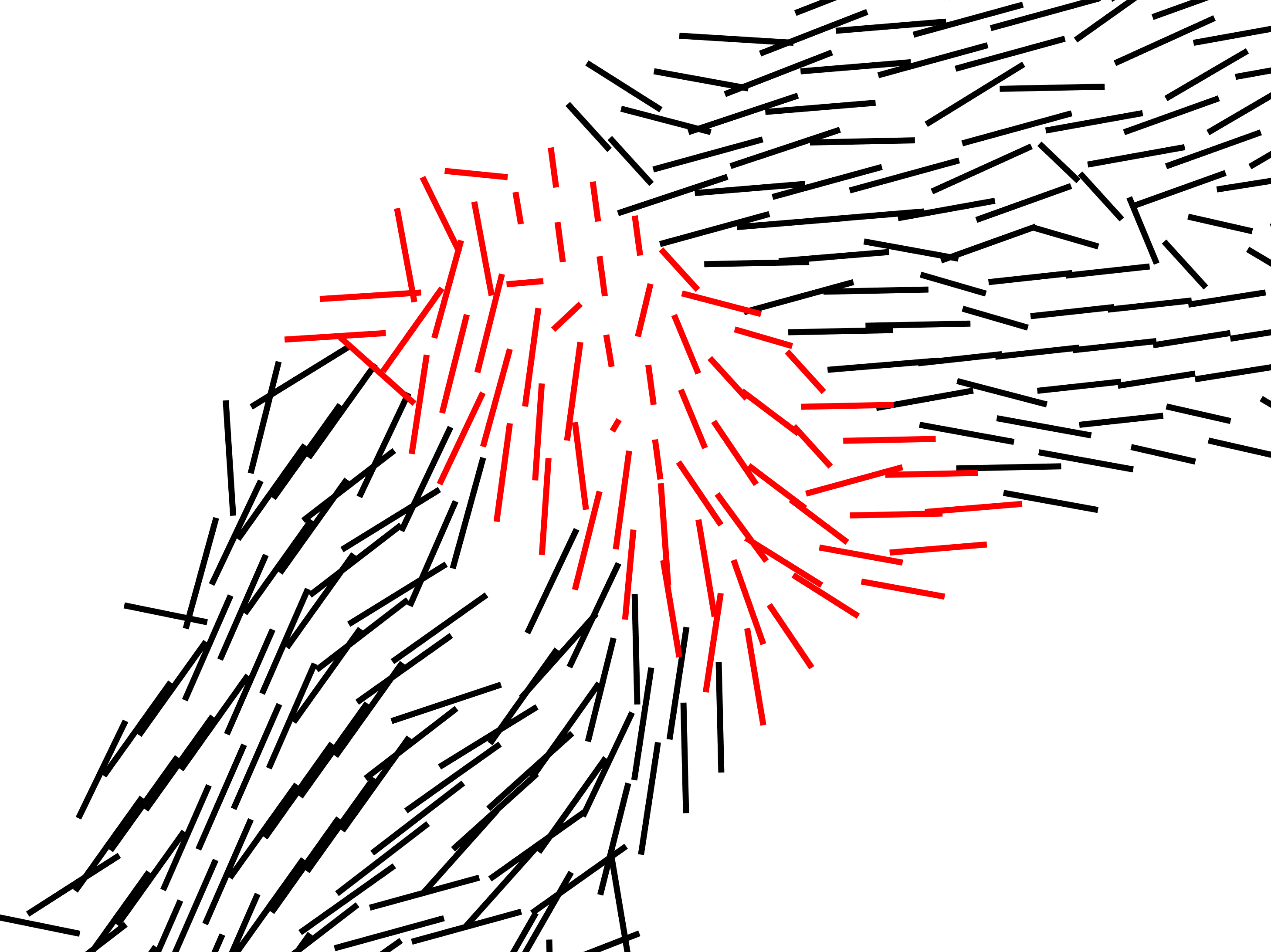}}
  		\centerline{AGS(200,33\%)}\medskip
\end{varwidth}
\begin{varwidth}{0.19\linewidth}
  		\centering
  		\centerline{\includegraphics[width=\linewidth]{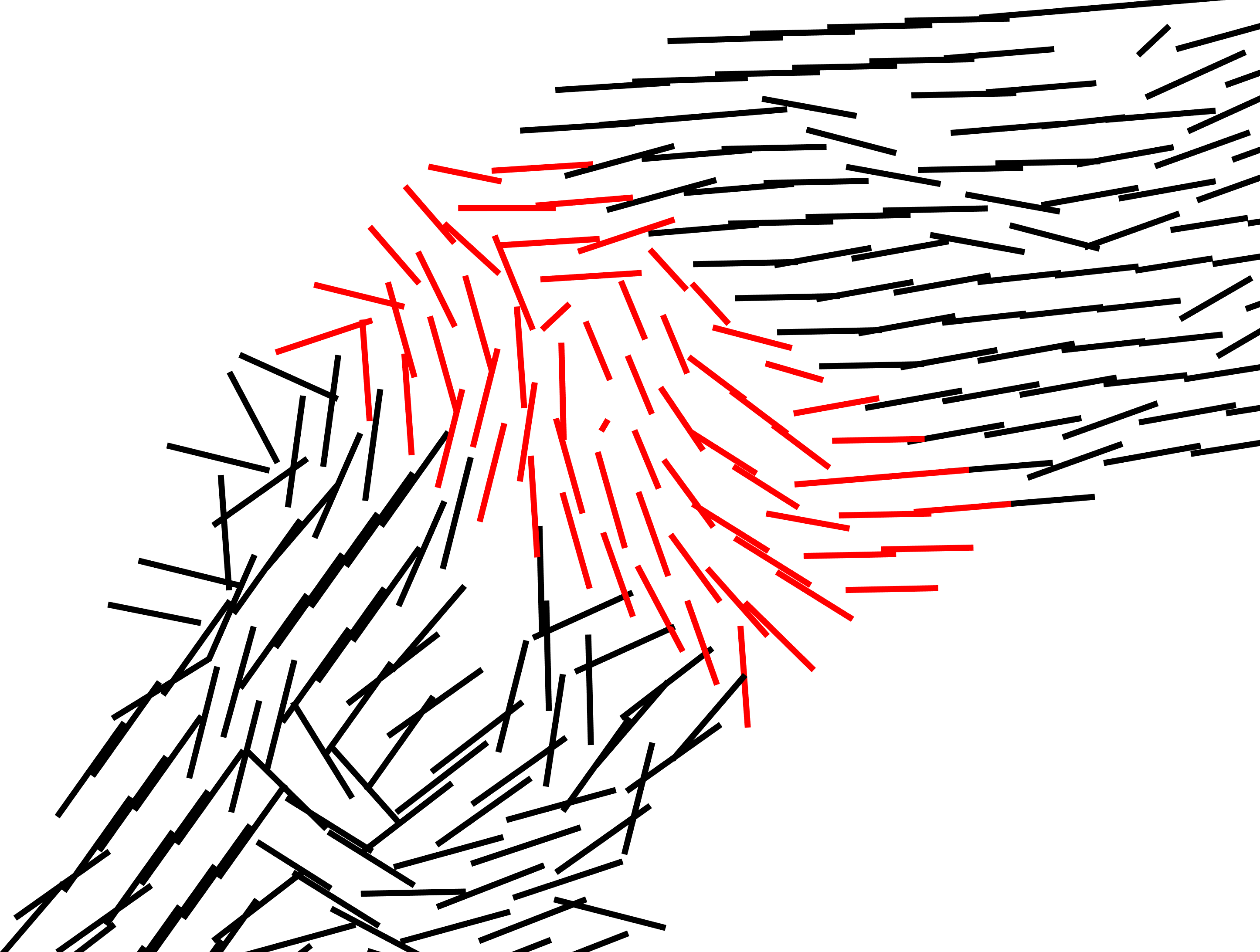}}
  		\centerline{AGS(250,33\%)}\medskip
\end{varwidth}
\begin{varwidth}{0.19\linewidth}
  		\centering
  		\centerline{\includegraphics[width=\linewidth]{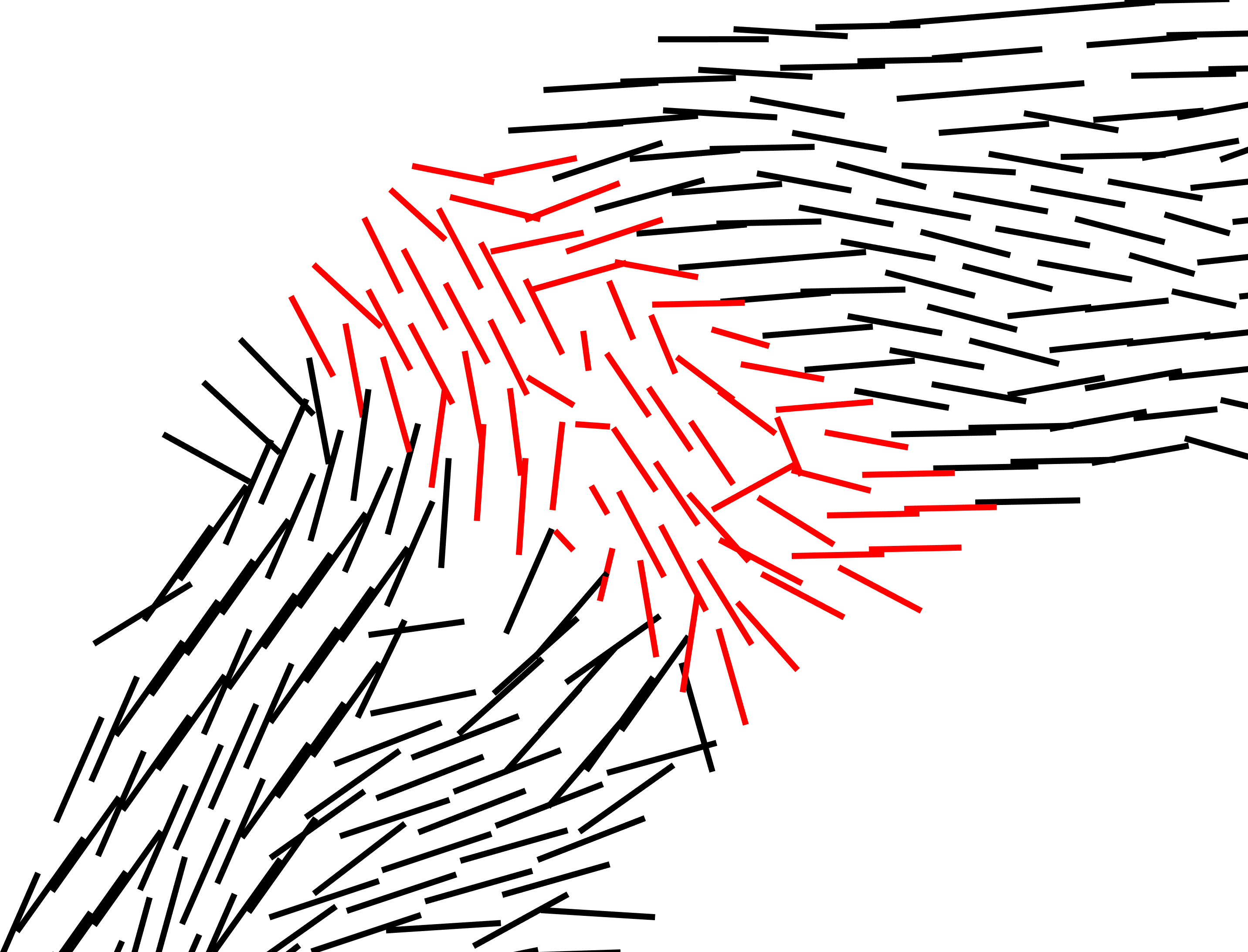}}
  		\centerline{AGS(300,33\%)}\medskip
\end{varwidth}
\\
\begin{varwidth}{0.19\linewidth}
  		\centering
  		\centerline{\includegraphics[width=\linewidth]{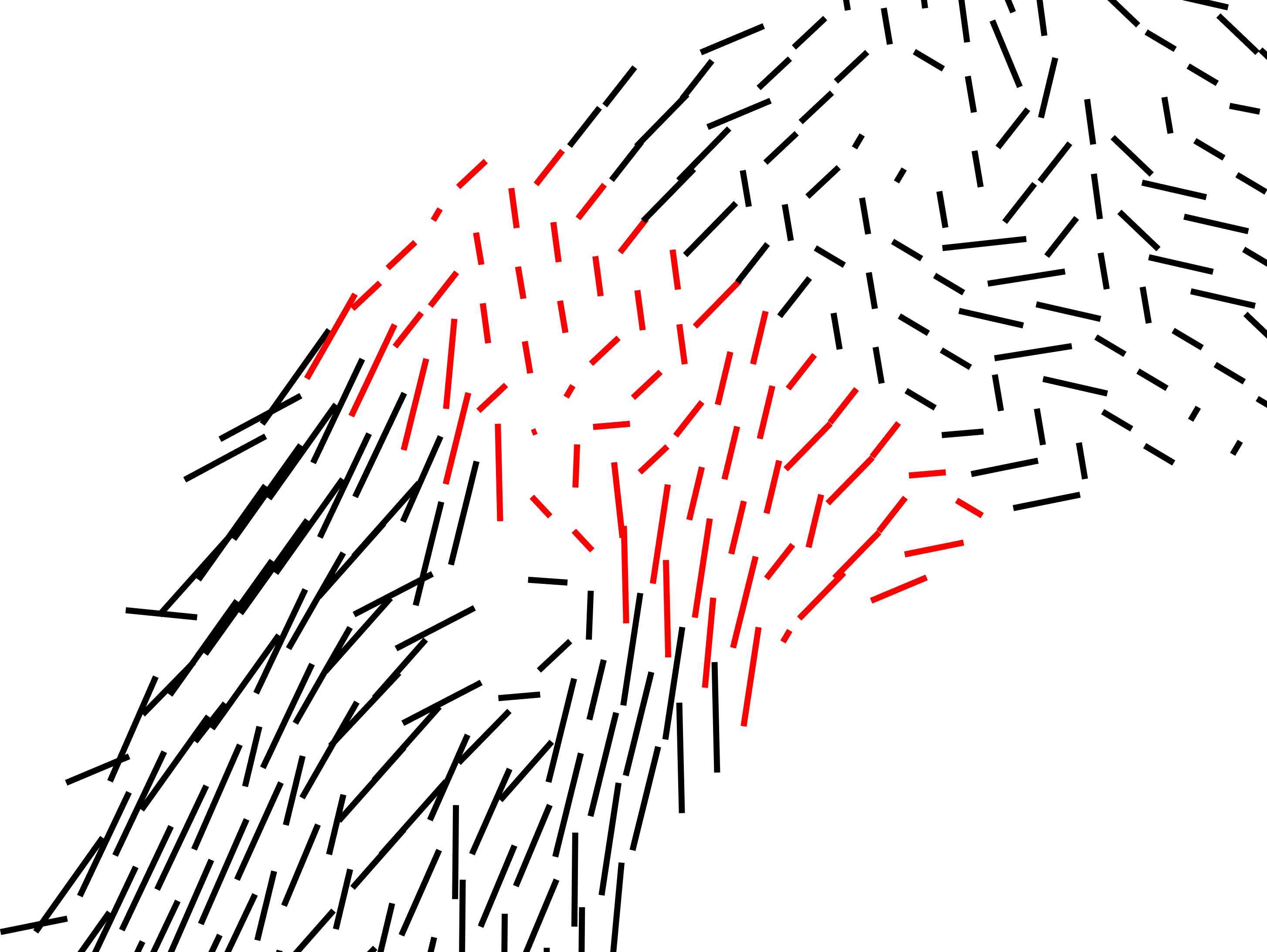}}
  		\centerline{AGS(100,25\%)}\medskip
\end{varwidth}
\begin{varwidth}{0.19\linewidth}
  		\centering
  		\centerline{\includegraphics[width=\linewidth]{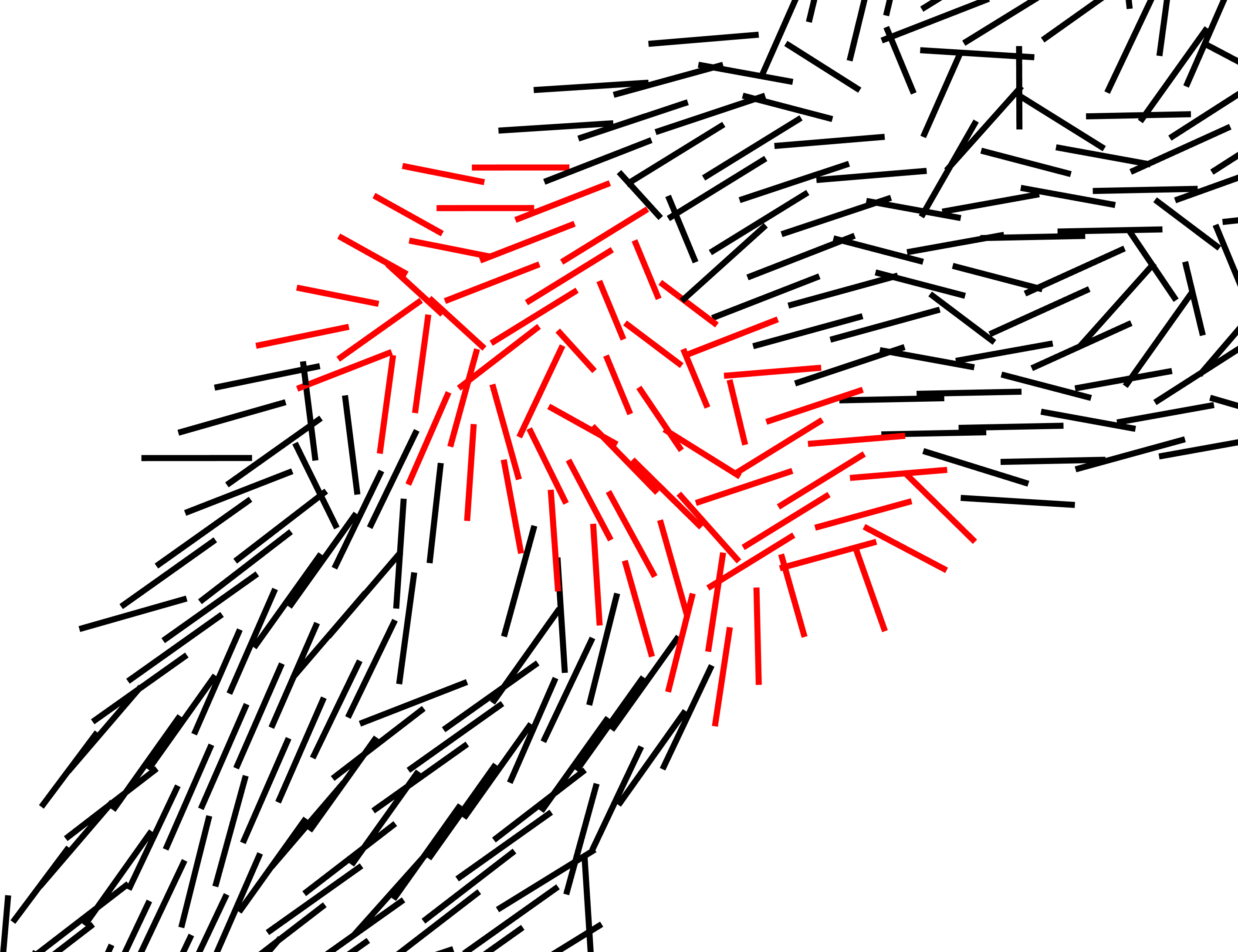}}
  		\centerline{AGS(150,25\%)}\medskip
\end{varwidth}
\begin{varwidth}{0.19\linewidth}
  		\centering
  		\centerline{\includegraphics[width=\linewidth]{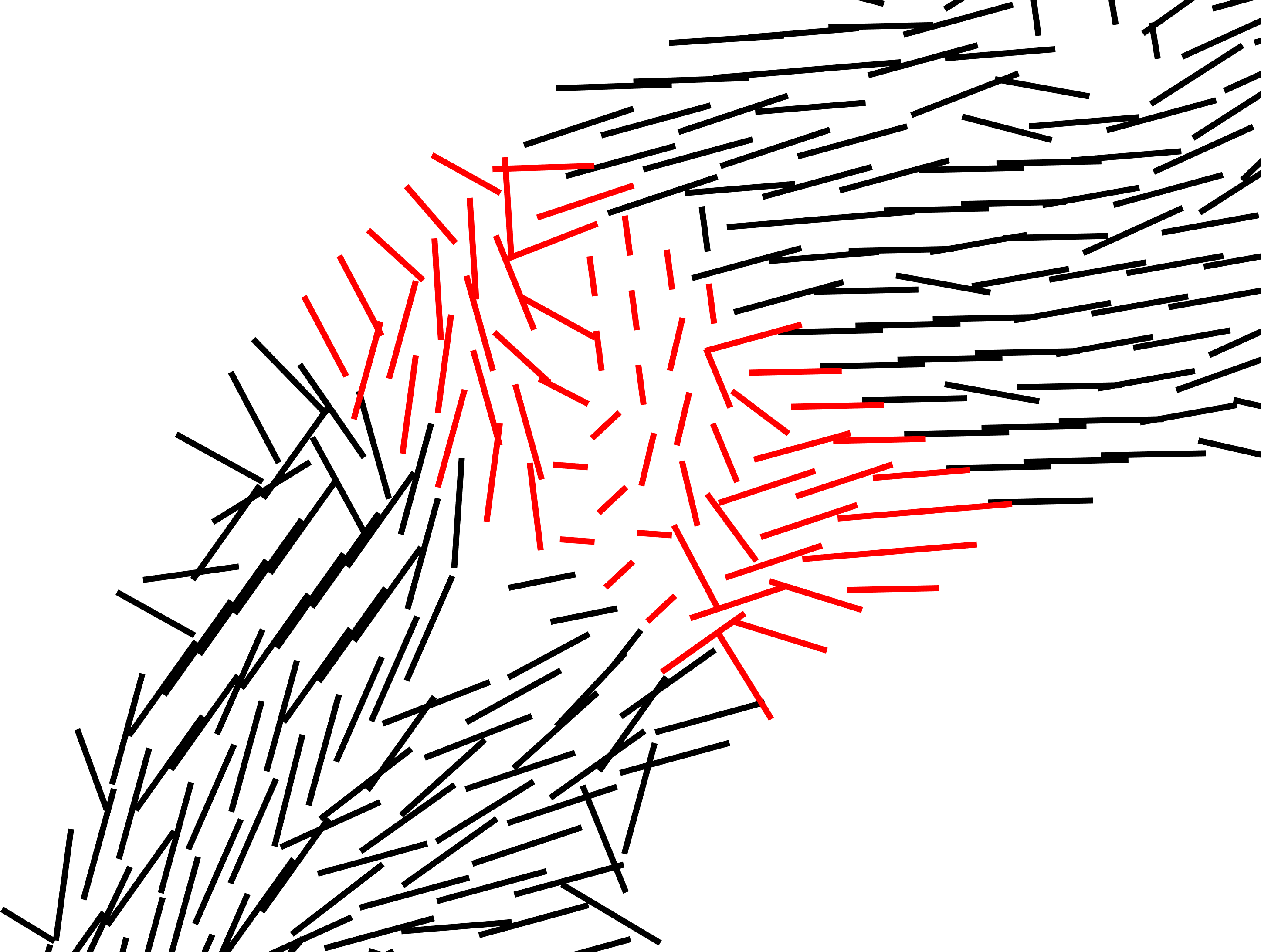}}
  		\centerline{AGS(200,25\%)}\medskip
\end{varwidth}
\begin{varwidth}{0.19\linewidth}
  		\centering
  		\centerline{\includegraphics[width=\linewidth]{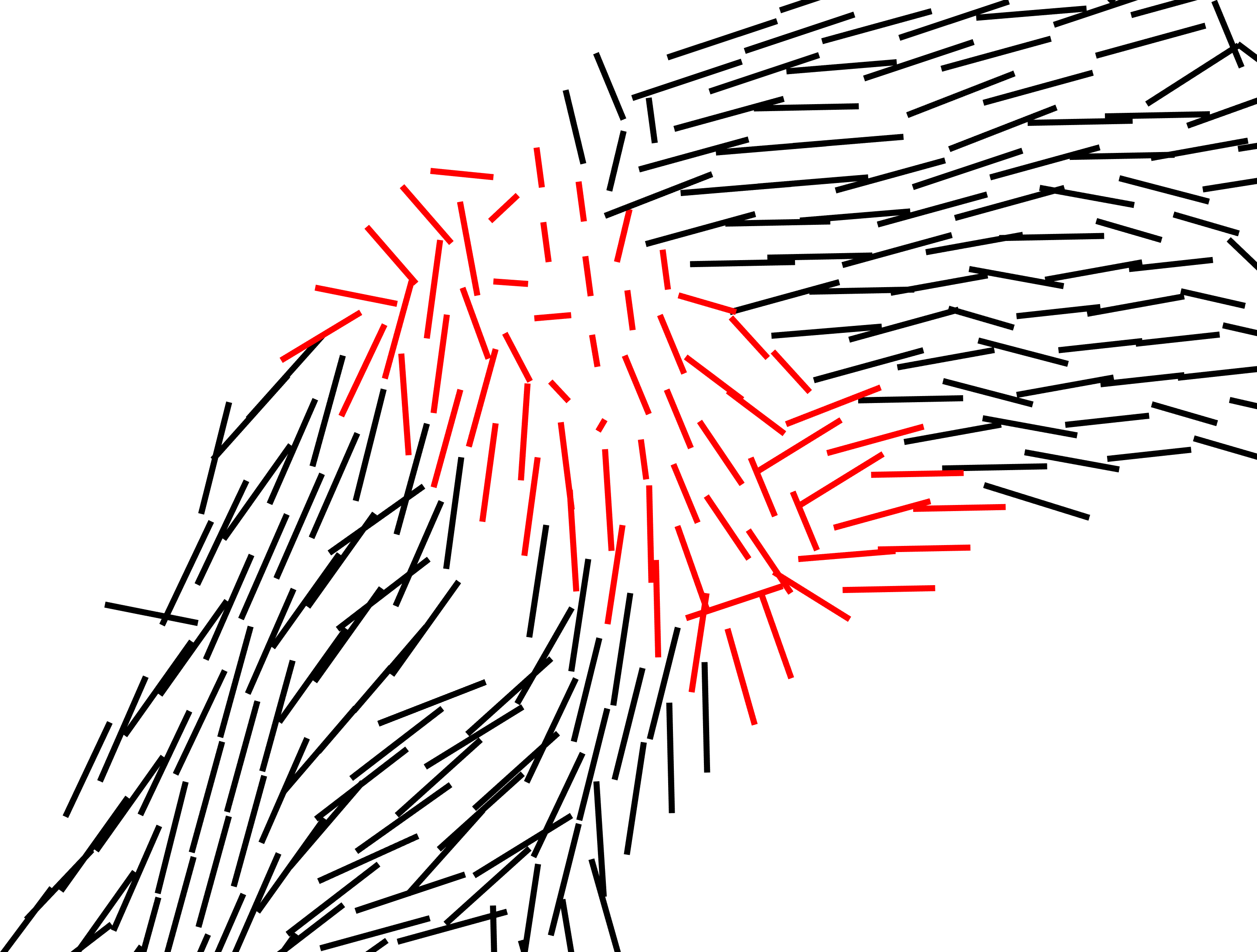}}
  		\centerline{AGS(250,25\%)}\medskip
\end{varwidth}
\begin{varwidth}{0.19\linewidth}
  		\centering
  		\centerline{\includegraphics[width=\linewidth]{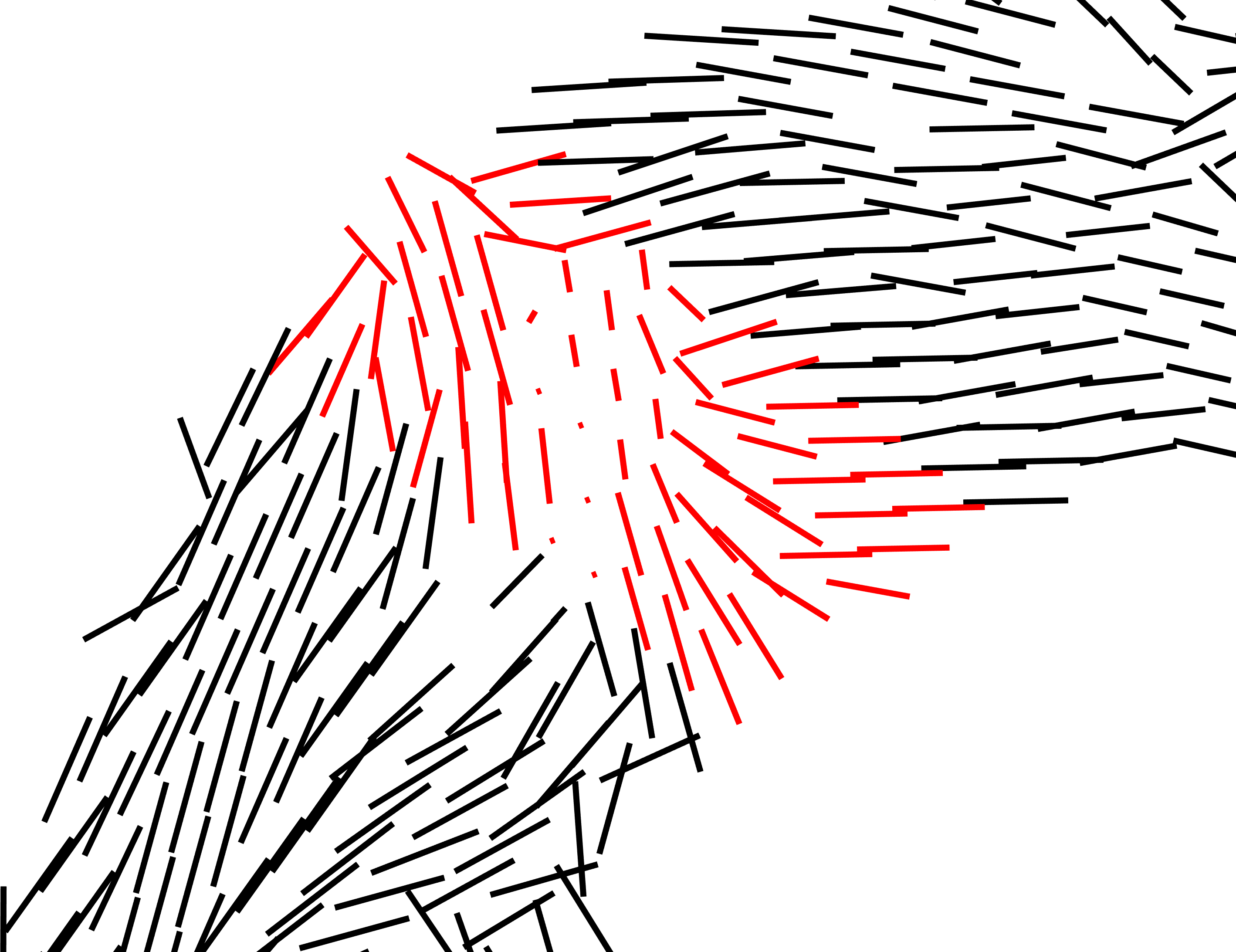}}
  		\centerline{AGS(300,25\%)}\medskip
\end{varwidth}
\\
\begin{varwidth}{0.19\linewidth}
  		\centering
  		\centerline{\includegraphics[width=\linewidth]{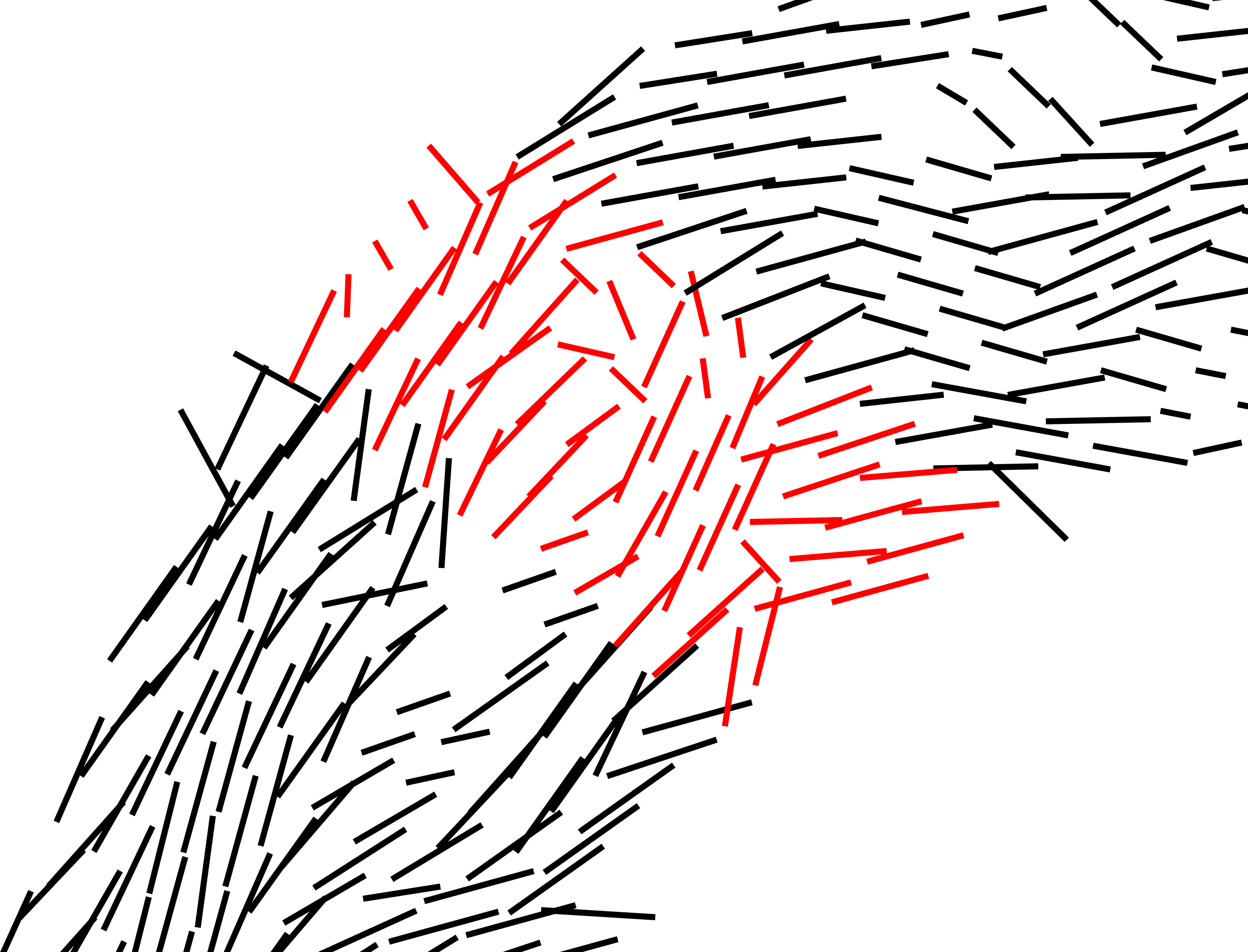}}
  		\centerline{AGS(100,20\%)}\medskip
\end{varwidth}
\begin{varwidth}{0.19\linewidth}
  		\centering
  		\centerline{\includegraphics[width=\linewidth]{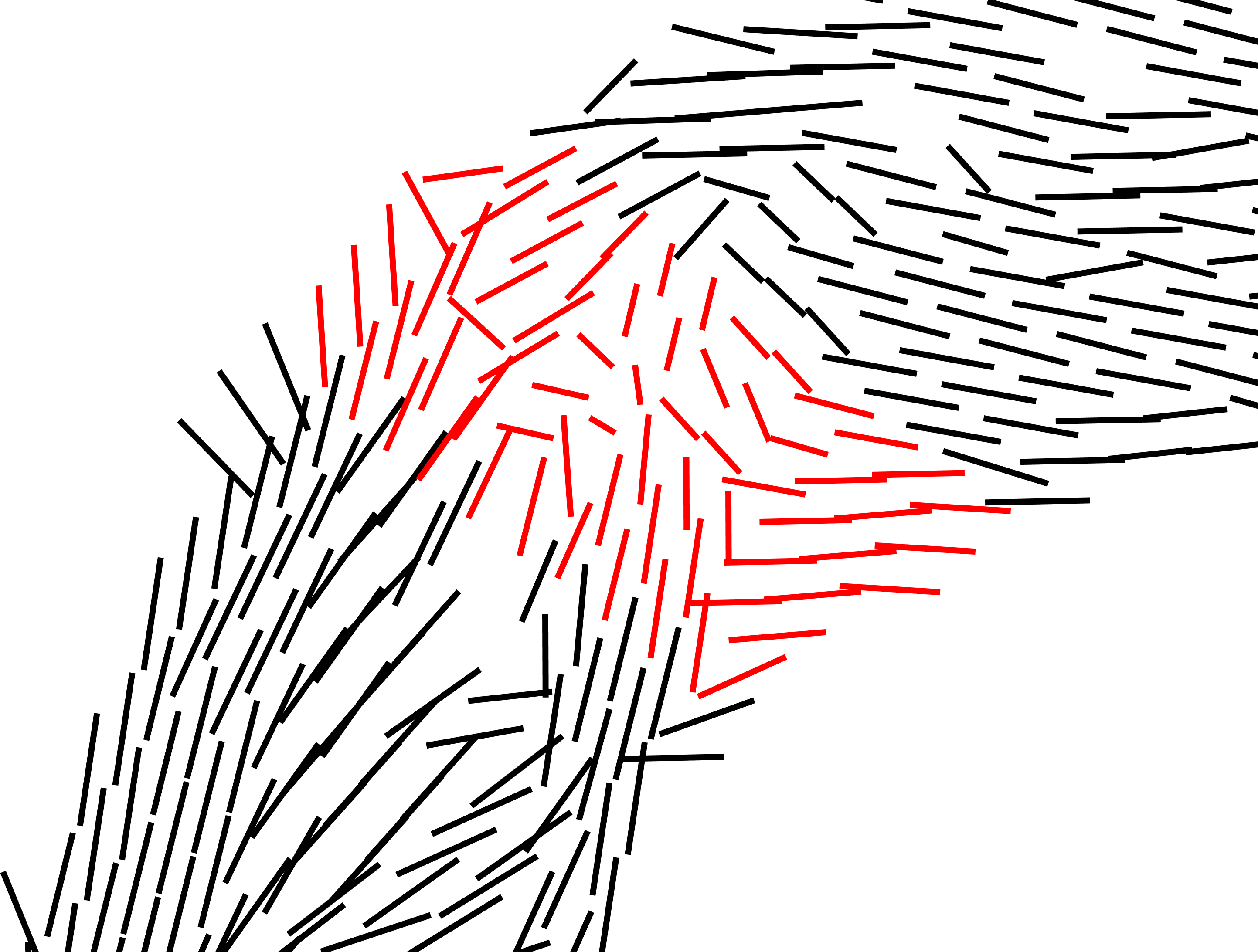}}
  		\centerline{AGS(150,20\%)}\medskip
\end{varwidth}
\begin{varwidth}{0.19\linewidth}
  		\centering
  		\centerline{\includegraphics[width=\linewidth]{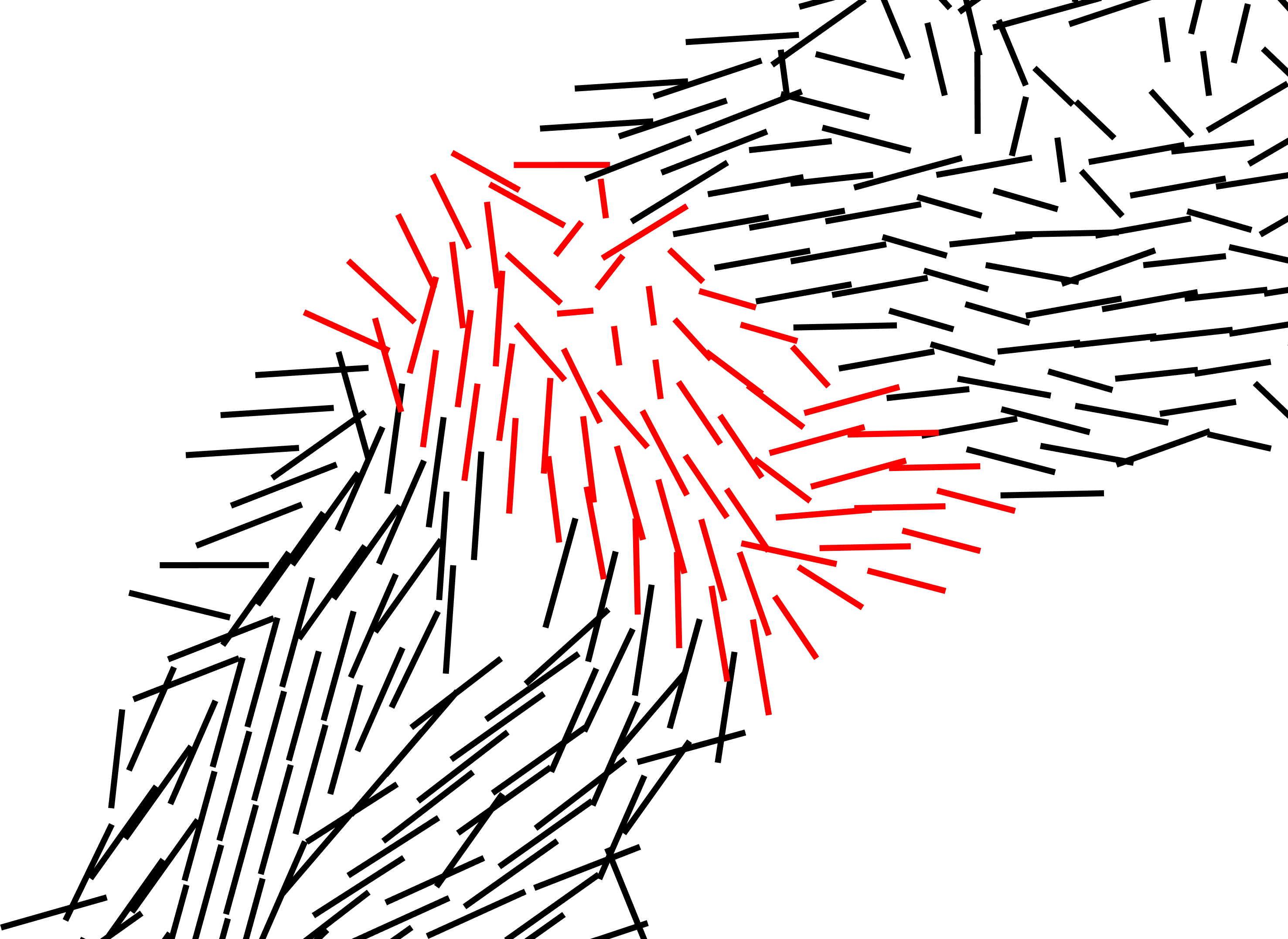}}
  		\centerline{AGS(200,20\%)}\medskip
\end{varwidth}
\begin{varwidth}{0.19\linewidth}
  		\centering
  		\centerline{\includegraphics[width=\linewidth]{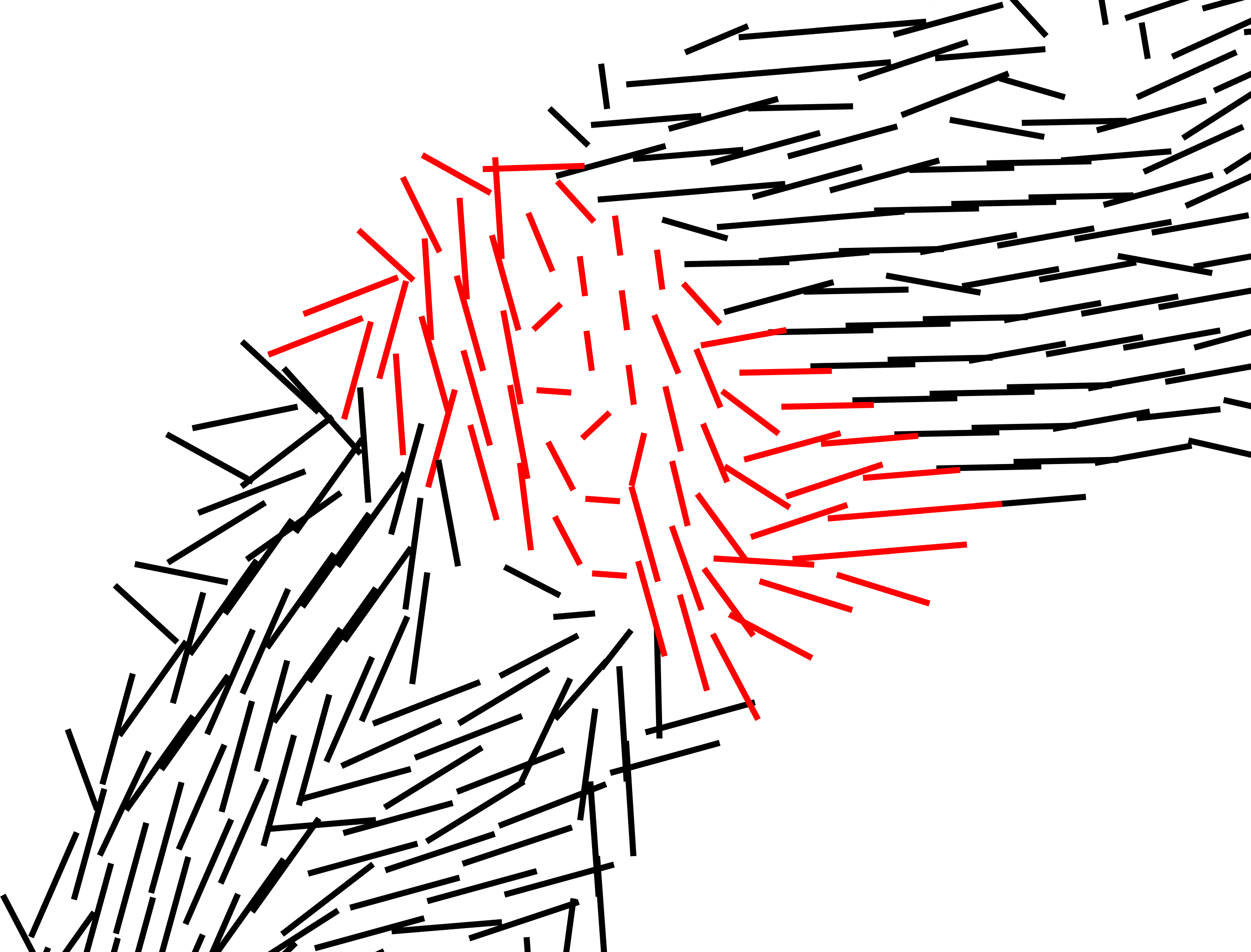}}
  		\centerline{AGS(250,20\%)}\medskip
\end{varwidth}
\begin{varwidth}{0.19\linewidth}
  		\centering
  		\centerline{\includegraphics[width=\linewidth]{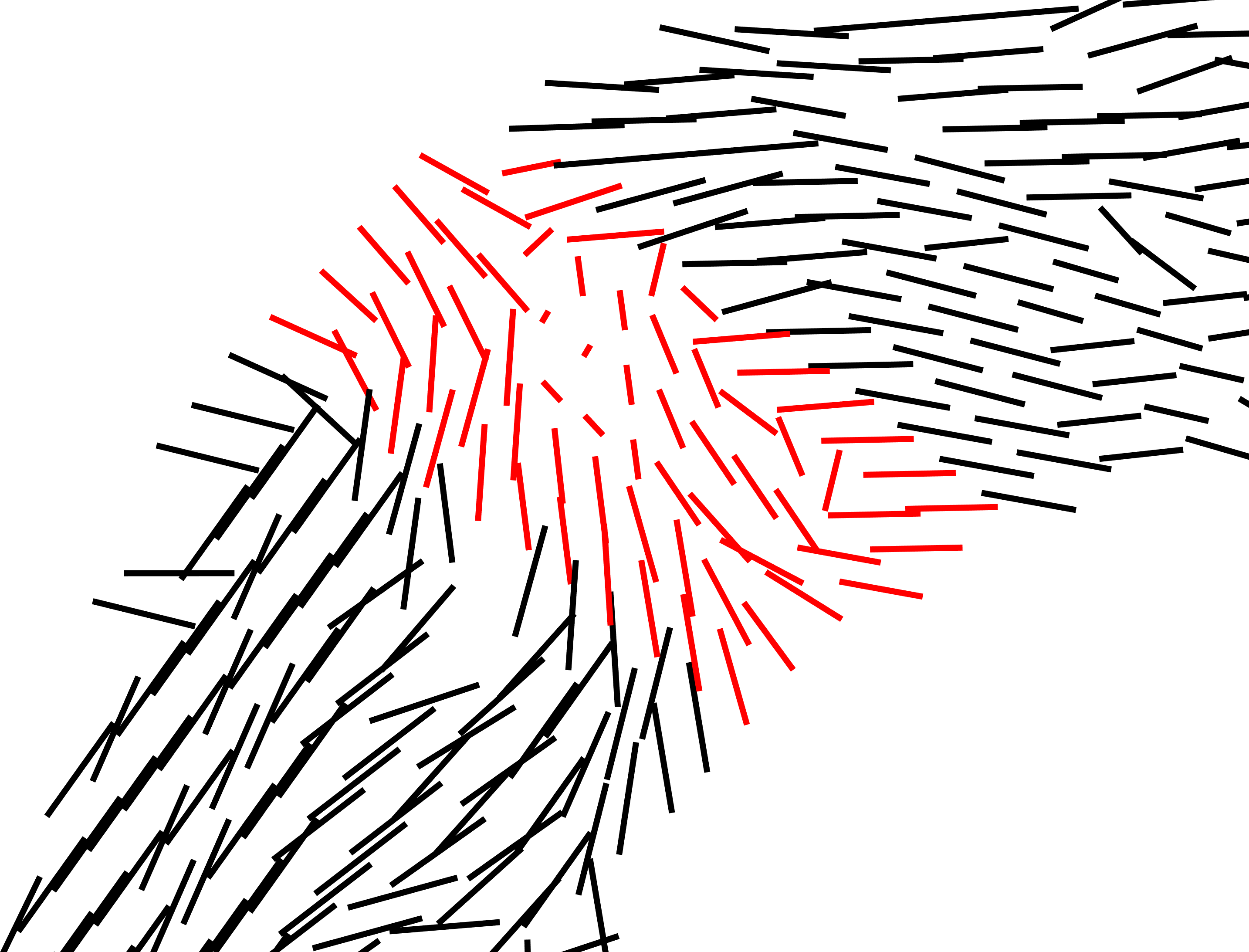}}
  		\centerline{AGS(300,20\%)}\medskip
\end{varwidth}
\\
\begin{varwidth}{0.19\linewidth}
  		\centering
  		\centerline{\includegraphics[width=\linewidth]{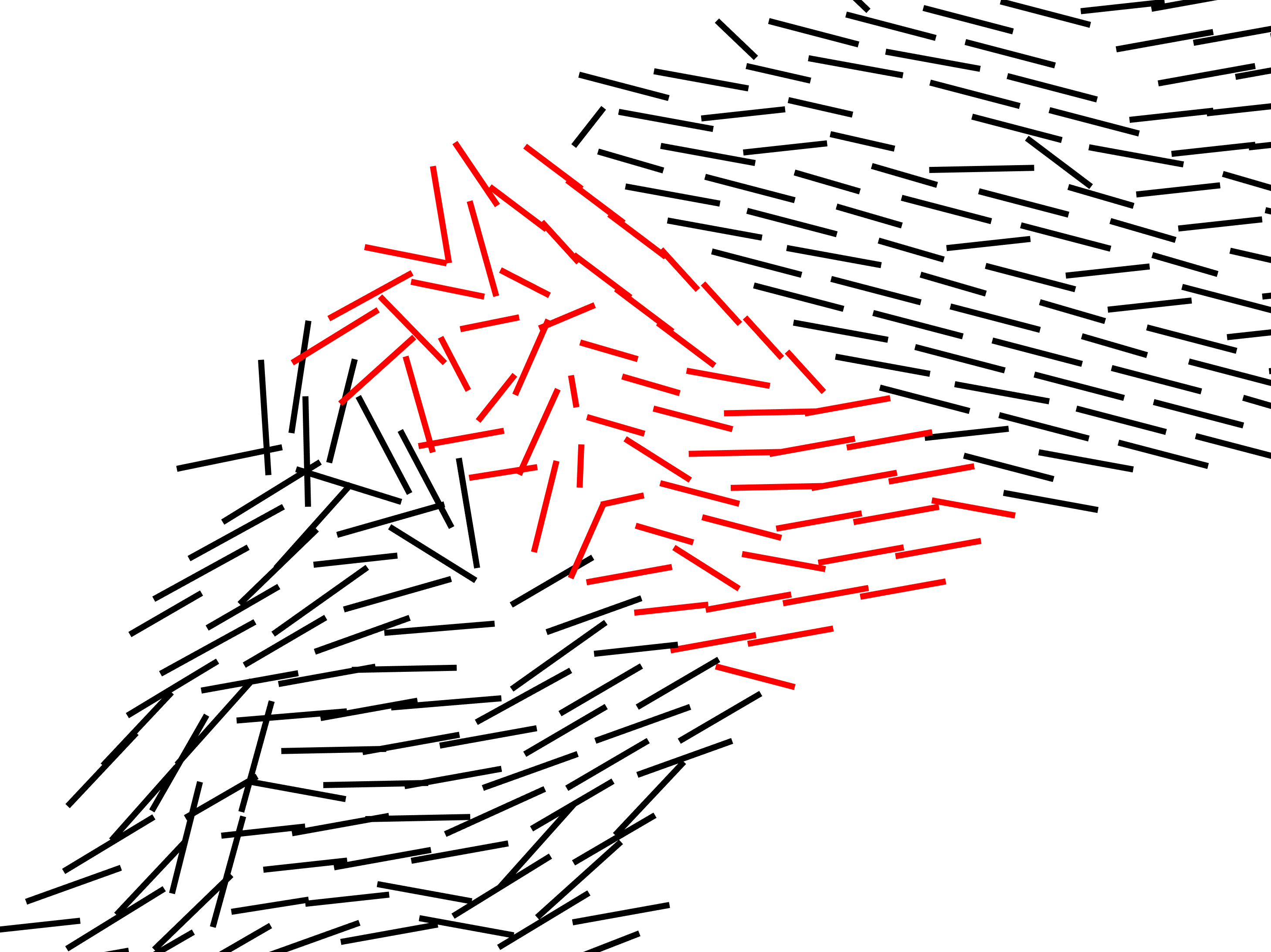}}
  		\centerline{AGS(100,10\%)}\medskip
\end{varwidth}
\begin{varwidth}{0.19\linewidth}
  		\centering
  		\centerline{\includegraphics[width=\linewidth]{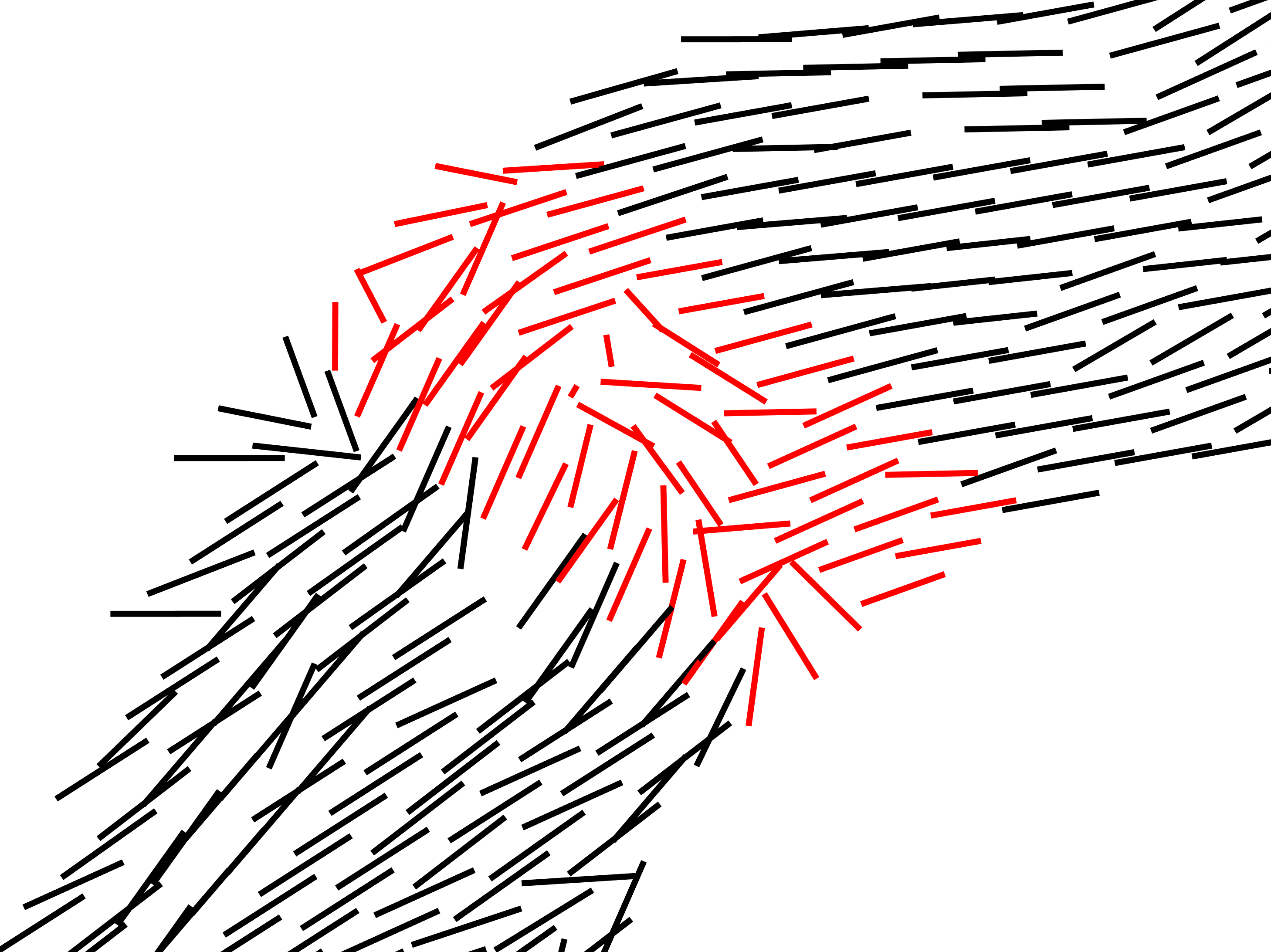}}
  		\centerline{AGS(150,10\%)}\medskip
\end{varwidth}
\begin{varwidth}{0.19\linewidth}
  		\centering
  		\centerline{\includegraphics[width=\linewidth]{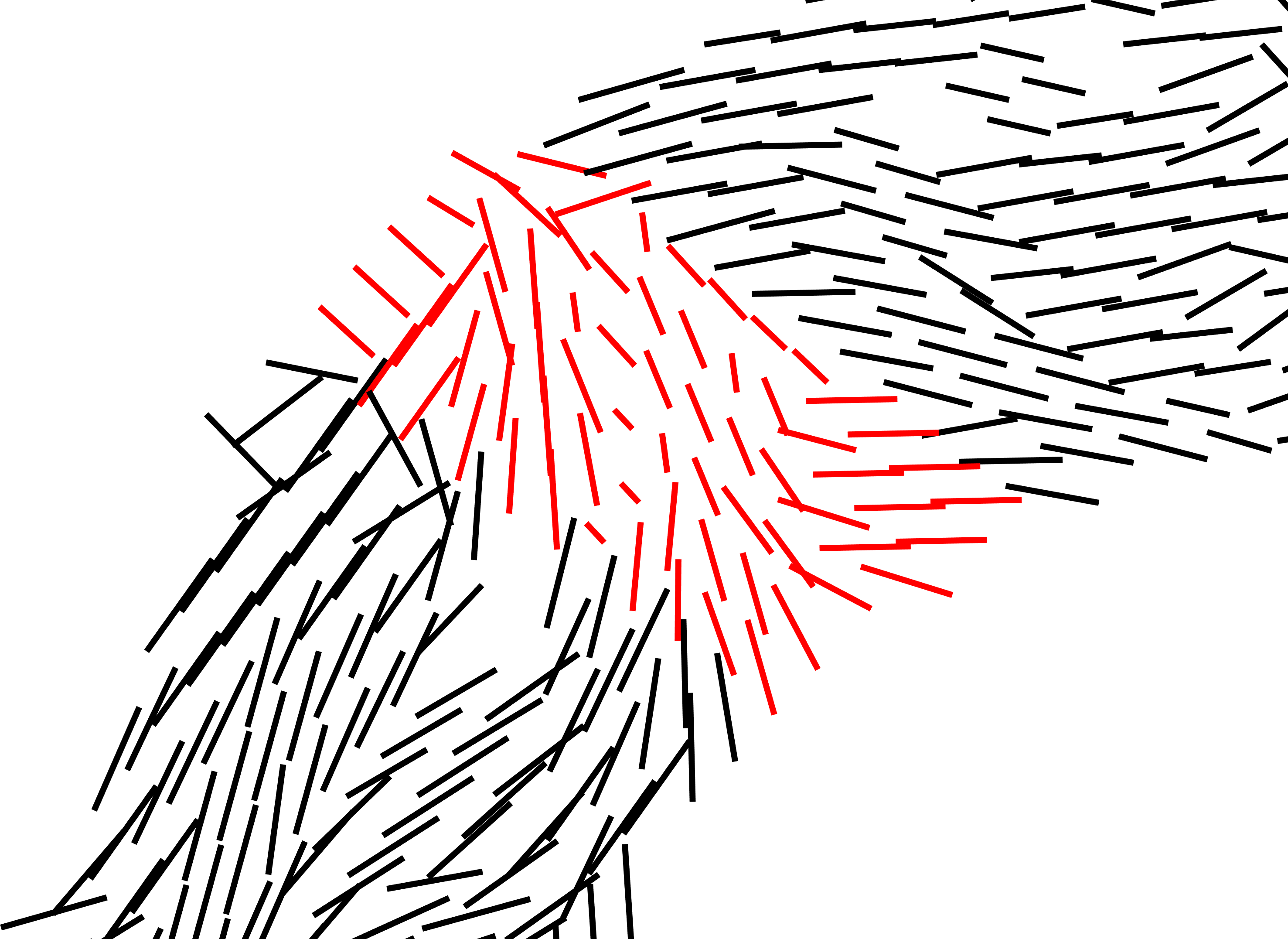}}
  		\centerline{AGS(200,10\%)}\medskip
\end{varwidth}
\begin{varwidth}{0.19\linewidth}
  		\centering
  		\centerline{\includegraphics[width=\linewidth]{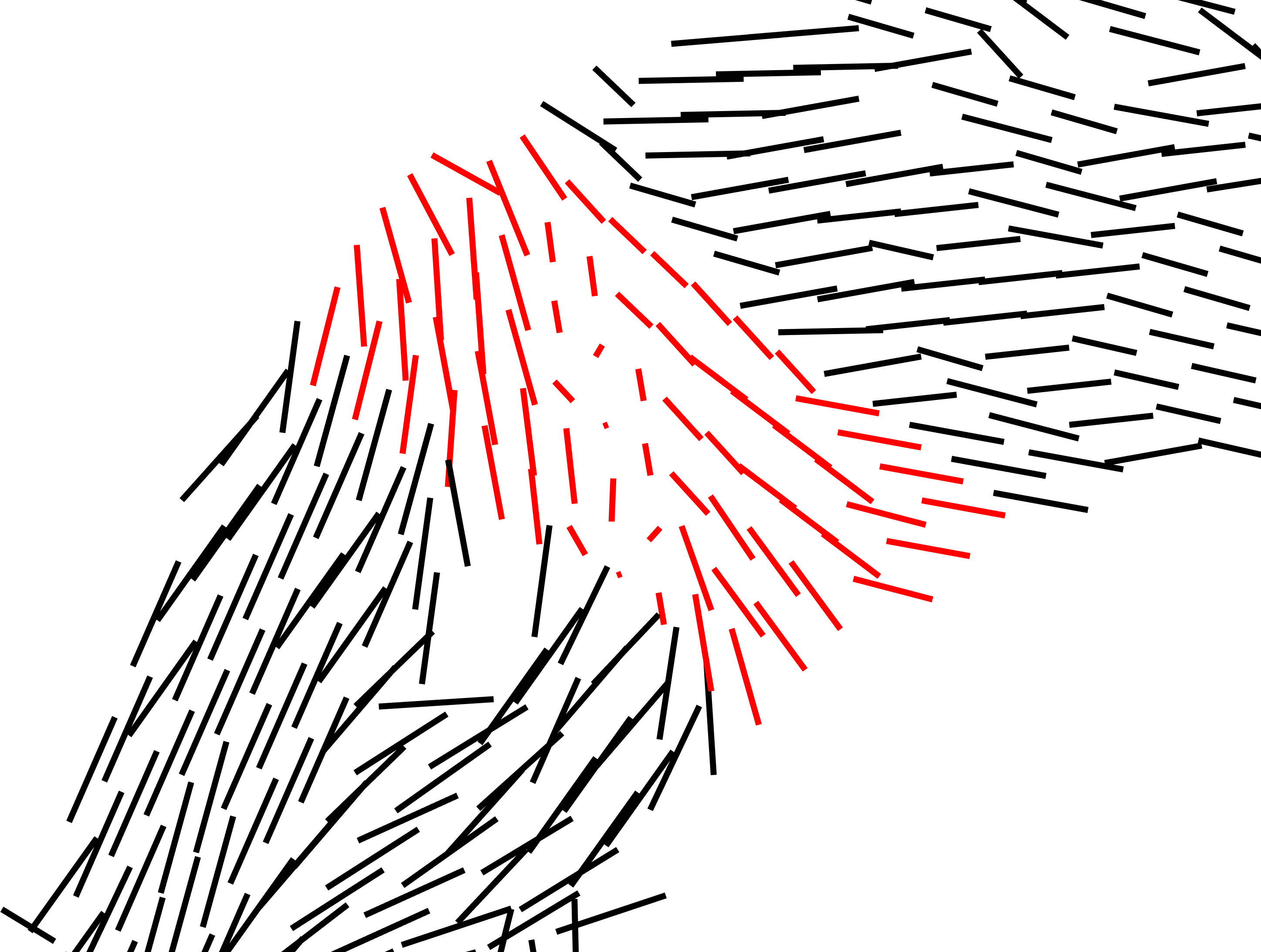}}
  		\centerline{AGS(250,10\%)}\medskip
\end{varwidth}
\begin{varwidth}{0.19\linewidth}
  		\centering
  		\centerline{\includegraphics[width=\linewidth]{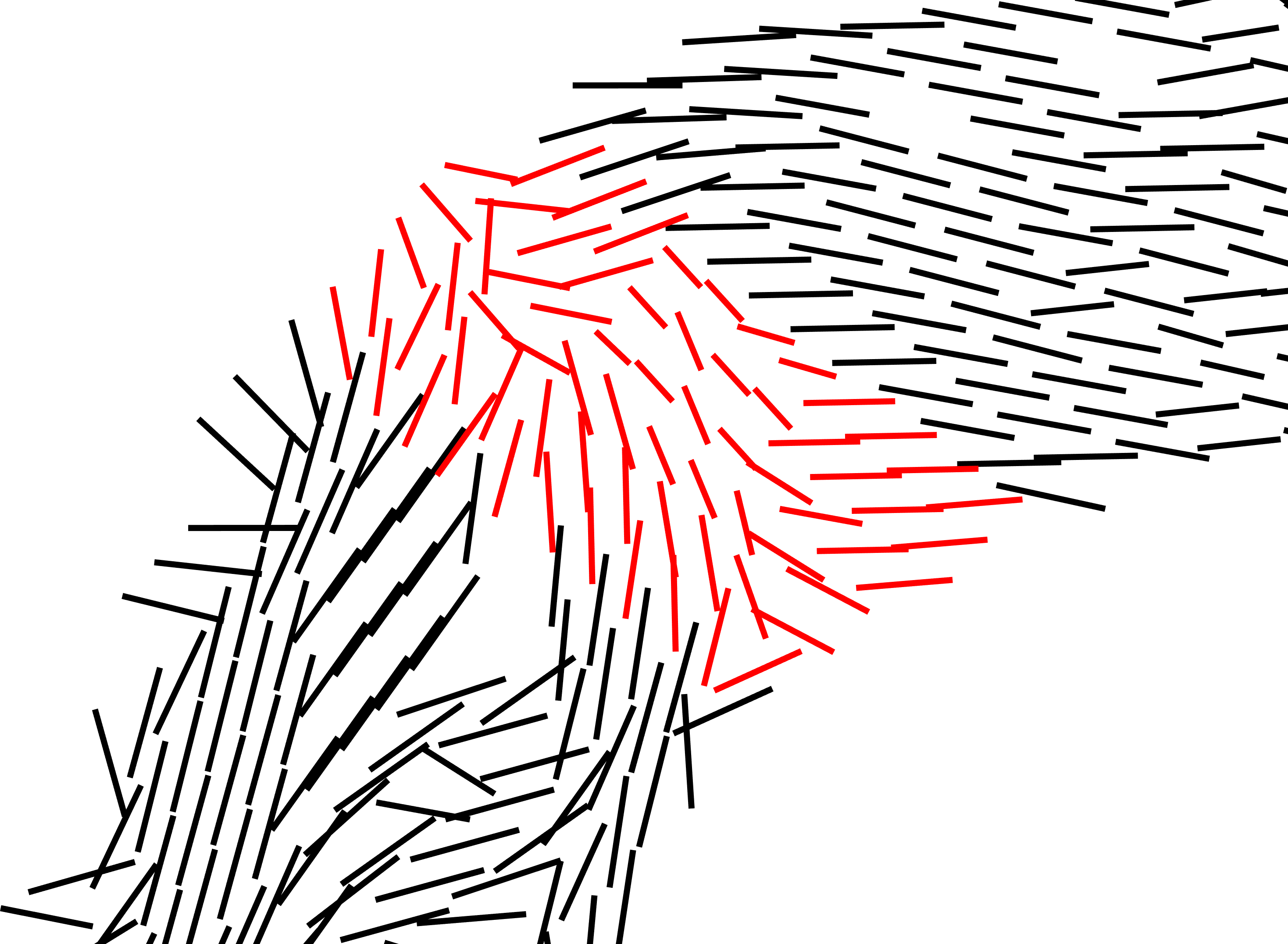}}
  		\centerline{AGS(300,10\%)}\medskip
\end{varwidth}
\caption{\reviewerthree{Results of the experimental data using the thermoplastic fiber mould sample.  
Shown are the \secondreview{main} fiber orientations extracted using AXDT in the region of interest for different acquisition trajectories computed using our proposed algorithm $\text{AGS}(N,b)$, where $N$ is the number of poses in the trajectory, and $bN$ the batch size. The \enquote{Reference} image used the full high-quality trajectory consisting of $966$ poses.}}
\label{fig:cfk_result}
\end{figure*}

Fig.~\ref{fig:cfk_result} and Fig.~\ref{fig:cfk_result_geometries} show a selection of the computed \enquote{optimal} trajectories for different parameters \reviewerthree{$b$ and $N$}, as well as slices of the fiber orientations extracted from the AXDT reconstruction computed from the corresponding acquisition trajectory.
Using the reconstruction of the high-quality trajectory as \enquote{Reference}, we also computed the experimental metric $\text{EM}(\text{AGS}(N,b))$ from eq.~(\ref{eq:expmetric}), to quantitatively evaluate the results of our proposed algorithm, see Table~\ref{table:cfk_em}.
\reviewermain{Additionally, Table~\ref{table:cfk_em} shows the RMSE according to eq.~(\ref{eq:rmse}) between the \enquote{Reference} reconstruction and the other reconstructions, computed for the region of interest (the weld line).}

\begin{table*}[t]
\resizebox{\textwidth}{!}{
\begin{tabular}{| c || c | c || c | c || c | c || c | c || c | c |}
\hline
AGS$(N, b)$ & \multicolumn{2}{c|}{$N=100$} & \multicolumn{2}{c|}{$N=150$} & \multicolumn{2}{c|}{$N=200$} & \multicolumn{2}{c|}{$N=250$} & \multicolumn{2}{c|}{$N=300$} \\
\hline
  	& RMSE 	& EM 		& RMSE 	& EM 		& RMSE 	& EM 		& RMSE 	& EM 		& RMSE 	& EM \\
\hline
$b=100$\%& 0.0549  		& 0.5776			& 0.0525 		& 0.6561 		& 0.0467 		& 0.6855 		& 0.0449 		& 0.6708 		& 0.0437 		& 0.6878 		\\
\hline
$b=50$\% &\textbf{0.0406} & \textbf{0.7509} & 0.0412 			& 0.7313 		& 0.0309 		& 0.8017 		& 0.0373 		& 0.8344 		& 0.0250 		& 0.8799 		\\
\hline
$b=33$\% & 0.0495  		& 0.5400 		& \textbf{0.0337}		& \textbf{0.8095} & 0.0324 		& 0.7925 		& 0.0386 	& \textbf{0.8647} 		& \textbf{0.0220} 		& 0.9076		 \\
\hline
$b=25$\% & 0.0488  		& 0.6085 		& 0.0374 		& 0.7565 		& 0.0275		& 0.8597 		& 0.0301 		& 0.8233 		& 0.0238 		& \textbf{0.9137} \\
\hline
$b=20$\% & 0.0444  		& 0.6669 		& 0.0420 		& 0.7726 		& \textbf{0.0264} 		& \textbf{0.8732} & \textbf{0.0248} 		& 0.8611 		& \textbf{0.0220} 		& 0.8912 		\\
\hline
$b=10$\% & 0.0467  		& 0.7163 		& 0.0389 		& 0.6739 		& 0.0283 		& 0.8668 		& 0.0283 		& 0.8572 		& 0.0254 		& 0.8493 		\\
\hline
\end{tabular}
}
\caption{\reviewerthree{Quantitative results of the experimental data of the thermoplastic fiber mould sample. 
The RMSE is computed according to eq.~(\ref{eq:rmse}) in the region of interest (weld line) between the high-quality \enquote{Reference} reconstruction and reconstructions using $\text{AGS}(N,b)$, where $N$ is the number of poses in the trajectory, and $bN$ is the batch size. 
The experimental metric (EM) is computed according to eq.~(\ref{eq:expmetric}) against the \enquote{Reference} reconstruction.
A value of $\text{EM}=1$ is best, while $\text{EM}=0$ is worst.
Highlighted in bold are the smallest values for RMSE and the highest values for EM for each individual geometry.  
}}
\label{table:cfk_em}
\end{table*}

\section{Discussion}

In this section we discuss the results from the previous section~\ref{sec:exp} as well as their implications for the practical implementation of AXDT.

\subsection{Detectability index with spherical impulses}

Previous works \cite{Stayman2013,Fischer2016} used the detectability index very successfully in scalar-valued conventional CT, where an impulse is a simple Dirac impulse, only encoding the location of the perturbation.
In spherical function-valued AXDT, an impulse has to encode both location and its anisotropy or shape.
Such impulses are shown in Fig.~\ref{fig:impulse}, where (A) and (B) are simple spherical Dirac impulses, and (C) and (D) relate to more complex scattering profiles.
Using these spherical impulses, in Fig.~\ref{fig:detindex_angle_map} we evaluated the detectability index values for a typical AXDT acquisition trajectory, with poses placed regularly on a sphere around the sample.

Fig.~\ref{fig:detindex_angle_map}(A) shows that the detectability index values are highest when the simple spherical impulse (A) is measured from inclination angles $ \phi \approx 90 \degree $, which matches the particular X-ray grating interferometer configuration that measures the strongest dark-field signal from these angles. 
For more extreme inclination angles towards the north and the south poles ($ \phi \to 0 \degree $) the detectability index decreases to almost $ 0 $, which again matches 
experimental data \cite{Vogel2015}. 
A matching effect is obtained for the same impulse rotated by $ 90 \degree $, see Fig.~\ref{fig:detindex_angle_map}(B). 
The two more elaborate \secondreview{and realistic} impulses (C) and (D) generate more complex detectability index maps, see Fig.~\ref{fig:detindex_angle_map}(C) and (D), but still matching experimental observations \cite{Wieczorek2016}.

Hence we conclude that the detectability index derived in eq.~(\ref{eq:detindex}) is also a useful metric in spherical function-valued AXDT, providing an accurate relative estimate of a feature's visibility that can be used to find \enquote{optimal} trajectories containing valuable acquisition poses.

\subsection{Algorithm validation with simulated data}

\reviewerthree{Using phantoms containing a single impulse, i.e. (A) and (B)} from Fig.~\ref{fig:impulse}, we now evaluated our proposed Algorithm~\ref{alg:pathopt}, the \textit{Accelerated Greedy Search} (AGS) using simulated data of \reviewerthree{those phantoms}.
Comparing to a \enquote{Reference} trajectory containing $589$ poses, we computed optimized trajectories $\text{AGS}(N,b)$ with a size $N=50$, where the batch size $b$ varied from \reviewerthree{$2\%$ to $100\%$}, see Table~\ref{table:simulation_geometry} \reviewerthree{for results using (B)}.

The reconstructions from those trajectories, see Table~\ref{table:simulation_reconstruction}, yield qualitatively similar results for the small batch sizes, with comparable values of RMSE as well as the sum of the detectability index values, see Fig.~\ref{fig:simulation_results}.
The bigger batch sizes \reviewerthree{($b=50\%,100\%$)} result in qualitatively better reconstructions, which is reflected in lower RMSE values,  despite having a lower detectability index overall (see Fig.~\ref{fig:simulation_results}), the optimum here being \reviewerthree{$b=50\%$} for impulse (B). 

The \enquote{non-optimal} trajectory (choosing poses with the lowest detectability index) and the \enquote{circular} trajectory (similar to conventional CT trajectories) yield the worst results, with the spherical impulse (B) reconstructed in completely the wrong orientation (see Table~\ref{table:simulation_reconstruction} and Fig.~\ref{fig:simulation_results}), while the \enquote{t-design}  trajectory with its uniform acquisition pose distribution yielded qualitatively good results, but quantitatively inferior to the optimized trajectories.

\reviewerthree{For impulse (A), the lowest reconstruction error was achieved when $b=100\%$. 
However, for this impulse the \enquote{Circular} trajectory performed much better than any other geometry, as  impulse (A) aligns exactly with the grating sensitivity of the simulated setup, and thus the \enquote{Circular} trajectory matches all the poses with the highest detectability index from Fig.~\ref{fig:detindex_angle_map}.
The generic \enquote{t-design} trajectory performs worse compared to the optimized trajectories for this impulse as well.}

\secondreview{For the complex spherical scattering profile (D), which was generated by two fibers, the highest combined EM score of the two extracted fibers was achieved for $b=50\%$.}

\secondreview{We conclude that our proposed AGS algorithm performs well in simulation both in a simple setting and  for more complex realistic scattering profiles, with the sorted batches not only drastically reducing the computational complexity of the algorithm, but also having beneficial effects on the reconstruction quality.}

\subsection{Algorithm validation with experimental data}

We studied a thermoplastic fiber mould sample with a notable weld-line feature, see Figs.~\ref{fig:cfkrender} and \ref{fig:cfk_render_slice}, to investigate the performance of our proposed algorithm in an experimental setting. 
Using a high-quality \enquote{Reference} trajectory with $966$ acquisition poses and the weld-line feature as a task, we computed optimized trajectories $\text{AGS}(N,b)$ containing $N=100$ up to $N=300$ poses, i.e. only a fraction of the poses of the high-quality trajectory, with varying batch sizes \reviewerthree{$bN$}.
The quantitative trends of the EM metric eq.~(\ref{eq:expmetric}) in \reviewerthree{Table ~\ref{table:cfk_em}} indicate that more acquisition poses $N$ yield better image quality \reviewerthree{(see the general increasing trend of EM values from left to right in the table)}, which is no surprise and is also qualitatively corroborated in Fig.~\ref{fig:cfk_result}.

More notably, the batch size \reviewerthree{parameter} $b$ of algorithm AGS plays \reviewerthree{a major role for} both the computational performance and the quality of the resulting reconstruction.
Previous algorithms correspond to our method with \reviewerthree{$b=\frac{1}{N}$ (in other words a batch contains only one pose)}, which have proven to be extremely computationally expensive in case of AXDT \cite{Boghiu2019}, while batch sizes \reviewerthree{$b>\frac{1}{N}$ lead to a $bN$-fold reduction in} computational complexity.
However, the fastest variant with \reviewerthree{$b=100\%$}, where only one iteration of AGS is run and then the $N$ poses with the highest detectability index are picked, restricts the generated trajectory to a cluster of poses from a similar perspective, with a hit on image quality, \reviewerthree{see Fig.~\ref{fig:cfk_result_geometries} where the clustering effect can be observed for $\text{AGS}(200,100\%)$}.  

\reviewerthree{Choosing smaller batch sizes, for example $20\% \leq b \leq 33\%$}, yields more algorithm iterations and trajectories with better coverage of the sample, while still reducing computational complexity drastically. 
The trends in \reviewerthree{Table~\ref{table:cfk_em}} indicate that in our experiment a \reviewerthree{smaller batch size in this $20\%$ to $33\%$ range} appears to be performing well consistently, which is also qualitatively confirmed in Fig.~\ref{fig:cfk_result}.

Overall, we conclude that it is possible to compute \enquote{optimal} trajectories with a fraction of the acquisition poses while still yielding comparable image quality, as evidenced, for example, by \reviewerthree{$\text{AGS}(300,25\%)$}, which uses less than a third of the poses of the high-quality trajectory \reviewerthree{ and is obtained after only 4 iterations of AGS}.

\subsection{Region-based user-defined tasks}

In our experiment with the thermoplastic fiber mould, the weld-line feature spanned a region of interest comprising of $288$ voxels, as marked in red in Fig.~\ref{fig:cfk_render_slice}.
Previous works, such as \cite{Stayman2013,Fischer2016} computed the detectability index only for a single location but multiple times for different locations, which in our experiments did not yield good results for AXDT.

Similar to Stayman et al. \cite{Stayman2019}, who explored the computation of the detectability index over a region of interest using three different choices of measure, mean, median, and minimum, we opted for a mean detectability approach defining the impulse directly as the whole region of interest. 
This has the benefit of computing the detectability index value for the whole region of interest only once, and directly getting an estimate for the whole region without the need of separately computing it for each voxel and then estimating the mean value for the index.

As the detectability index quantifies how good the spectrum of a signal can be resolved from the noise, using the whole region of interest as an impulse yields a good approximation for the local resolution properties, and also encodes complex spatial information about the whole region in one single detectability index value.

\subsection{Future work}

We demonstrated that the non-prewhitening matched filter observer (NPWM) \reviewerthree{is a useful predictor for task-based observer performance for} the spherical function-valued AXDT reconstruction problem. 
\reviewerthree{Nevertheless, it should be worthwhile to study more suitable observer models for AXDT detection tasks, such as the channelized hotelling observers \cite{Richard2008, Brankov2013} or prewhitening model observers, both of which could be extended with filters in the spherical function domain, similar to the eye filters used for conventional X-ray tomography applications. 
Studying other observer models} that may suit the complex nature of AXDT even better are subject of future work.

The statistical model of the anisotropic dark-field signal that is assumed in eq.~(\ref{eq:likelihood}) is an approximation, and more accurate models have been described in Schilling et al.~\cite{Schilling2017}.
Once those more accurate models also have matching practical reconstruction algorithms, the resolution properties would have to be re-derived for those models.

A limitation of our proposed algorithm is the requirement of prior knowledge.
In AXDT in particular, CAD models as prior knowledge as in \cite{Fischer2016}, are not applicable.
What is applicable though, is a prior high-quality scan, similar to \cite{Stayman2013}, as was done in our thermoplastic fiber mould experiment.
For industrial samples, for example in quality control, this is a very feasible scenario.
For potential medical applications \cite{Wieczorek2018}, a prior high-quality scan might not exist, so a hybrid approach of using a short generic scan to generate prior knowledge, and then running the proposed algorithm on top of that to generate more optimal poses, might be required.

With the increasing viability of deep learning techniques in conventional X-ray CT, a data driven approach could provide novel solutions to estimate the resolution properties, see for example Gang \textit{et al.}~\cite{Gang2019}.

\section{Conclusion}

In this work we propose a detectability index for AXDT along with a batched trajectory optimization algorithm, in order to address the required long acquisition times and corresponding high radiation dose. 
We validated the detectability index and the trajectory optimization algorithm using simulations and demonstrated their good performance on experimental data of thermoplastic fiber mould sample.
This approach of optimizing an acquisition trajectory using a task-based performance metric appears very promising to moving AXDT towards more practical applications.

\section*{Acknowledgment}
The authors thank O. Focke (Faserinstitut Bremen), S. Zabler (Fraunhofer EZRT) and M. Willner (MITOS GmbH) for providing the sample used in this study. 
We acknowledge financial support through the Munich-Centre for Advanced Photonics (MAP), the DFG (Gottfried Wilhelm Leibniz program) and the European Research Council (AdG 695045). 
This work was carried out with the support of the Karlsruhe Nano Micro Facility (KNMF, www.kit.edu/knmf), a Helmholtz Research Infrastructure at Karlsruhe Institute of Technology (KIT).

\renewcommand{\bibfont}{\footnotesize}
\printbibliography

\end{document}